\documentclass{article}
\usepackage{jheppub}
\usepackage{graphicx} % Required for inserting images

\usepackage{appendix}
\usepackage{amsmath}
\usepackage{amssymb}
\usepackage{amsfonts}
\usepackage{braket}
\usepackage{leftindex}
\usepackage{hyperref}
\usepackage %[sort&compress,elide, numbers, square]
{natbib}

\usepackage{xcolor}

\newcommand{\gfrak}{\mathfrak{g}}
\newcommand{\pfrak}{\mathfrak{p}}

\newcommand{\hfrak}{\mathfrak{h}}

\newcommand*{\rbb}{\mathbb{R}}
\newcommand*{\cbb}{\mathbb{C}}
\newcommand*{\hbb}{\mathbb{H}}

\newcommand{\ccal}{\mathcal{C}}

\newcommand{\kkcal}{\mathcal{K}}

\newcommand*{\acal}{\mathcal{A}}
\newcommand*{\bcal}{\mathcal{B}}

\newcommand{\p}{\partial}
\newcommand{\s}{\Sigma}
\newcommand{\Tr}{\text{Tr}}

\title{Chern-Simons corner phase spaces in BF-BB theories}
\author{Simon Langenscheidt}
\affiliation{Perimeter Institute, 39 Caroline St. Waterloo, Ontario, Canada}
\emailAdd{slangenscheidt@perimeterinstitute.ca}

\date{\today}
\abstract{
    We investigate an approach to determine corner Poisson brackets of fields restricted to codimension 2 and 3 surfaces in 4D theories, without making use of boundary conditions. Instead, within the example of BF-BB type theories, we show that the constraint/generator algebra on a partial Cauchy slice is enough to determine the Poisson brackets and a corner symplectic form, useful in extended phase space constructions or holography. The codimension 2 phase space turns out to be of Chern-Simons type, but lacking a flatness constraint. In a setting with codimension 3 surfaces, there is also a further Wess-Zumino-type current algebra. We apply this approach also to a specific formulation of full 4D gravity, based on the Maxwell algebra $\gfrak=\mathfrak{so}(1,3)\ltimes(\rbb^{1,3}\tilde\oplus \mathfrak{so}(1,3)^\ast)$. This realises the corner Poisson bracket of the spin connection for the first time and shows it is off-shell commutative, while the corner metric is noncommutative.
}

\begin{document}

\maketitle

\tableofcontents
\section{Introduction}
The purpose of this paper is twofold. \\
One motivation for this work is the observation that regardless of boundary conditions, the kinematics of boundary degrees of freedom in a gauge theory, meaning a `corner phase space' and its Poisson brackets, seem to always follow from the bulk's Poisson brackets and the constraints it is subject to. \\
Boundary degrees of freedom and their dynamics are known under many names, such as `singletons'~\cite{maldacena_d-brane_2001} in the context of AdS-type spacetimes, `boundary gravitons', `soft gravitons' or `reparametrisation modes' in the context of gravity~\cite{carlipDynamicsAsymptoticDiffeomorphisms2005,cotler_theory_2019,nguyenCelestialIRDivergences2021,Asante:2018kfo,mertensSolvableModelsQuantum2023,rielloQuantumEdgeModes2018,joungExtractingEdgeModes2026}, or `edge modes' both in the context of condensed matter systems~\cite{balachandranConformalEdgeCurrents1992,cohenNewPlatformEdge2018,asbothShortCourseTopological2016} such as topological insulators, or gauge theories more generally~\cite{Ball:2024hqe,ballDynamicalEdgeModes2024,Carrozza:2021gju,carrozzaEdgeModesDynamical2024,blommaertEdgeDynamicsPath2018,geillerExtendedActionsDynamics2020,araujo-regadoSoftEdgesMany2025}. What they all have in common is that they arise in the context of boundary-supported gauge transformations, and their kinematics seem to be tied to the gauge structure of the theory. \\
This splits the study of boundary degrees of freedom into two parts: one given by a kinematics, which appears to be determined by the choice of gauge group, and another given by the dynamics, for example through a boundary condition on the bulk, or equivalently a Hamiltonian on the phase space of the boundary degrees of freedom.\\
\textit{In this paper, we argue for this observation by explicitly determining the corner kinematics of a class of gauge theories in 4D by using only kinematical data of the bulk.}\\

The second purpose of this paper concerns specifically gravity in a similar regard.\\
In the study of gravity, particularly quantisations~\cite{dewittQuantumTheoryGravity1967} of it, it has been historically rare to employ symmetry principles outside of gauge invariance for diffeomorphisms, or special situations like cosmology~\cite{bojowaldQuantumCosmologyReview2015,Oriti:2024qav}. As a more recent phenomenon, in contrast, global symmetries in gravity have at the very least been recognised to be present and of significant interest in asymptotically flat sectors~\cite{bondiGravitationalWavesGeneral1962,Strominger:2013jfa,Pasterski:2021raf,Bagchi:2022emh,Fuentealba:2022xsz,geillerPartialBondiGauge2022}, but the presence of symmetries for general spacetime regions with boundaries is now a popular subject as well~\cite{Giesel:2024xtb,freidelEdgeModesGravity2020a,freidelEdgeModesGravity2020,carrozzaEdgeModesDynamical2024,rielloQuantumEdgeModes2018,Carrozza:2021gju,Ball:2024hqe,Donnelly:2020xgu,rielloEdgeModesEdge2021,araujo-regadoSoftEdgesMany2025,Ciambelli:2022cfr}. For regions of finite extent, these symmetries are particularly interesting as they allow for a refined perspective on the subdivision of larger regions into pieces, and how these smaller chunks may be recombined, in rough analogy to the cutting and glueing of topological field theories~\cite{Ciambelli:2024qgi,Ciambelli:2021nmv,rielloHamiltonianGaugeTheory2023,Freidel:2023bnj}. 
In this work, we will be interested in studying the impact of these symmetries to its logical conclusion at cutting surfaces and intersections of them, meaning at codimension 2 and 3 surfaces. \\

Our goal will be to deduce appropriate Poisson brackets of fundamental fields when these are restricted\footnote{We will make this more precise in a moment.} to such surfaces, without making use of more than the kinematical Poisson brackets (in the codimension 1 bulk) and the constraints defining the physical phase space. In this way, we can make kinematical statements about the boundary degrees of freedom of gauge theories (where kinematical refers to us not choosing any particular Hamiltonian on the physical phase space). \\
These Poisson brackets may be a priori different from the ones on codimension 1 surfaces, and we will treat them as entirely separate. They are really more analogous to Barnich-Troessaert brackets~\cite{Barnich:2011mi}, which are defined through actions of gauge transformations on corner charges. The Poisson bracket we propose is essentially a Barnich-Troessaert bracket on the space of all pullbacks of fields to the codimension 2 surfaces.\\ 
When this procedure is successful, it provides a starting point for reconstruction of the bulk theory from the boundary point of view, at least for certain sectors of it. We will see an example of this in 3D gravity shortly. However, our overall goal is to suggest that by analogies with topological theories in 4D, we can motivate a similar corner Poisson algebra in 4D gravity. \\
Such a phase space would certainly not present us with the data of a full-fledged holographic dual of gravity, but capture at least certain low-energy aspects of it.
For example, low-energy universal sectors of holographic duals (analogous to the Schwarzian~\cite{blommaertSchwarzianTheoryWilson2018} or the Alekseev-Shatashvili model~\cite{alekseev_path_1989,alekseev_geometric_1990,cotler_theory_2019}) in 4D could then be realised as particular dynamics on this corner phase space.
But beyond this, it has applications more generally for the formalisation of subregion phase spaces, where we can make use of it in the following manner.

\paragraph{Subregions and splitting}\, \\
To begin, let us state how we split a scalar field's classical phase space on a Cauchy slice of codimension $1$ (so a $3$D slice in a $D=4$ spacetime) that is a union of two subregions, $\s_L\cup \s_R$. Said phase space is described by the field $\phi\in \Omega^0(\s_L \cup \s_R)$ with its conjugate partner $\Pi \in \Omega^{3}(\s_L \cup \s_R)$, and a symplectic potential/Poisson bracket
\begin{equation}
    \Theta_{\s_L \cup \s_R}=\int_{\s_L \cup \s_R}\Pi\delta\phi \; \Longleftrightarrow \{\Pi(x),\phi(y)\}_{\s_L \cup \s_R} = \delta(x,y) d^{D-1}x.
\end{equation}
Disregarding subtleties of functional analysis, the splitting of the fields themselves is simple. One simply replaces the fields, e.g. $\phi$, by independent $\phi_L,\phi_R$ satisfying continuity $\phi_L \stackrel{\p \s_L}{=}\phi_R$, and the symplectic form splits into two pieces
\begin{equation}
    \Theta_{\s_L \cup \s_R}=\int_{\s_L }\Pi_L\delta\phi_L + \int_{ \s_R}\Pi_R\delta\phi_R. 
\end{equation}
In other words, at the level of description of relevance for us, there are no interesting effects on the cutting surface for the scalar field. In particular, to restrict to a subregion $\s_L$, we simply drop the $\s_R$ pieces everywhere. \\
The situation is much more interesting for a gauge field, for which we take an abelian 1-form as an example. We call it $A= A_i dx^i\in \Omega^1(\s_L \cup \s_R)$ and its conjugate partner $B= B_{ij}dx^i dx^j\in \Omega^{2}(\s_L \cup \s_R)$, matched through the symplectic potential
\begin{equation}
    \Theta_{\s_L \cup \s_R}=\int_{\s_L \cup \s_R}B \wedge \delta A \; \Longleftrightarrow \{B_{ij}(x),A_k(y)\}_{\s_L \cup \s_R} = \delta(x,y) \epsilon_{ijk}.
\end{equation}
If the fields stay as they are, then there are no changes to the splitting. However, usually, such 1-form phase spaces are subject to a differential constraint like the Gauss constraint $dB\approx 0$. The phase space on-shell of it is then the zero level set of the family of functions
\begin{equation}
    J_\alpha = \int_{\s_L \cup \s_R} B\wedge d\alpha \qquad\alpha\in \Omega^0(\s_L \cup \s_R)
\end{equation}
which are a functionally differentiable version of the Gauss constraint.\footnote{By this we mean that $\delta J_\alpha$ has the form of a contraction $-I_V\Omega$, where $\Omega=\delta\Theta$ is the symplectic form. This is a nontrivial condition when boundaries are present. E.g. $-d\delta B\alpha= d(-\delta B\alpha) + \delta B d\alpha$ has a total derivative term which cannot be matched with the symplectic form. This notion is therefore always with respect to a given symplectic form.} These functions also generate the gauge transformations $X_\alpha [A] = d\alpha$, which must be declared as redundancies on the phase space once one goes on-shell. This is because when the functions $J_\alpha$ vanish, so does the contraction $I_{X_\alpha}\Omega$ of the transformations with the symplectic form. The transformations are then in the kernel of the symplectic form, and therefore to have a \textit{symplectic}\footnote{The on-shell phase space is Poisson, so without the demand of symplecticity we can in principle leave it at that.} phase space again we must quotient away the action of these transformations. This leaves us with the physical phase space.\\
The splitting problem now becomes more interesting: everywhere within $\s_L,\s_R$, the Gauss constraint must hold, independently of what happens on the other side of the cut. This means we can also split the functions defining the on-shell phase space into $J_{\alpha_L}+J_{\alpha_R} $, and the matching equations are again $A_L \stackrel{\p \s_L}{=}A_R$, $B_L \stackrel{\p \s_L}{=}B_R$ together with $\alpha_L \stackrel{\p \s_L}{=}\alpha_R$. We can split the symplectic form again as in the scalar field case, but we also have a slightly different route: We can go on-shell of the Gauss constraint on, say, $\s_R$, giving the partially reduced functions
\begin{equation}
    J_{\alpha_L,\alpha_R} = \int_{\s_L} B_L\wedge d\alpha_L - \oint_{\p\s_L} B_R\alpha_R
\end{equation}
and seeing them as constraints imposes the matching equation for $B_L,B_R$ at the interface as $J_{\alpha_L,\alpha_R} \approx  \oint_{\p\s_L} (B_L- B_R)\alpha_L$. This shows a simple way to treat the subregion $\s_L$ in relation to its complement $\s_R$. 
An important phenomenon arises then when we naively describe the subregion $\s_L$ in isolation with the naive symplectic potential, and the function $J_{\alpha_L}$. Precisely, \begin{equation}
    J_{\alpha_L} = \int_{\s_L } B_L\wedge d\alpha_L \approx \oint_{\p\s_L } B_L \alpha_L,
\end{equation}
so the function does not vanish on-shell (the classic case for this is for asymptotic diffeomorphisms in general relativity (GR)~\cite{Regge:1974zd,Brown:1986nw, iyerPropertiesNoetherCharge1994}), and imposing it to vanish for all $\alpha_L$ would artificially restrict the phase space. This shows two points: First, in order to properly account for the relation to the complement region $\s_R$, we need to include an additional field $B_R$ in the phase space of the isolated $\s_L$, which we will do by means of a phase space extension soon. Second, taken literally, the functions $J_{\alpha_L}$ generate the gauge transformations on $\s_L$ all by themselves already, so the nonvanishing of these functions indicate that not accounting for $B_R$ (so the complement region), gauge transformations of $A_L$ are \textit{nontrivial} transformations on phase space, so not redundancies. The on-shell values $B_L\alpha_L$ are then called the \textit{corner charges}, and they are the generators of the \textit{corner group}. 

\paragraph{ Phase space extensions }\, \\
There is, then, a seeming mismatch between the action of gauge transformations on the whole slice and on the subregion. However, if one sees the field $B_R$ as additional data attached to the subregion $\s_L$, this mismatch is little more than a problem of finding the right description. What we will now explain is the logic of phase space extensions on $\s_L$ which provide such a description. We note upfront that these extensions are not \textit{mandatory}\cite{rielloEdgeModesEdge2021}, and rather provide useful local formulations of relevant data (in the same sense that the use of gauge fields for gauge theory instead of other gauge invariant observables is optional to a large extent, but preferable due to locality).\\
The general idea is to mimic the imprint of the piece $-\oint_{\p\s_L} B_R\alpha_R $ in the phase space defining function, but on the subregion space $\s_L$ alone. For this purpose, one extends the phase space on $\s_L$ by a fields corresponding to the corner group. In the example from just now, this is generically a $D-2$ form (boldface notation)  $\mathbf{B}_L \in \Omega^{D-2}(\partial\s_L) $ and a function $\lambda_L\in \Omega^0(\p\s_L) $\footnote{This is for gauge group $\rbb$, rather than $U(1)$.}, and one gives the space the extended symplectic potential
\begin{equation}
    \Theta_{\s_L} = \int_{\s_L} B_L\wedge\delta A_L + \oint_{\p\s_L}\mathbf{B}_L\delta\lambda_L.
\end{equation}
While this simple abelian example misses a lot of structure, key elements of the idea are already present. The main point is that the field $\mathbf{B}_L$ is a stand-in for the field $B_R$, but is not conceptually identical to it. In the non-abelian case, it will be a \textit{gauge-transformed} or \textit{dressed} version of $B_L$. Due to this stand-in role, it features in an extended defining function for the phase space,
\begin{equation}
    J_{\alpha_L} = \int_{\s_L} B_L\wedge d\alpha_L - \oint_{\p\s_L} \mathbf{B}_L\alpha_L
\end{equation}
which generates on $\lambda_L$ the transformation $X_{\alpha_L}[\lambda_L]= -\alpha_L $, i.e. a (left) translation with inverted parameter. In other words, $\lambda$ transforms as (the opposite of) a \textit{gauge frame of reference}, meaning it is in a faithful representation of the gauge group. This is the defining property of this field, which we will refer to as the \textit{gauge frame} in all its upcoming incarnations, though it is often referred to as a \textit{Stückelberg field} or an \textit{edge mode}. The field $\mathbf{B}_L$, then, more generally fulfils the role of being the \textit{charge aspect} which generates the right translations of $\lambda_L$. In this abelian case, of course, there is no difference between left and right translations, but in the nonabelian case this difference will be crucial.\\
In this extended phase space, we can now make a distinction between two sets of transformations: The extended gauge transformations
$X_{\alpha_L}[A_L]=d\alpha_L\quad X_{\alpha_L}[\lambda_L]= -\alpha_L $
generated by $J_{\alpha_L}$ and the \textit{frame reorientations} $\tilde{X}_{\tilde\alpha_L}[\lambda_L] = \tilde\alpha_L$ generated by
\begin{equation}
    Q_{\tilde\alpha_L}= \oint_{\p\s_L} \mathbf{B}_L\tilde \alpha_L.
\end{equation}
The logic of the extended phase space is as follows: the pair $(\mathbf{B}_L,\lambda_L)$ captures relational information about $\s_L$ and whatever its environment may be. Therefore, when it is included in the phase space, and jointly transformed by $J_{\alpha_L}$, the resulting transformations mirror those of the whole slice $\s$ and therefore are pure redundancy (making the generator $J_{\alpha_L}$ vanish). However, one can still model the gauge behaviour in the absence of the complement, by considering the frame reorientations. These transformations change the gauge relational information, and nontrivially change, from the point of view of $\s_L$ viewed in isolation, the state of the system. For this reason, frame reorientations $\tilde X_{\tilde \alpha_L}$ are nontrivial transformations from the point of view of the subregion $\s_L$, while gauge transformations $X_{\alpha_L}$ are not.\\
In technical terms, the phase space is then defined by the vanishing of $J_{\alpha_L}$ \textit{for all $\alpha_L$}, including those supported on $\p\s_L$. $Q_{\tilde\alpha_L}$ then captures the corner charge algebra. The extended phase space thus transparently disentangles gauge redundancies from nontrivial corner transformations.\\

In the nonabelian case for some structure group $G$\footnote{For simplicity of exposition we assume a compact matrix group like $SU(N)$.}, a generic change is that the field $\mathbf{B} \in \Omega^{D-2}(\p\s,\gfrak)$ (dropping the subscript L simplicity) will transform as well. To make the distinction to the abelian case clear, let us denote the gauge reference frame by $\phi\in\Omega^0(\p\s,G)$ now. In this case, we replace the corner symplectic potential contribution by
\begin{equation}
    \oint_{\p\s} \Tr [ \mathbf{B} \, \phi^{-1}\delta\phi ]
\end{equation}
which to the trained eye has the form of a standard symplectic potential on $T^\ast G$, but at every point of $\p\s$, which we can think of as the cotangent bundle to $G^{\p\s} \hat{=} \Omega^0(\p\s,G)$. This is a simple phase space extension, and works the same way as in the abelian case, with a crucial difference: the gauge transformation generator is now
\begin{equation}
    J_\alpha = \int_\s \Tr[B \wedge d_A\alpha] - \oint_{\p\s} \Tr[ \mathbf{B} \phi^{-1}\alpha\phi]
\end{equation}
while the corner charges are again
\begin{equation}
    Q_{\tilde\alpha}= \oint_{\p\s} \Tr [ \mathbf{B} \tilde\alpha ].
\end{equation}
The main difference is that $B$ now transforms as $X_\alpha[B] = [B,\alpha]$ under gauge transformations, while $X_\alpha[\mathbf{B}]=0$ does not transform. We see $\mathbf{B}$ as a version of $B$ \textit{dressed} by $\phi$, as becomes explicit when imposing the bulk constraint $d_A B \approx 0$:
\begin{equation}
    J_\alpha \approx \oint_{\p\s} \Tr[ ( \phi^{-1} B \phi -\mathbf{B}) \phi^{-1}\alpha\phi]
\end{equation}
On the on-shell phase space, then, $\mathbf{B}$ is equal to the dressed $\phi^{-1} B \phi$, which does not transform under gauge transformations, as
\begin{equation}
    X_\alpha[A]= d_A \alpha \quad X_\alpha[\phi] = -\alpha\phi.
\end{equation}
Importantly, under joint gauge transformations of the bulk and corner fields, $\phi$ transforms by a \textit{left} translation, $\phi \mapsto e^{-\alpha}\phi$. The situation is the \textit{opposite} for the corner charges $Q_{\tilde\alpha}$, which generate the right translation frame reorientations
\begin{equation}
    \tilde X_{\tilde\alpha}[\phi] = \phi \tilde\alpha\quad \tilde X_{\tilde\alpha}[\mathbf{B}] = [\mathbf{B},\tilde \alpha].
\end{equation}
Again, the corner symplectic extension captures the part of the gauge structure of the theory that imprints on the corner $\p\s$. We only discussed 0-form parametrised transformations here, but the same story can be applied to $p$-form symmetries, and we will see this applied later.\\

\paragraph{Glueing}\, \\
Once we have two extended phase spaces in place, we can ask how to glue them together. We do not need too many details of this in this work, but point out two crucial elements. \\
First, that each subregion phase space, so both on $\s_L$, $\s_R$ are individually defined through the vanishing of their respective $J_{\alpha_L},J_{\alpha_R}$, for all $\alpha_L,\alpha_R$. The remaining glueing conditions then arise from a matching of corner charges. \\
Second, this matching of corner charges is not `on-the-nose'. What is required is that the bulk fields $B_L \stackrel{\p\s_L}{=} B_R$ match continuously on the interface, and this translates into the condition on the corner fields
\begin{equation}
  \phi_L \mathbf{B}_L \phi^{-1}_L  \stackrel{\p\s_L}{=} \phi_R \mathbf{B}_R \phi^{-1}_R.
\end{equation}
In words, the corner fields are not matched continuously, but only up to a gauge transformation. This means that we can conceptually identify the combination $\phi_L \mathbf{B}_L \phi^{-1}_L$ as $B_R$, so a field coming from the outside of the subregion $\s_L$ we are considering. The corner phase space of $(\mathbf{B}_L,\phi_L)$ is then akin to describing an isolated `1-sided surface' whose intrinsic field is given by $\mathbf{B}_L$ and its extrinsic relation to the environment is given by $\phi_L$. Glueing together two copies of these 1-sided surfaces in the above fashion then recovers an ordinary 2-sided surface with a single value $B$. \\
Let us now recapitulate our interest in this and move on to a relevant example.

\paragraph{Recap and goals} \, \\
We are thus generally interested in situations where some kinematical phase spaces are equipped with gauge transformations and their generators. When such a generator vanishes when evaluated on physical configurations of the theory, we call the transformation a redundancy. However, generically for gravity and other transformations in gauge theories like Yang-Mills, instead we find a nonzero generator when evaluated on on-shell configurations~\cite{Regge:1974zd,Brown:1986nw, iyerPropertiesNoetherCharge1994,harlowCovariantPhaseSpace2020}. This means that diffeomorphisms supported on the boundary are not redundancies, and instead (depending on boundary conditions) they become genuine symmetries of the theory. The consequence is that, when these generators are indeed conserved, we can call them corner symmetry charges, and they can aid the construction of the quantum theory, the scattering matrix~\cite{Strominger:2013jfa} or more generally inform us about the infrared behaviour of the theory~\cite{stromingerLecturesInfraredStructure2018}.\\
\textit{In this work, we want to take the presence of these charges to a logical extreme, and argue for the form of a phase space associated purely to corners $\p \s$ which captures the corner symmetry aspects of a gauge theory manifestly.} \\
By this we mean the following: the bulk theory already induces a number of degrees of freedom on corners, but the precise Poisson brackets may be subtly different from the bulk, like those of $B$, which is Poisson commutative, versus $\mathbf{B}$, which fulfils a Lie algebra $\cong \gfrak$. Such changes are reflected in the corner charges. We might therefore associate to corners a separate Poisson algebra whose quantisation governs the corner symmetry aspects, and we will study precisely this algebra. Note that this is not in principle the same as the corner symmetry algebra itself - the symmetry charges are, generally speaking, a subalgebra of this Poisson algebra. For one example, there may be fields which are not part of any corner charge, such as $A$ in the examples before. Alternatively, like in Einstein-Cartan 4D gravity, the corner charges are quadratic in the fundamental bulk fields, whereas the Poisson algebra we want to consider is linear in them.\\
A central question to address in this angle of `corner algebras' is how one determines the corner symplectic extension $\Theta_{\p \s}$, and how one determines the Poisson relations of not just the charges, but all the bulk fields of the theory. When this is found, we can then study the phase spaces of corner independently from the bulk, and even further subregions of the corner itself. \\
\textit{An important further question is whether this can be done without invoking any particular boundary condition on the system.} By this we mean that we may invoke aspects of the bulk dynamics such as the constraints defining the physical phase space, or the Hamiltonian of the bulk, but we should not have to use e.g. a specific boundary Lagrangian to determine the phase space structure. \\
There is evidence for the latter in the fact that in examples like BF theory, different boundary conditions still lead to the same kinds of kinematics~\cite{maldacena_d-brane_2001}. For example, a $p-$form higher gauge theory of BF type in $D$ spacetime dimensions always has a boundary dynamics for a $(p-1)$-form and a $D-p-2$ form, for multiple different boundary conditions. We will return to this point in section \ref{sec:Examples} This leads us to think that the \textit{kinematics} of these boundary dynamics is determined by the gauge structure of the bulk alone, while the \textit{dynamics} corresponds to a choice of boundary condition for the bulk, or a Hamiltonian on the corner phase space. In what follows, we will only concern ourselves with the kinematical aspects.\\
Let us now understand how to find the corner Poisson algebra in an example.

\paragraph{Corner Poisson algebra of 3D gravity}\, \\
Gravity in first-order variables~\cite{geillerMostGeneralTheory2021,freidelQuantumGravityDisk2021} has two 1-form fields, the triad $e^I$ and the spin connection $\omega^{IJ}$. The bulk Poisson brackets for these two, on a codimension 1 slice $\Sigma$ with corner $\p\s$, simply say that these two are canonically conjugated,\footnote{Only within this section, we work in $D=3$. Here, $i,j$ denote 2-dimensional indices on $\s$, and $I,J,A$ denote internal 3D indices. These will be 4D indices in the rest of this work.}
\begin{equation}
    \{ \omega^{IJ}_i, e^A_j  \}_\s = \epsilon_{ij} \epsilon^{IJA}.
\end{equation}
The theory has two corner charges, associated to internal Lorentz and shift transformations,\footnote{Here, $\alpha^{IJ}$ is an $so(1,2)$ and $\phi^A$ an $\rbb^{1,2}$ gauge transformation parameter which may depend on position. $F_{\omega} $ denotes the curvature of $\omega$.} which become on-shell localised in codimension 2,
\begin{equation}
\begin{gathered}
        J_\alpha =-\int_\s \epsilon_{IJK}e^I d_\omega\alpha^{JK} \approx \oint_{\p\s}\epsilon_{IJK}e^I \alpha^{JK}\\ 
        P_\phi = -\int_{\s} \epsilon_{IJK} F_\omega^{IJ}\phi^K + \oint_{\p\s} \epsilon_{IJK}\omega^{IJ}\phi^K  \approx \oint_{\p\s} \epsilon_{IJK} \omega^{IJ}\phi^K,
\end{gathered}
\end{equation}
\textit{and which, importantly, are \textit{linear} in the fundamental fields of the theory}. The algebra of the full off-shell generators is e.g.
\begin{equation}
    \{ J_\alpha, J_\beta \}_\s = J_{[\alpha,\beta]} \approx \oint_{\p\s}\epsilon_{IJK}e^I [\alpha,\beta]^{JK}.
\end{equation}
However, this does not make sense when evaluating the charges on-shell \textit{before} applying the Poisson bracket, where one would find
\begin{equation}
\begin{gathered}
        \{ \oint_{\p\s}\epsilon_{IJK}e^I \alpha^{JK}, \oint_{\p\s}\epsilon_{ABC}e^A \beta^{BC} \} 
    _\s \\
    =
    \oint_{\p\s\times \p\s}\epsilon_{IJK} \alpha^{JK} \epsilon_{ABC}\beta^{BC} \{e^I,e^A\}_\s  = 0.
\end{gathered}
\end{equation}
Instead, this suggests that \textit{on-shell} corner charges (so the corner pieces of the gauge generators) are subject to a different Poisson structure $\{,\}_{\p\s} $, which can be read off to be\footnote{Here, the star $\star$ stands for the internal 3D Hodge dual. Later on, we will use it for the 4D dual.}
\begin{equation}\label{eq:3DGravPoissonAlg}
    \begin{gathered}
        \{\star e^{IJ}\alpha_{IJ},\star e^{KL}\bar\alpha_{KL}\}_{\p\s} = \star e^{AB}[\alpha,\bar\alpha]_{AB}\\
        \{\star e^{IJ} \alpha_{IJ},\omega^{KL} \beta_{KL}\}_{\p\s} = \omega^{IJ}[\alpha,\beta]_{IJ} + \beta^{IJ} d\alpha_{IJ}
    \end{gathered}
\end{equation}
In particular, in this algebra $\star e$ does not commute, instead forming a $\mathfrak{so}(1,2)$ algebra. Similarly, the Poisson bracket of $\star e,\omega$ does not produce only central terms, but also terms linear in $\omega$. We give an embedding of this structure in the same language as used in the rest of the paper in appendix \ref{app:3D}.\\
One can identify this Poisson algebra as a \textit{Kac-Moody} algebra of currents $J^a= (\star e, \omega)$ for the Lie algebra $\gfrak=\mathfrak{so}(1,2)\ltimes\mathfrak{so}(1,2)^\ast$ parametrised by $(\star e,\omega)$\footnote{The identification requires an implicit choice of bilinear form on $\mathfrak{so}(1,2)$.}. These affine Lie algebras are based on Loop algebras $\mathcal{L}\gfrak$ of a Lie algebra $\gfrak$, so maps $S_1\rightarrow \gfrak$, and have central pieces given by an invariant bilinear form $g^{ab}$ on $\gfrak$:
\begin{equation}
    [J^a_m,J^b_n] = f^{ab}{}_c J^c_{n+m} + m \delta_{m+n,0}g^{ab}
\end{equation}
The indices $n,m$ refer here to the Fourier mode number of $\star e,\omega$ when decomposed on the circles that make up $\p\s$. The appearance of this loop algebra is not surprising because 3D gravity can be understood as a Chern-Simons theory for the gauge algebra $\gfrak$, and these theories generically feature Kac-Moody boundary algebras.\\
In the case of 3D gravity, then, the invariant form between $(X,u),(Y,v)\in\gfrak$ is given by the diagonal evaluation,
\begin{equation}
    g((X,u),(Y,v))=u(Y)+v(X).
\end{equation}
This gives $\gfrak$ the structure of a Manin triple, (equivalently, a Drinfel'd double) and the phase space at the level of the zero modes becomes a Heisenberg double. This directly uncovers the relevance of quantum group structures for 3D gravity, already at the classical level~\cite{Dupuis:2020ndx}. Here, the statement is that the corner $\p\s$ naturally carries a Kac-Moody algebra with this invariant bilinear form. The zero-mode subalgebra of $J^a_0$s then is subject to the representation theory of this Manin triple. With this knowledge, one can then give different kinds of dynamics to these currents\cite{freidelQuantumGravityDisk2021}, and approach quantisation from the corner angle. In particular, the corner is naturally equipped with an action of the Virasoro algebra, a quantum analogue of the diffeomorphism algebra\footnote{The energy-momentum tensor in this case is roughly given by $T(z)\sim \epsilon_{IJK}\omega^{IJ}e^K$.}. We will see that this can be done in 4D gravity as well, but we will need to move away from Manin triples into the class of quadratically extended Lie algebras.\\
Note also that the phase space here contains only the dressed fields $\omega,e$, and is symplectic by itself. This will not generalise to 4D as-is. This can be easily changed by doubling: by including frames for the gauge transformations, which will be called $q,\phi$ in what follows. The details can be found in appendix \ref{app:3D}. The phase space obtained this way is closer to what we will work with in the following. \\
The main point is that 3D gravity naturally realises the notion of a corner phase space, there given by the Poisson-Lie loop group $L\exp(\gfrak)$. Since 3D gravity has no bulk degrees of freedom, this boundary data is already sufficient to describe the full theory. The same logic should then tell us crucial information about 4D gauge theories, or 4D gravity as well. To do so, we will try to find an appropriate corner Poisson algebra. \textit{What the 3D example shows us is that the presence of charges linear in the fields is important, and that no boundary conditions are a priori necessary to determine the correct structure.}

\paragraph{Previous studies}\,\\
The conceptual key point here is that if we want to work with on-shell charges as a guide for quantisation, we cannot use the naive bulk Poisson brackets - instead, there is an alternative structure which better describes the degrees of freedom. In the case of 3D gravity, the corner charges, due to being linear in the fields, fully determine this corner Poisson structure. A similar phenomenon happens in all Chern-Simons theories, where the charge aspect is given by the gauge field $A_i$ on the corner. If one now wants to quantise the theory based on corner data, this corner Poisson structure is the most interesting algebraic structure to consider. The (ambitious) hope is then that (some) bulk structures can be defined constructively from the corner. \\
This is in particular the point of view taken in the 4D gravity construction of the loop gravity string ~\cite{freidelLoopGravityString2017} and Poincare networks ~\cite{freidelGravitationalEdgeModes2019}. In their setup, one considers a finite region of gravity, and employs the corner Poisson structure to study the degrees of freedom that the boundary symmetries pertain to. To do so, one uses the pullback of bulk equations of motion to the corner, and studies the gauge structure of the resulting constraints in its own right, independently from the bulk. The result is a structure quite naturally very similar to that of an $\rbb^3$ Chern-Simons/Wess-Zumino-Witten\footnote{In this work, we use the terms `Chern-Simons' and `Wess-Zumino-Witten' in a loose sense, mostly making note of the phase space structures, rather than specific dynamics. We also include in the terms in principle the case of noncompact gauge groups.}(WZW) setup, situated on the corner $\p\s$, which provides a Cauchy slice for this boundary theory.\\
A main limitation of these works is that they break internal Lorentz covariance by selecting a particular internal time direction\footnote{This is known as the `time gauge' in the literature, though it really amounts to a restriction of states when performed on the corner.} $\delta e_t =0$, as well as fixing the connection on the corner. Generalising this requires significant effort, and we found that the best way to present the result is to begin from a top-down approach with a specific class of 4D theories. The main reason for difficulties is the fact that the corner charges in 4D gravity are \textit{quadratic} in the fundamental fields. These are the tetrad $e^I$ and spin connection $\omega^{IJ}$ again\footnote{Here, of course, the indices are 4-dimensional instead.}, but the corresponding corner charges are, unlike in the 3D case,
\begin{equation}
    J_\alpha \approx \oint_{\p\s} \frac{1}{2}\epsilon_{IJKL}e^I\wedge e^J \alpha^{KL} \qquad P_\phi \approx \oint_{\p\s} \epsilon_{IJKL}\omega^{IJ}e^K \phi^L.
\end{equation}
The former, $J_\alpha$, generates internal Lorentz transformations, while $P_\phi$ generates complicated curvature-dependent shifts of the tetrad $e\mapsto e+ d_\omega\phi + \dots$~\cite{langenscheidtNewEdgeModes2025,Langenscheidt:2025lha}. These two are symmetries of the theory, and their algebra can be computed, but it is not clear that it would determine a corner Poisson algebra of $e,\omega$ fully. In fact, we have reason to believe it does not: Since the two generate gauge transformations, any gauge-invariant function of $e,\omega$ will Poisson commute with them. While in 3D gravity, there is no nontrivial gauge-invariant function which is on-shell nonzero, in 4D the large on-shell phase space carries many such functions. So, we have an ambiguity: what is the corner Poisson bracket of the corner metric\footnote{Note that in ~\cite{freidelEdgeModesGravity2020,freidelEdgeModesGravity2021}, an argument is made for the corner metric to form an $\mathfrak{sl}(2;\rbb)$ Poisson algebra. This relies on a specific corner Poisson structure for the triads $e^i$ which is not implied by the bulk \textit{per se}. Furthermore, no statement is made about the connection $\omega$. We will find the same algebra, derived from BF-BB theory.} $q_{ab} = e^I_a \eta_{IJ} e^J_b $ or the gauge invariant parts of $\omega_a$? \\
In the present work, we will follow one strategy to fix this ambiguity, which is useful for quantisation attempts. 
We will present the analogy of the corner Poisson analysis for a general class of theories where the corner charges are linear in the fields, like in 3D gravity. We then select among this general class one case which reduces, to a very large degree, to 4D gravity, through the use of constraints. This informs our choice of corner Poisson structure. With this in hand, we can then perform the corner analysis in its own right.\\
It is also important to mention work on 2-Chern-Simons theory~\cite{zucchini_4-d_2021} as the BF-BB theories we study will be very similar in structure to those theories. The cited work exclusively deals with topological theories, but already investigates many similar questions such as the algebra of charges, and the corner symplectic form. It also deals with global issues of the gauge group and topology, which we do not attempt in the present work. What existing work on this side does \textit{not} do is attempt to continue the chain of corner algebras further, and similarly no analogies to 4D gravity or other nontopological theories are made. This is where our paper expands upon the applications to pure topological models.

\paragraph{Puncture algebras}\,\\
We can take the logic of above even further. Say we have arrived at a natural Poisson structure and phase space on corners $\p\s$ of finite regions. The nature of these phase spaces is that they are localised entirely on codimension 2 surfaces in spacetime. We can then think of the degrees of freedom within them as living entirely on these surfaces and their (time) evolutions, \textit{isolated} from the rest of spacetime, not part of any foliation or similar extended concepts. In this way, the assignment of corner phase spaces to \textit{any} closed codimension 2 surface makes sense and defines, to some extent, a theory of its own (at least the kinematics).\\
We can then run the same logic again in one dimension lower: if we have a closed surface $\p\s$ and the phase space associated to it, we can try and split the surface into two parts, $S\cup \bar S$, and consider the subregion phase space on $S$. If we want to think of the corner phase spaces as encoding some low energy holographic information about the bulk, then this new step corresponds to studying subregions of the holographic dual.\\
Another way that this situation can arise is when we consider the intersection of two overlapping finite codimension 1 regions, $\s_1, \s_2$. They both have codimension 2 boundaries $\p\s_1, \p\s_2$, and they may have overlap as well, which generically is a surface $\p\s_1\cap \p\s_2$ of codimension 3 in spacetime. This codimenson 3 surface divides $\p\s_1$ in the same form $S\cup\bar S$, and we can consider the corner degrees of freedom that live only there.\\
In what follows, we will use the word \textit{puncture} to refer to the codimension 3 surfaces like $\p\s_1\cap \p\s_2$ or $\p S$. This is to distinguish them from `boundaries' and `corners', which carry the connotation of being codimension 1 and 2, respectively. Furthermore, the word `puncture' is particularly accurate in 4 dimensions, where these surfaces are generically circles which remove disks from a closed 2D surface, in direct analogy to punctured Riemann surfaces.\\
The phase spaces we associate to a codimension 2 region $S$ with boundary $\p S$ are subject to the same issues as codimension 1 regions $\s$ with boundary $\p \s$. Particularly, if there are differential constraints which we impose on the corner phase space on $S$, then they will come with their own corner charges and gauge transformations, and so can give rise to their own `corner algebras', which for distinction we refer to as `puncture algebras'.\\
In the case of 4D BF-BB theory, we will see that the flat subsector of the corner phase space (as in, subject to an additional constraint $F_\acal=0$) admits such a puncture algebra, and it is of course a current Kac-Moody algebra, as for all Chern-Simons theories.\\
With this out of the way, we now summarise our methodology and results.

\paragraph{Methodology}\,\\
Our method of determining a corner Poisson algebra can be summarised as follows:
\begin{enumerate}
    \item Find a formulation of the theory of interest in terms of a gauge theory with one corner charge per bulk field, and where the charges are linear in the same fields. (We consider BF-BB theories as such a class.)
    \item Set up a starting `kinematical' corner phase space corresponding to that formulation of the theory, for example by analogy to a topological theory in the same class. 
    \item On this kinematical corner phase space, impose and reduce by all constraints of the bulk that can be pulled back to the corner. (In our case, algebraic relations between 2-forms). 
\end{enumerate}
The result is a `physical' corner phase space which captures the corner charge relations in the Poisson bracket of the bulk fields, and which also respects all relations between these fields that come from the bulk equations of motion. 

\paragraph{Results}
Our methodology provides us the following statement: on codimension 2 surfaces of 4D BF-BB theories, the `physical' corner phase space is that of a Chern-Simons theory \textit{without flatness constraint}. When one further restricts the phase space to the flat sector by imposing the flatness constraint by hand or boundary condition, there arise puncture algebras which are Kac-Moody algebras.\\
We apply this logic to a particular first-order tetradic formulation of 4D gravity as well, where it is sensible for the fields $\omega,e$, seen as connections, and an auxiliary connection $S^{IJ}$, to form a Chern-Simons like phase space. These three fields are arranged into a connection $\acal$ for the \textit{Maxwell algebra}
\begin{equation}
    \gfrak =\mathfrak{so}(1,3)\ltimes(\rbb^{1,3}\tilde\oplus \mathfrak{so}(1,3)^\ast)
\end{equation}
where $\omega$ is associated to Lorentz transformations, $e$ to translations and $S$ to dual Lorentz transformations. Their phase space is extended by an $SO(1,3)$ gauge frame 0-form field $\phi$, and is given a symplectic form
\begin{equation}\label{IntroSP}
    \Omega_S = \int_{\p \s} \frac{1}{2} \delta S_{IJ}\wedge \star_r\delta\omega^{IJ} + \frac{1}{2t }\delta e_I\wedge\delta e^I +\delta(\frac{1}{4}(\star_\beta e^2)_{IJ}(\phi^{-1}\delta\phi)^{IJ}).
\end{equation}
For relevant notation we direct the reader to section \ref{sec:Gravity} on gravity. Importantly, we find that $e$ becomes self-conjugate and $S,\omega$ form a connection for the double $\mathcal{D}\mathfrak{so}(1,3)$, analogous to 3D gravity. Finally, we can also find a puncture Kac-Moody algebra for the Maxwell algebra in 4D gravity. This happens when we impose flat boundary conditions on codimension 2 surfaces apart from some small disks, the boundaries of which then carry the Kac-Moody currents. \\
Two caveats are in order: \textit{first}, the result in gravity is the product of an analogy to the topological BF-BB case, where we start from an extended kinematical corner phase space whose structure is inherited from the topological case. The \textit{gravitational} bulk constraints are then imposed on this kinematical corner phase space, and the result differs meaningfully from the same result in the topological case by the addition of the third term in \ref{IntroSP}. \textit{Second}, the presented phase space is not completely the physical one yet, and should be further subjected to the torsion constraint $d_\omega e\approx 0$. These constraints are more subtle to incorporate explicitly and we discuss their imposition in the conclusion of this work.

\paragraph{Structure of this paper}
In section \ref{sec:BFBB}, we set up BF-BB theory, in particular its kinematical phase space and charge algebra. We then induce a corner Poisson algebra for it, finding it to be a Chern-Simons phase space. Repeating the same procedure on the corner towards punctures, we recover a WZW phase space. \\
Section \ref{sec:Gravity} then argues for the same procedure in 4D gravity, presented in a BF-BB like form. We first spend some time exposing this formulation, and then find a heuristic corner phase space by restricting the BF-BB result to gravity. We highlight differences to the topological case here. \\
Section \ref{sec:Outlook}, an outlook, discusses several open questions and applications of the corner phase space.\\

The section on gravity is supplemented by several technical appendices. The symplectic reduction in the topological case is done in detail in appendix~\ref{App:TopReduction}. For illustrative purposes, we include a comparison to the 3D gravity case in appendix~\ref{app:3D}. We discuss abelian 4D examples in appendix~\ref{sec:Examples}, including standard choices of dynamics for the corner.~\ref{App:BulkAlgebra} describes the expanded set of codimension 1 gauge transformations in 4D gravity, together with the charge algebra.~\ref{App:CornerAlg} displays in full the BF-BB corner Poisson algebra of the BF-BB theory used as the basis for section~\ref{sec:Gravity}.~\ref{App:CornerSympForm} then imposes a number of constraints on the associated corner symplectic form, in order to obtain the corresponding one for 4D gravity. As additional supplements, appendix~\ref{App:SecondClassSchematic} describes some of the aspects that are involved in the logic of defining a corner phase space. It investigates how the induction of a corner Poisson algebra works if there are second-class constraints present in the bulk. %The latter then applies this to the 4D gravity case and argues how the result from this perspective matches up with the one from the main text.

\section{Corner and puncture data in BF-BB theories}\label{sec:BFBB}
We will first introduce our class of theories in a simple form, as well as their symmetries. 

\subsection{Codimension 0: BF-BB and its symmetries}

A large class of field theories in 4D can be formulated as \textit{BF-BB theories} for a connection $\acal$ and conjugate partner 2-form $\bcal$, with Lagrangian in codimension 0 given by
\begin{equation}
    L_0 = \bcal\wedge F_\acal - \frac{1}{2}\bcal\wedge\Phi(\bcal)
\end{equation}
where $\Phi$ is not a field but a constant invertible linear map on the Lie algebra $\gfrak$ in which the fields are valued. We assume throughout this section that all expressions are wrapped in a trace $\Tr[-]$ on the same Lie algebra, which we omit to make notation simpler. At a certain point, we will need to replace this trace again by a general bilinear form $g$ on $\gfrak$. We will assume that with respect to the inner product/trace, the homomorphism $\Phi$ is self-adjoint. Also, \textit{for this section only}, we will also assume a special property for $\Phi$, which ensures that the theory described has only first-class constraints and is thus topological:
\begin{equation}\label{eq:TopProperty}
    [X,\Phi(Y)]=\Phi([X,Y])
\end{equation}
In particular, this makes $d_\acal\Phi(Y)=\Phi(d_\acal Y)$. This property is equivalently the statement that $Tr[(-)\Phi(-)]$ is an invariant inner product on $\gfrak$, or that $\Phi$ is an equivariant map. In the section on gravity, this property will no longer hold, making the theory nontopological.  The equations of motion are simply
\begin{equation}
    E_0 = \delta\bcal\wedge (F_\acal - \Phi(\bcal)) - d_\acal \bcal \wedge\delta\acal
\end{equation}
which lets us integrate out the $\bcal$ field to give the effective `topological Yang-Mills' Lagrangian
\begin{equation}
    L_{0,eff} = \frac{1}{2}F_\acal \Phi^{-1} (F_\acal).
\end{equation}
In principle, we only need the local structure of the theory for what we want to do, i.e. the equations of motion and the symplectic potential in codimension 1 derived from the Lagrangian,
\begin{equation}
    \theta_1 = \bcal\wedge\delta\acal
\end{equation}
which allow us to define Poisson structures on codimension 1 surfaces of any spacetime $M_0$ we care about.

We have three natural local symmetries of the bulk theory: Internal `Yang-Mills' transformations with $\gfrak$-valued 0-form parameter $\alpha$, `Kalb-Ramond' transformations\footnote{This name is chosen in analogy to the 1-form gauge symmetry of the Kalb-Ramond field in string theory.} with $\gfrak$-valued 1-form parameter $\mu$, and diffeomorphisms parametrised by vector fields $\xi$. We will exclusively be concerning ourselves with the first two in this paper, as their charges are all we need to infer the corner phase space.\\
These symmetries act by the transformations
\begin{equation}
    \begin{gathered}
        X_\alpha[\acal]= d_\acal\alpha\qquad X_\alpha[\bcal]= [\bcal,\alpha]
    \end{gathered}
\end{equation}
\begin{equation}
    \begin{gathered}
        Y_\mu[\acal]= \Phi(\mu) \qquad Y_\mu[\bcal]= d_\acal\mu
    \end{gathered}
\end{equation}

Overall, we have the symmetry structure:
\begin{equation}
    %\mathfrak{diff}(M_0;M_1)\ltimes 
    (
    \Omega^0(M_0)\ltimes\Omega^1(M_0) )\otimes\gfrak
\end{equation}
where the Kalb-Ramond transformations are abelian and acted on by the Yang-Mills transformations. %All of these are further acted on by diffeomorphisms which restrict to the boundary in the correct way. 
The symmetry algebra is equipped with the Lie bracket
\begin{equation}
    [(\alpha,\mu),(\bar\alpha,\bar\mu)]
    = ([\alpha,\bar\alpha],  [\alpha, \bar \mu] - [\bar \alpha, \mu]).
\end{equation}
While we described a dynamics here, we really only need the spatial part of it in what follows, meaning the pullbacks of the equations of motion to codimension 1 and 2 slices. In particular, the equation $F_\acal=\Phi(\bcal)$ imprints on any codimension 2 slice, which will be crucial in arriving at the physical corner phase space.\\
At this point we also want to make a short comment about the relation to higher categorical structures. While naively, the BF-BB theories we work with share some symbolic similarity with \textit{2-Chern-Simons} theories\cite{zucchini_4-d_2021}, and indeed for the topological case one can identify the former as a subset of the latter, we need to be careful. When $\Phi$ does not obey the condition (\ref{eq:TopProperty}), then the algebraic structure does not form a strict 2-Lie algebra anymore. In particular, one can choose $\gfrak_0=\gfrak_1=\gfrak$, $m(-,-)=[-,-]$ and $\Phi$ as the $t$-map, but then the compatibility conditions are not met in general. Hence, we refrain from invoking higher categorical language in what follows, though certain analogies may in principle be possible.

\subsection{Codimension $1\rightarrow 2$: Generators and corner charge algebra}
We will exclusively be concerned with the Yang-Mills and Kalb-Ramond transformations, and from now on work purely on a fixed slice $\s$, with no mention of boundary conditions or corner contributions to the symplectic form\footnote{Our convention is $I_V\Omega + \delta F_V = 0$ for a Hamiltonian vector field $V$ and its generator $F$.} $\Omega_\s$. The transformations can be given the generators\footnote{We will repeatedly refer to differentiability of generators. What is meant is that the field space derivative $\delta$ of the generator must take the form of $-I_V\Omega$ for some field space vector field $V$.}
\begin{equation}
\begin{gathered}
        J_\alpha = \int_\s \bcal\wedge d_\acal\alpha \approx \oint_{\p\s} \bcal \alpha \\
    K_\mu = -\int_\s (F_\acal-\Phi(\bcal))\wedge\mu + \oint_{\p\s}\acal\wedge \mu \approx \oint_{\p\s}\acal\wedge \mu.
\end{gathered}
\end{equation}
As advertised, our class of theories has corner charges linear in the fundamental fields, which is the key point in the induction procedure we want to perform. Their algebra is given by
\begin{equation}
\begin{gathered}
        \{J_\alpha,J_{\bar\alpha}\}= J_{[\alpha,\bar\alpha]}, \{J_\alpha,K_\mu\}=K_{[\alpha,\mu]} + \oint_{\p\s} d\alpha\wedge\mu
    \\
    \{K_\mu,K_{\bar\mu}\}= \oint_{\p\s} \mu\wedge\Phi(\bar\mu)
\end{gathered}
\end{equation}
and contains a central term for $K$, which is a harbinger of an Atiyah-Bott Poisson structure on the corner. 
By imposing the bulk constraints $d_\acal \bcal=0=F_\acal-\Phi(\bcal)$ both inside and outside the Poisson brackets above, we determine the corner Poisson bracket fully. The result is the following:
\begin{equation}
    \{ \bcal \alpha , \bcal  \bar\alpha \}_{\p\s} = \bcal [\alpha,\bar \alpha]
\end{equation}
\begin{equation}
    \{ \acal\wedge\mu , \acal\wedge\bar \mu\}_{\p\s} = \mu\wedge \Phi(\bar\mu)
\end{equation}
\begin{equation}
    \{ \bcal \alpha , \acal\wedge\mu \}_{\p\s} = d_\acal\alpha\wedge\mu = \acal\wedge[\alpha,\mu] + d\alpha\wedge\mu
\end{equation}
All these equations hold pointwise on the codimension 2 boundary $\p\s$ and carry traces over the Lie algebra implicitly. Note in particular that there are central terms in the Poisson brackets involving $\acal$, even with itself
%\footnote{As further expanded on in appendix \ref{App:CornerAlg}, these central terms are ambiguous by codimension 3 terms. On closed codimension 2 surfaces, this is no issue, but when considering punctured surfaces as we do later, we make the choice presented here, given by realising $\bcal$'s action on $\acal$ through the bulk gauge transformation, e.g. $\{\bcal\alpha,F\}=X_\alpha[F]$.}
. This is a key point - it is the reason we can associate a Chern-Simons type phase space to the corner. For comparison, in Chern-Simons theory, the Atiyah-Bott symplectic form induces a Poisson bracket of the form
\begin{equation}
    \{A_a^i, A_b^j\} = \delta^{ij}\epsilon_{ab}
\end{equation}
which is essentially the same as the one we have for $\acal$ once we choose a basis of the Lie algebra that diagonalises $\Phi$.\\ 

Note, however, that we can also give a symplectic form $\omega_{\p\s}$ to the corner fields which gives the same Poisson relations as above, and it does reduce to the Atiyah-Bott form under certain conditions. By itself, the algebra above does not come from an invertible Poisson bivector. Instead, then, we perform a doubling of the phase space variables. We need to introduce extra fields $\phi,q$ which function as corner reference frames for the gauge transformations, and which will be canonical conjugates to $\bcal,\acal$. 
More precisely, suppose we extend the corner phase space manually by fields
\begin{equation}
    (\phi,q)\in \Omega^0(\p\s,G)\times\Omega^1(\p\s,\gfrak)
\end{equation}
and introduce 1-forms on phase space (field space connections~\cite{gomesUnifiedGeometricFramework2018,gomesUnifiedGeometricFramework2019})
\begin{equation}
    \chi = \phi^{-1} \delta \phi  \qquad \psi = \delta q - [q,\chi] = \phi^{-1} \delta(\phi q \phi^{-1} )\phi
\end{equation}
These then fulfil the critical properties
\begin{equation}
    \delta\chi + \frac{1}{2}[\chi,\chi]=0=\delta\psi+[\psi,\chi]
\end{equation}
which state that the field-space curvature of these connections is zero.
Then a corner symplectic form which reproduces the Poisson brackets is given by
\begin{equation}
   \begin{aligned}
       \Omega_{\p\s}=\delta \Theta_{\p\s} &=\int_{\p\s}
        \delta\bcal\chi - \bcal\frac{1}{2}[\chi,\chi] 
         + \delta\acal\wedge\psi 
        - 
        \psi \wedge d_\acal\chi - \frac{1}{2}\psi\wedge\Phi(\psi) 
   \end{aligned}
\end{equation}
where the corresponding symplectic potential is\footnote{For the reformulation, a useful identity is $\delta \acal^\phi=\phi( \delta\acal - d_\acal\chi)\phi^{-1} $.}
\begin{equation}\label{eq:codim2BFBBBsymppot}
    \Theta_{\p\s}=\int_{\p\s} \bcal\chi + \acal^\phi\wedge\delta(\phi q\phi^{-1}) - \frac{1}{2}\Phi(\phi q\phi^{-1})\wedge \delta(\phi q\phi^{-1}).
\end{equation}

Here, we let $ \acal^\phi := \phi \acal\phi^{-1} + \phi d\phi^{-1}$. With the background of defining extended phase spaces for finite regions, we can recognise this as a corner symplectic potential that accounts for the relation between a region $\s$ and its environment. We should therefore interpret the fields $\bcal,\acal$ in it as dressed versions of the bulk fields, which do not transform under bulk gauge transformations. Similarly, the frames $\phi,q$ encode the relation between the slice and its environment as far as gauge properties are concerned, and should transform both under bulk gauge transformations and corner frame reorientations, which will be the focus in what follows. \\
The first term in this potential is well-known from a number of setups~\cite{freidelEdgeModesGravity2020,rielloEdgeModesEdge2021,geillerLorentzdiffeomorphismEdgeModes2018,araujo-regadoSoftEdgesMany2025,Goeller:2022rsx} in slight variations, and is the canonical symplectic potential on $T^\ast G$. It describes the data relating to Yang-Mills transformations and corner charges. The other term is less standard, and does the same thing for the 1-form parametrised Kalb-Ramond transformations. This is also the only part where $\Phi$ enters, and \eqref{eq:TopProperty} ensures this is a $\delta$-closed 2-form independently of the other terms.\\
The additional fields present here parametrise the orbit of gauge transformations, and are referred to somewhat equivalently as Stückelberg fields, Goldstones, or (gauge) edge modes. They may be interpreted as extrinsic reference frames~\cite{Carrozza:2022xut,Carrozza:2021gju}, which relate the region $\s$ under consideration with some hypothetical complement in the full spacetime. We will be interested in this symplectic potential as it can be straightforwardly reduced by the corner constraint $F_\acal=\Phi(\bcal)$, circumventing a lengthy Dirac bracket discussion.\\
%In particular, here once again $\Phi$ appears, and on closed $\p\s$ it is not necessary to assume that it is invertible. We will, however, also be interested in placing this same potential on subsets $S\subseteq \p\s$ with boundary. In these codimension 3 boundaries, one acquires extra terms which will potentially involve $\Phi^{-1}$. \\
%A somewhat technical remark is in order at this point. Without the addition of the $\chi$-frames and $\psi$-frames, the corner Poisson algebra is not really realised as coming from a symplectic form. The easiest case to see this is when just looking at the $\bcal$-fields. There is no symplectic form of the $\bcal$ alone that inverts to the given Poisson algebra. The symplectic form we give, then, as we said, is analogous to the canonical one on the cotangent bundle $T^\ast C$ of any configuration space $C$, with the fields $\bcal,\acal$ functioning as momenta. In certain settings, we will be able to remove the frames again, yielding symplectic forms for only $\acal$. This means that we need to strictly distinguish the Poisson versus symplectic phase spaces in this work. Optimally, we would only work with Poisson algebras, but it will be highly convenient to invoke the extension by frames in order to facilitate calculations and a straightforward reduction of the phase space by constraints later on, in the context of gravity.\\  
In effect, we will now consider not just this kinematical corner phase space, but also the action of residual bulk constraints on it, which will bring further structure with it. Before we move on, we note however that the $q$/$\psi$-frames can be consistently removed without affecting invertiblity or the Poisson algebra of $\acal,\bcal$: by imposing $ \acal^\phi= \Phi(\phi q\phi^{-1})$, we can reduce the previous form to
\begin{equation}
   \begin{aligned}
       \Omega_{\p\s}' =\int_{\p\s}
        \delta\bcal\chi - \bcal\frac{1}{2}[\chi,\chi] 
         +\frac{1}{2} \delta_\chi\acal\wedge\Phi^{-1} \delta_\chi\acal
         \qquad
         \delta_\chi\acal:= \delta \acal - d_\acal \chi.
   \end{aligned}
\end{equation}
So strictly speaking, only the $\phi$/$\chi$ frames are necessary to have invertibility. In the gravity case, this will allow us to remove the $\psi$ frames and find a minimal corner Poisson algebra.
This presentation clearly shows the relation to the Atiyah-Bott symplectic form of Chern-Simons theory.

\subsection{Codimension $2\rightarrow 3$: Corner phase space analysis}
The Poisson structure on $\bcal,\acal$ we gave before is \textit{kinematical} in the sense that, while it reproduces the corner charge Poisson relations correctly, it does not necessarily account for the presence of residual corner constraints. What do we mean by this? Consider how constraints arise, first in the bulk. There, we had an equation $d_\acal\bcal=0$, which must be satisfied by physical histories of the theory. Because it is a $3$-form, and codimension 1 slices $\s$ are 3-dimensional, one gets a nontrivial pullback of this equation to $\s$, and thus a constraint on initial data of the physical histories. Then, from this constraint, one can reverse engineer a weakly vanishing, differentiable function $J_\alpha$ on the phase space of $\s$, and use the kinematical Poisson bracket to see it generates gauge transformations, and in particular redundancies in the bulk.\\
Once we restrict our attention to the corner, the $3$-form $d_\acal\bcal$ has trivial pullback - it does not provide any restriction on initial data \textit{on the corner}. This is however not the case for the other equation of motion, $F_\acal=\Phi(\bcal)$, which is a 2-form and can thus be pulled back into a nontrivial constraint on the corner.\\
We therefore can repeat the same procedure as with the bulk slice: We use this constraint on physical corner data to check whether there are corner redundancies or changes in the on-shell Poisson structure, and if there are puncture charges that inform a puncture Poisson algebra. \\
For this, we will now study 2-dimensional subsets $S\subseteq \p\s$ of the corner. These may themselves have dimension 1 boundaries, which we refer to as punctures (a disjoint union of circles).
We will treat these 2-dimensional regions $S$ as partial Cauchy slices in their own right, and assign the corner symplectic potential to them.
When we do so, we need to first fix a slight technical issue.\\
The fundamental property of the Poisson algebra on the corner derived from $\Omega_{\p\s}$ is that the functions $\acal,\bcal$ obey the Poisson brackets of the bulk corner charges. This includes the property that these functions, when smeared arbitrarily, are functionally differentiable. This is no longer the case when we work on a generic $S\subset \p\s$ with boundary $\p S$.\footnote{ I.e. the function $\int_S \acal\wedge\mu $ generates a transformation $V$ in the interior of $S$ which changes $q,\phi$. If in turn we contract $V$ with the symplectic form when punctures are present, then puncture terms involving $\phi$ appear which cannot be matched with the original function that generates $\acal$.}  \\
To fix this, we need to modify the symplectic form \eqref{eq:codim2BFBBBsymppot} by a puncture term
\begin{equation}
    \begin{gathered}
        \Omega_S =\int_{S}
        \delta\bcal\chi - \bcal\frac{1}{2}[\chi,\chi] 
         + (\delta\acal-d_\acal \chi)\wedge\psi 
         - \frac{1}{2}\psi\wedge\Phi(\psi)
        -\oint_{\p S} \psi \chi + q \frac{1}{2}[\chi,\chi]
    \end{gathered}
\end{equation}
which is exact, as it can be written as $\delta(q\chi)$. This additional term makes the smeared fundamental variables $\int_S \acal\wedge\mu$ differentiable in the presence of punctures. It has, however, the effect of making the differentiable version of $\bcal\alpha$ slightly different. So, in this phase space, the differentiable basis generators of the Poisson algebra are instead 
\begin{equation}
    \acal(\mu) := \int_S \acal\wedge\mu
    \qquad
    \bcal(\alpha) := \int_S \bcal \alpha - \oint_{\p S} q\alpha
\end{equation}
which reduce to the usual smearings of $\acal,\bcal$ when no punctures are present. They satisfy the same corner Poisson algebra as before:
\begin{equation}
    \{ \bcal (\alpha) , \bcal  (\bar\alpha) \}_{S} = \bcal ( [\alpha,\bar \alpha])
\end{equation}
\begin{equation}
    \{ \acal(\mu) , \acal(\bar \mu)\}_{S} = \int_S \mu\wedge \Phi(\bar\mu)
\end{equation}
\begin{equation}
    \{ \bcal (\alpha) , \acal(\mu) \}_{S} =  \acal([\alpha,\mu]) + \int_S d\alpha\wedge\mu
\end{equation}
This sets up the phase space in which we will work on codimension 2 corners $S$ with codimension 3 boundaries $\p S$. As already stated, there is an imprint of $F_\acal - \Phi(\bcal)=0$ on all such surfaces, and we need to impose it to find the physical phase space. Importantly, the constraint $d_\acal\bcal=0$ does not need, and can not be imposed here. For the purpose of defining the physical corner phase space, let us consider the differentiable function obtained by smearing the constraint against a parameter $\alpha\in\Omega^0(S)\otimes\gfrak$:
\begin{equation}
    \bar\kkcal_\alpha = -\int_S (F_\acal-\Phi(\bcal))\alpha + \oint_{\p S} (\acal-\Phi(q))\alpha.
\end{equation}
This function generates the transformation on phase space
\begin{equation}
    \begin{gathered}
        \{\bar\kkcal_\alpha,\bcal\}_S =  - [(F_\acal-\Phi(\bcal)),\alpha] \quad 
        \{\bar\kkcal_\alpha,\acal \}_S = 0
    \end{gathered}
\end{equation}
\begin{equation}
   \{\bar\kkcal_\alpha,\phi\}_S= \phi\Phi(\alpha)\qquad  \{\bar\kkcal_\alpha,q\}_S= [q,\Phi(\alpha)] -d_\acal\alpha 
\end{equation}
and in particular we have that
\begin{equation}
    \{\bar\kkcal_\alpha, \bar\kkcal_{\bar\alpha} \}_S = \bar\kkcal_{\Phi[\alpha,\bar\alpha]} + \oint_{\p S} d\alpha \Phi \bar\alpha .
\end{equation}
Here, $\Phi$ explicitly appears in the algebra. 
The function $\bar\kkcal_\alpha$ is a constraint for all $\alpha$ vanishing on $\p S$. Note that these transformations are \textit{not} the gauge transformations of the bulk theory. \\
The precise Poisson relations between the constraints show they are \textit{first class}, i.e. the Poisson bracket of constraints vanishes on the constraint surface. In the symplectic setting, the subset where they vanish is coisotropic in the phase space, and with a quotient by the redundant transformations, we again end up with a symplectic phase space as we know it. \\ 
In more pedestrian terms, \textit{we impose the constraints, and then need to further quotient by gauge transformations, or impose a gauge fixing}. On the level of observables, this means that observables must weakly Poisson commute with the constraints (weak Dirac observables). This is the case for $\acal$, which is even a strong Dirac observable (Poisson commutes exactly), but also for $\bcal$, as its Poisson bracket gives the constraint $F_\acal-\Phi(\bcal)$. Both $q,\phi$ in the interior of $S$ are neither weak nor strong Dirac observables, though. However, there is a family of Dirac observables $\bar q$ (\ref{eq:qDiracObs}) made from combinations of $q,\acal,\phi$ that replace $q$ on any gauge fixed surface.   \\
As by standard theory of constrained systems, the Poisson brackets of the Dirac observables $\bcal,\acal, \bar q$ descend to the on-shell phase space as-is, and we can impose the constraints in a straightforward way as no Dirac bracket is necessary. The reduction also makes $\phi$ pure gauge, and it can be gauge fixed away \textit{in the interior of $S$}. That is, we can keep variations $\chi$ at $\p S$ by gauge fixing $d\phi \phi^{-1}=0$. We give details for this reduction in appendix (\ref{App:TopReduction}).\\
The reduction to the physical phase space proceeds by setting $\phi$ to a spacetime constant configuration, which technically is a gauge-fixed phase space. There, the symplectic form is best parametrised through $\bar \acal = \phi\acal\phi^{-1}, \bar q = \phi q \phi^{-1}$ and the constant $\phi$,
\begin{equation}
    \begin{gathered}
        \Omega_S =\int_{S}
         \delta \bar\acal\wedge\delta \bar q 
         - \frac{1}{2}\delta \bar q\wedge\Phi \delta \bar q + 
         \Omega_S^\chi.
    \end{gathered}
\end{equation} 
Here,
\begin{equation}
    \Omega_S^\chi = \delta \left(  ( \int_S \Phi^{-1}F_{\bar\acal} -\oint_{\p S}  \bar q) \;\delta\phi \phi^{-1} \right).
\end{equation}
This physical symplectic form can be rewritten in an illustrative form by defining a second connection (we have $\{\bar\kkcal_\alpha, \acal^{op}\} = d_{\acal^{op}} \Phi(\alpha)$),
\begin{equation}
    \bar\acal^{op}:= \bar\acal -   \Phi(\bar q),
\end{equation}
\begin{equation}
    \begin{gathered}
   \Omega_{S} 
         = 
         \int_{S} \frac{1}{2}\delta  \bar\acal \Phi^{-1} \delta \bar\acal - \frac{1}{2}\delta \bar\acal^{op} \Phi^{-1} \delta \bar\acal^{op} + \Omega_S^\chi \\
         = \Omega_{\bar\acal} - \Omega_{\bar\acal^{op}} + \Omega_S^\chi.
    \end{gathered}
\end{equation}

This shows that the resulting symplectic form is, in fact, just the difference of two Atiyah-Bott forms, giving in particular
\begin{equation}
    \{\bar\acal\wedge\mu,\bar\acal\wedge\bar\mu\}_S = \mu\wedge\Phi(\bar\mu)
\end{equation}
and the same for $\bar\acal^{op}$. In fact, this shows $\bar\acal^{op}$ is an alternative Dirac observable to $\bar q$ which is more natural for the physical symplectic form. Rewriting the relation between the two, $\Phi(\bar q) = \bar \acal-\bar \acal^{op}$ and reminding ourselves that the frame $q$ functions as a relation between the inside and outside of the bulk regions, we can think of $\bar\acal^{op}$ as the connection on the complement of the bulk region $\s_L$. Then, this difference in signs in the symplectic form is simply from the difference in orientations.\\ 
To make this precise, recall that $\acal$ is really the dressed version of a bulk field $\acal^b$, and does obeys a continuity equation up to gauge transformation. I.e. if $\acal_R$ is the corresponding field of the complement region of our stating slice $\s_L$, the continuity should read\footnote{The exact details depend on how $q$ is parametrized, i.e whether one transforms the bulk field with $q$ or $\phi q\phi^{-1}$.}
\begin{equation}
    \phi_L(\acal_L- \Phi(q_L))\phi_L^{-1}  - d\phi_L \phi_L^{-1}  
    = \acal_L^{b} = \acal^b_R=
    \phi_R(\acal_R- \Phi(q_R))\phi_R^{-1}  - d\phi_R \phi_R^{-1}.
\end{equation}
Now we have set the fields $\phi$ constant in our reduction process, so the differential terms drop. Then, we can read the continuity equation as
\begin{equation}
    \bar \acal_L - \Phi( \bar q_L) = \bar\acal_R- \Phi(\bar q_R)
\end{equation}
which, in the glueing of corner phase spaces, we interpret as a matching of corner charges. Only the difference $\bar q_L-\bar q_R$ appears.
We can take this one step further, as this matching constraint, on two copies of the corner phase space, induces a redundancy which shifts\cite{donnellyLocalSubsystemsGauge2016} $(\bar q_L,\bar q_R)\mapsto (\bar q_L+\mu,\bar q_R+\mu)$. (This invariance is analogous to the group algebra modular tensor product $\otimes_{\cbb[G]}$ found in Lattice Gauge Theories for finite groups.) With this invariance, we can fix $\bar q_R=0$, and then the interpretation of $\bar\acal^{op}$ is as follows:\\
The $d\phi\phi^{-1}$ gauge fixed "2-sided" corner phase space coordinatised by $\bar\acal,\bar q$, which features the extension by the gauge frame $\bar q$, is equivalently two un-extended "1-sided" corner phase spaces coordinatised by $\bar\acal_L,\bar\acal_R$ each where $d\phi_L\phi_L^{-1}=d\phi_R\phi_R^{-1}=0$ and $q_R=0$ have been imposed together with the matching/continuity constraint.

Summarising for now, we have found that BF-BB theory induces a corner phase space described by Atiyah-Bott symplectic forms, so the corner Poisson algebra is that of Chern-Simons theory. We did not have to use any timelike equations of motion for this, showing that the kinematics of the phase space, including the constraint structure, are enough to conclude this.\\
We will now go even one step further and consider not codimension 2, but codimension 3 strata of spacetime, by the same exact logic as before.

\subsection{Codimension 3: Punctures}\label{sec:BFBB:Puncture}
Repeating the same procedure, we notice that the Poisson bracket of $\bar\kkcal_\alpha$ carries a central term on punctures. This signals that the Poisson relations of $\acal^{op} = \acal-\Phi(q)$ must change on punctures. This can be encoded in the \textit{puncture brackets}
\begin{equation}
    \{ \acal^{op} \alpha, \acal^{op}\bar\alpha\}_P =  \acal^{op} \Phi [\alpha,\bar\alpha] + \bar\alpha \Phi d\alpha,
\end{equation}
which fully encode the Poisson brackets of this fields on the punctures. Strictly speaking, this is only the Poisson bracket of the component $\acal^{op}_\theta$ of $\acal^{op}$ along the punctures.
While $\bcal$ is not included here, it is not a function one can even evaluate on the punctures in the first place, and instead arises from the gradient of $\acal$ away from the punctures.\\
More importantly, though, this is not a bracket for $\acal$. It is tempting to look at the form of the phase space and state that $\acal$ should also obey the same Poisson relation as $\acal^{op}$. In particular, consider the functions
\begin{equation}
    \tilde{\mathcal{K}}_\alpha := \int_S F_\acal \Phi^{-1}\alpha - \oint_{\p S} \acal \Phi^{-1}\alpha
\end{equation}
which are on-shell equal to $\bcal(\alpha)$ and thus generate Yang-Mills transformations on $\acal$. They obey the same algebra as the $\bar\kkcal_\alpha$, only difference being $\Phi^{-1}$ appears,
\begin{equation}
    \{ \tilde{\mathcal{K}}_\alpha, \tilde{\mathcal{K}}_{\bar\alpha} \}_S =
    \tilde{\mathcal{K}}_{\Phi^{-1}[\alpha,\bar\alpha]}- \oint_{\p S} \bar\alpha\Phi^{-1} d\alpha.
\end{equation}
\\
Let us pause here for a moment to point out a curiosity of the corner algebra logic: there are off-shell three distinct charges on the corner phase space which have slightly different status.
{\begin{enumerate}
    \item $\bcal(\alpha)$: Smeared fundamental variable, but also codimension 1 corner charge for Yang-Mills gauge transformations of $\bcal$. Generates rotations of the $\chi$-frame on the corner. Nonvanishing, thus not redundancies.
    \item $ \bar \kkcal_\alpha$: Corner constraint, does not generate any usual gauge transformations. Vanishing, thus redundancies.
    \item $\tilde\kkcal_\alpha$: Off-shell corner generator for Yang-Mills transformations on $\acal,\psi$. Nonvanishing, thus not redundancies.   
\end{enumerate}
}
The statement $\bar\kkcal_\alpha\approx0$ implies $\bcal(\alpha)\approx \tilde\kkcal_\alpha$, i.e. on-shell of the corner constraints, the generators coming from codimension 1 and 2 agree. \textit{Perhaps surprisingly, this property is special to the topological case where \eqref{eq:TopProperty} holds.} This is because when $\Phi$ does not pass through Lie brackets, the constraint $\bcal-\Phi^{-1}F_\acal$ is no longer covariant under Yang-Mills transformations. Instead then, the charges 1 and 3 will in principle be different even on-shell of charges 2 vanishing, and charges 3 will involve additional terms. We will see this explicitly in the gravity case in section \ref{sec:Gravity:CornerConstraints}. \textit{In other words, the generators of gauge transformations in the non-topological case are not necessarily the same in the codimension 1 and codimension 2 phase spaces, even on-shell. In fact, there is no guarantee that on the corner phase space, all gauge transformations are integrable/symplectic.} \\

However, in a special sector of the phase space, there is of course still an algebra for $\acal$. 
\textit{If we take the functions} $\tilde{\mathcal{K}}_\alpha$ \textit{as constraints} for $\alpha$ vanishing on $\p S$, this gives a puncture bracket for $\acal$, in the form
\begin{equation}
    \{ \acal \alpha, \acal\bar\alpha\}_P =  \acal \Phi^{-1} [\alpha,\bar\alpha] - \bar\alpha \Phi^{-1} d\alpha.
\end{equation}
However, this is not warranted without reason; the bulk equations of motion do not imply that they are constraints. This is an additional, manual input in the form of boundary conditions on the bulk region, e.g. $\s$.
\textit{We therefore take as a strict result that there is a corner charge algebra for $\acal^{op}$, and regard the one for $\acal$ as an additional, special case result.} Of course, this particular result is not surprising: It is just the statement that phase spaces with Atiyah-Bott form and a flatness constraint carry current algebras on their boundaries.\\
However, as far as kinematics go, the symmetry of the phase space under switch of $\acal,\acal^{op}$, may provide additional motivation for considering a current algebra also for $\acal$.\\

Unlike in lower codimensions, there are no more constraints to satisfy, and this is really the final phase space sitting on punctures. We can immediately recognise its structure: the Poisson brackets describe two copies of the Kac-Moody algebra for the gauge algebra $\gfrak$. This comes with many benefits, and is expected given the appearance of Chern-Simons structures before.\\
In particular, a construction of an energy-momentum tensor and associated diffeomorphism generators on the punctures is provided by the Sugawara construction, which exists for all Lie algebras $\gfrak$ carrying an invariant bilinear form. All the Lie algebras we will consider carry such a form, though it may not be equal to the Killing form. \\
We especially point out the interpretation of this algebra as the Poisson algebra of a non-chiral Wess-Zumino-Witten model~\cite{eberhardtWessZuminoWittenModels,gawedzkiCanonicalQuantizationBoundary2004}, via the identification with the current $\tilde J$ of the WZW as
\begin{equation}
    %i_P^\ast \acal = J\quad 
    i_P^\ast \acal^{op} = \tilde J
\end{equation}
which is physically sensible as we do interpret $\acal^{op}$ as a puncture charge current. \\
%From the point of view of the puncture, then, there are uncoupled left and right moving sectors (kinematically speaking). What does it mean in terms of dynamics, though? We have so far not specified boundary conditions $\bcal_t,\acal_t$ which would allow us to directly state things about this. However, we can reverse the question and ask what kind of dynamics would give rise to a manageable on the punctures. e.g. we could ask for the dynamics on the punctures to say
%\begin{equation}
%    \partial_t J = 0
%\end{equation}
%which gives an infinite number of conserved currents, and thus is particularly easy to quantise. This is the condition $\partial_t A_\theta=0$, translating into $\Phi(\bcal_{t\theta}) + D^\acal_\theta \acal_t = 0 $. Therefore, this condition actually only constrains one component of $\bcal_t$, and $\acal_t$ is not restricted. The example $\acal_t = \Phi(\alpha_t), \bcal_t = -d_\acal\alpha_t , \delta\alpha_t\stackrel{P}{=}0$, which closes the system, fulfils this condition in particular. However, we stress that this is not the only option, and rather a choice. \\
Let us again summarise the result so far. Starting from BF-BB theory in 4D, we work on a codimension 1 slice $\s$ and induce a Chern-Simons like phase space in codimension 2, with symplectic form \eqref{eq:codim2BFBBBsymppot}. By pulling back the constraints to codimension 2, we can further repeat this procedure and induce a codimension 3 bracket. This one is not fully determined, and instead gives rise to an algebra for $\acal^{op}$, but in special sectors the punctures carry two copies of a Kac-Moody current algebra. In all this, we used that the constraints that appear are all first class.\\
We give examples of how to apply these kinematics in abelian BF-BB theories in appendix \ref{sec:Examples}, where we also discuss different kinds of dynamics. Moreover, one may compare the results of this section to the 3D case, see appendix \ref{app:3D}.\\
We now move on to the more interesting case of study, which is 4D gravity. The same first-class property will no longer hold there.

\section{Corner and puncture data in 4D gravity}\label{sec:Gravity}
So far, we have treated a  generic BF-BB type theory whose gauge group we left unspecified. The relation to gravity becomes apparent when we make a specific choice, which at first seems peculiar, but which will make sense from a phase space point of view soon.
The choice of gauge group/algebra we make is
\begin{equation}
    \gfrak=\mathfrak{h}\ltimes (\mathfrak{p}\tilde\oplus \mathfrak{h}^\ast)
\end{equation}
with $\pfrak=\rbb^{1,3}, \mathfrak{h}=\mathfrak{so}(1,3) $ and the twisted sum such that $[\pfrak,\pfrak]\subseteq \hfrak^\ast$. In the literature, this is known as a \textit{double extension}~\cite{figueroa-ofarrillNonsemisimpleSugawaraConstructions1994}  of $\pfrak$ by $\mathfrak{h}$, or \textit{quadratic extension}~\cite{cattaneoGravityTorsionDeformed2024}  of $\hfrak$ by $\pfrak $, and this specific example is also known under the name of the \textit{Maxwell algebra}~\cite{Bacry:1970ye,Schrader:1972zd}, so named for its appearance in the system of a point particle in a fixed electromagnetic background. \\
The above gauge algebra allows for a close approximation of gravity by a BF-BB theory; in a certain kinematical presentation~\cite{deAzcarraga:2010sw,Durka:2011nf}, the symplectic structure is the same, differences arising in the form of some of the constraints being second class. The precise equivalences between deformations of this type of BF-BB theory and gravity have been explored elsewhere~\cite{cattaneoGravityTorsionDeformed2024}, and we are here instead more interested in using this specific kinematical setup as a means to understand the corner structure of gravity in a more straightforward way\footnote{A similar but contrasting approach is given for example via $BF^n V$~\cite{canepaCornerStructureFourDimensional2024}, which yields not a Poisson but a Poisson-$\infty$ structure on corners for first-order gravity, and does not make use of embeddings into BF-BB, but other substantial machinery.} in close analogy to the topological case.\\
\textit{Our strategy will be to use the same off-shell kinematical corner structure as BF-BB, but use the actual gravitational constraints of the bulk, which have distinct behaviour from that of the topological theory. We then proceed again to codimension 3 and check what changes. }
\paragraph{First versus second class constraints}\,\\
An obvious caveat, which we point out pre-emptively, is that the logic of the corner Poisson structure refers to corner charges, i.e. codimension 2 terms in generators living on codimension 1 surfaces. The split between corner terms and bulk constraints is the same regardless of the nature of the constraints; however, in the gravitational case we will encounter second class constraints, in codimension 1 and 2. This means that corner terms still make sense, but the interpretation of these terms as generating gauge transformations is no longer accurate. Furthermore, the second-classness of the constraints means that the algebra no longer localises in codimension 2 on-shell. \\
Technically, what one should do is determine the set of second-class constraints, switch to the Dirac bracket, and then work with the effective algebra structure of the resulting first-class theory. In what follows, we do not follow this strategy, and instead impose the bulk constraints on the corner algebra of BF-BB. This, if anything, would be the more amenable strategy in quantisation and/or discretisation in the sense of state sum/spin foam models, and so it is more than worthwhile checking what the result of this path is and if it is reasonable. We begin by motivating the BF-BB type Lagrangian for 4D gravity, and then explore the corner structure that BF-BB would suggest for it.\\
The result will indeed be qualitatively different from the purely first-class case, in that the $(B,\phi)$ canonical pair on the corner phase space stays independent from the $(\acal,q)$ pair. 
Instead, in appendix \ref{App:SecondClassSchematic}, we will explore the logic of removing the second class constraints and see an expectation for the resulting algebra structure. %Then, in appendix \ref{App:SolveSecondClass}, we will attempt this for the same gravitational case.\\

\subsection{First-order gravity in BF-BB like form}
Our starting point is the Einstein-Cartan-Holst (ECH) Lagrangian for independent 1-forms tetrad $e^I$ and spin connection $\omega^{IJ}$, again with $I,J$ etc being 4D internal indices,
\begin{equation}
    L_0 = \frac{1}{2}\star_\beta e^2 \wedge F_\omega
\end{equation}
which has the form of an $\hfrak=\mathfrak{so}(1,3)$ BF Lagrangian for $B=\frac{1}{2}\star_\beta e^2 := \frac{1}{2}(\star+\beta)e^2 $ with the inverse Barbero-Immirzi parameter\footnote{The self-dual case $\beta=\pm i$ requires a separate treatment as usual, which we do not cover.} $\beta=\frac{1}{\gamma}\in \rbb$. We fix here again a bilinear form $\kappa(-,-)$ on the Lie algebra $\hfrak$ that we leave implicit. Starting from an $\hfrak$-BF theory, the form of $B$ is imposed usually via Lagrange multipliers either in a quadratic manner (Plebanski)~\cite{buffenoirHamiltonianAnalysisPlebanski2004} or a linear form~\cite{gielenClassicalGeneralRelativity2010}. However, a simpler version of this is possible, by extending the space of fields of BF theory by a tetrad explicitly, e.g. taking a Lagrange multiplier $c$ and writing
\begin{equation}
    L_0 = B \wedge F_\omega + c \wedge (\frac{1}{2}e^2 -{\star}^{-1}_\beta B).
\end{equation}
Here, $\star^{-1}_\beta:= \frac{\beta-\star}{1+\beta^2}$ denotes the inverse of $\star_\beta:= \star + \beta$.
This Lagrangian has the property of not having a canonical conjugate for the Lagrange multiplier $c$ and the tetrad $e$, which means the pullbacks of the equations of motion do not give phase space constraints. One can now manually extend the phase space by the Lagrange multiplier, as is standard. However, in this specific case this can be done alternatively  and more naturally by adding kinetic terms for $e$ and $c$. We first add a kinetic term with a new field $\tau$,
\begin{equation}
    L_0 = B \wedge F_\omega + \tau_I\wedge d_\omega e^I + c \wedge (\frac{1}{2}e^2 - \star^{-1}_\beta B),
\end{equation}
freeing up the variations of $e$ without constraining $c$, and another for $c$,
\begin{equation}\label{eq:expandedGRLagr}
     L_0 = B \wedge F_\omega + \tau_I\wedge d_\omega e^I + c \wedge (d_\omega S + \frac{1}{2}e^2 - \star^{-1}_\beta B),
\end{equation}
which soon allows us to pull back the equations of motion into phase space constraints (this is a phase space construction a la Kijowski-Tuczijew~\cite{kijowskiSymplecticFrameworkField1979}). Note that the latter addition $c d_\omega S$ is essentially just a Kalb-Ramond shift in $B$ of the previous Lagrangian with parameter proportional to $S$. This, mainly due to the Bianchi identity $d_\omega F_\omega=0$, makes it so that the codimension 0 dynamics is not affected by $S$ at all. \\
That this is gravity follows in a naive fashion from varying $\tau,c$ which impose torsion-freeness and a relaxed version of the simplicity constraint. Going on-shell of these two equations reduces the effective Lagrangian to that of ECH, with the subtlety that one should take $e$ to be nondegenerate to really correspond to classical GR. The equivalence can be shown in much more rigour~\cite{cattaneoGravityTorsionDeformed2024}, however. The main point of this reformulation is that this leads to a set of kinetic terms which can be resorbed into a BF-term again, for the previously mentioned Maxwell algebra. Therefore, the theory will feature corner charges which are linear in the fields, and will have one charge for each field. This then allows us to apply the corner Poisson algebra logic to a reasonable extent and compare to the BF-BB construction. In particular, our plan will be to impose some of the constraints on the BF-BB corner symplectic form in order to get an effective one for the gravity case.\\ 
The $\rbb^{1,3}$ part of the Maxwell algebra is easy to recognise as the pair $(\tau,e)$. The other pieces are slightly less obvious and we will show where they are in the next section. The curvature of the new `dual spin connection' $S$ is given as $Q:=d_\omega S + \frac{1}{2}e^2$. The structure of the curvature is what informs the Lie algebra structure in the Maxwell algebra: there is no commutator of the form $[S,S]$, showing that $S$ is valued in an abelian Lie algebra ($\hfrak^\ast$, the vector space dual of $\hfrak$). The additional $e^2$ piece is then the result of a commutator $[e,e]=e\wedge e$, giving the nontrivial part of the extension $\mathfrak{p}\tilde\oplus \mathfrak{h}^\ast$. We will now proceed by giving a concrete form that reduces to gravity.

\paragraph{Algebraic structures}\,\\
We will need some structures on the Maxwell algebra $\gfrak =\mathfrak{h}\ltimes (\mathfrak{p}\tilde\oplus \mathfrak{h}^\ast)$ to proceed. We use the fact that this is a matrix Lie group to simplify some notions.\\
We will, for completeness, allow for some freedom in the choice of inner products on the relevant Lie algebras, but it will not affect most of the final results due to judicious choice of normalisations of fields.\\
Let $\eta$ be the standard Minkowski metric on $\pfrak=\rbb^{1,3}$, which we use as a reference inner product. The definition of the Maxwell algebra involves a choice of an inner product on $\pfrak$, which we choose to be
%\footnote{The letter $k$ is chosen to illicit associations with the level of a Chern-Simons theory. In the current algebra, it will appear and play the role of the level of the Kac-Moody algebra.} 
$\Omega_\pfrak= k_\pfrak \eta$, with $0\neq k_\pfrak\in\rbb$. Let $P_I$ be an orthonormal basis for $\eta$.
%, so that $\tilde P_I:=\frac{1}{\sqrt{k_\pfrak}}P_I$ is one for $\Omega_\pfrak$
Let $\kappa$ be the inner product on $\hfrak=\mathfrak{so}(1,3)=\Lambda^2\pfrak$ induced from $\eta$, again used as a reference, and $M_{IJ}=P_I\wedge P_J$ an orthonormal basis for $\kappa$,
\begin{equation}
    \kappa(M_{IJ},M_{KL})=-\frac{1}{2}\Tr[M_{IJ} M_{KL}] = \eta_{IK}\eta_{JL}-\eta_{JK}\eta_{IL}.
\end{equation}
Then, $k_\pfrak^2\kappa$ is induced from $\Omega_\pfrak$.
The Lie bracket on $\gfrak$ is then
\begin{equation}
\begin{gathered}
        [(X,a,u),(Y,b,v)]:= \\(
        [X,Y]_\mathfrak{h}, 
        X\cdot b-Y\cdot a, 
        u([Y,-]_\mathfrak{h})- v([X,-]_\mathfrak{h}) + k_\pfrak^2\, \kappa(a\wedge b,-)  )
\end{gathered}
\end{equation}
where we denote matrix-vector multiplication by a dot $\cdot$. In particular,
\begin{equation}
    [(X,0,u),(Y,0,v)]_{\hfrak^\ast}(Z) = u([Y,Z]_\mathfrak{h})- v([X,Z]_\mathfrak{h}) +  k_\pfrak^2\, a\cdot Z\cdot b
\end{equation}
and we can stop working with linear functionals by introducing an independent invariant bilinear form $\Omega_\hfrak$ on $\hfrak$. We choose $\Omega_\hfrak=k_\hfrak \kappa$, $0\neq k_\hfrak\in \rbb$. Importantly, we have the freedom to choose $k_\hfrak$ independently from $k_\pfrak$. 
%Then, $\tilde M_{IJ}:=\frac{1}{\sqrt{k_\hfrak}}M_{IJ}$ is an orthonormal basis for $\Omega_\hfrak$.
This allows us to write $u(-)= \Omega_\hfrak(U,-)$ and we pick a reference dual basis  $N_{IJ}(-):=\kappa(M_{IJ},-)$.
%$ \tilde N_{IJ}:= \Omega_\hfrak(\tilde M_{IJ},-)$
This then presents the Lie bracket equivalently as
\begin{equation}
\boxed{
\begin{gathered}
    [(X,a,U),(Y,b,V)]= \\(
        [X,Y]_\mathfrak{h}, 
        X\cdot b-Y\cdot a, 
        [X,V]_\mathfrak{h}-[Y,U]_\mathfrak{h} + \varepsilon a\wedge b  ) \quad {\varepsilon:=\frac{k_\pfrak^2}{k_\hfrak}}.
\end{gathered}
}
\end{equation}
Then our choice of $k_\pfrak$ determines a 1-parameter family of invariant bilinear form on $\gfrak$,
\begin{equation}
    \langle (X,a,u),(Y,b,v)\rangle = 
    k_\pfrak \eta_{IJ}a^I b^J + u(Y) + v(X)
\end{equation}
or equivalently by dualising $u,v$,
\begin{equation}
   \boxed{ \langle (X,a,U),(Y,b,V)\rangle =    k_\pfrak a^I \eta_{IJ}b^J + k_\hfrak \kappa(U,Y) + k_\hfrak \kappa(X,V)).}
\end{equation}
This will be our `trace' throughout what follows. It is of course not the same as the Killing form on $\gfrak$, which is degenerate on the $\pfrak$ parts. For this reason, it is important that we keep $\Omega_\pfrak$ to have something nondegenerate.

\textit{From here on fully identify $\hfrak^\ast$ with $\hfrak$ through $\Omega_\hfrak$.} We can now introduce the BF fields
\begin{equation}
    \bcal = (\frac{1}{k_\hfrak}c,\frac{1}{k_\pfrak}\tau, \frac{1}{k_\hfrak} B)  \quad \acal = (\omega, e,S).
\end{equation}
This normalisation makes the $k$-prefactors drop out of most expressions.
Now consider the curvature of $\acal$:
\begin{equation}
    \begin{gathered}
        \frac{1}{2}[\acal,\acal] = ( \frac{1}{2}[\omega,\omega], \omega^I_J\wedge e^J P_I, [\omega,S] + \frac{1}{2}e\wedge e )\\
        F_\acal = (F_\omega, T:= d_\omega e,Q:= d_\omega S + \frac{1}{2}e\wedge e  ).
    \end{gathered}
\end{equation}
We recognise respectively the curvature of the spin connection, the torsion of $\omega$ with respect to $e$ and the curvature of the dual spin connection $S$. 

\paragraph{BF term}\,\\
We now have the ingredients necessary to expand the BF action (with the diagonal pairing $\gfrak\times\gfrak^\ast\rightarrow \rbb$):
\begin{equation}
    \langle \bcal, F_\acal\rangle = \kappa(B,F_\omega) + \tau_I\wedge T^I + \kappa(c, Q).
\end{equation}
This is already quite close to what we seek, meaning \eqref{eq:expandedGRLagr}, and only misses the $c B$ term. However, to have a nondegenerate bilinear $\Phi$ in the BF-BB theory, we will also need an additional $\tau\tau$ term.

\paragraph{BB term}\,\\
We now specify the map $\Phi$,
\begin{equation}
    \boxed{\Phi((X,a,U)) := ( k_\hfrak\star^{-1}_r {X} ,  t k_\pfrak a , k_\hfrak \star^{-1}_r {U})}
\end{equation}
which is invertible if $r \neq \pm i$ and $t\neq 0$, and which explicitly depends on the normalisation scales $k_\pfrak,k_\hfrak$. Our choice of normalisation for the $\bcal$ fields ensures that the scales will not appear elsewhere in what follows.
As a reminder, we use the notation
\begin{equation}
    \star^{-1}_r{U} := \frac{1}{1+r^2}(r-\star) U
\end{equation}
which denotes the inverse of the $\star_r U = (\star+r)U$ Lie algebra map.
It is important to note that the map $\Phi$ does not pass through commutators,
\begin{equation}\label{XiBulkMatrix}
    \begin{gathered}
    \Xi(\mathcal{X},\mathcal{Y}):= \Phi([\mathcal{X},\mathcal{Y}])-[\Phi(\mathcal{X}),\mathcal{Y}] \\
       = \Phi([(X,a,U),(Y,b,V)])-[\Phi(X,a,U),(Y,b,V)]\\
    =
    \Big( 0,
     (tk_\pfrak X- k_\hfrak\star^{-1}_r  X)\cdot b ,
     -\varepsilon (t k_\pfrak (a\wedge b) - k_\hfrak \star^{-1}_r  (a\wedge b))
    \Big),
    \end{gathered}
\end{equation}
i.e. it is not an equivariant map of Lie algebras, and so $\langle-,\Phi(-)\rangle$ is not an invariant inner product on $\gfrak$. This explicitly breaks the $\gfrak$-symmetry of the BF-term down to $\hfrak$ and leads to the appearance of second class constraints in the gravity case.\\
This gives the additional BB-term
\begin{equation}
  -\frac{1}{2}  \langle \bcal,\Phi(\bcal)\rangle = -  \kappa(c , \star^{-1}_r {B}) - \frac{t }{2}\tau_I \wedge\tau^I
\end{equation}
The result is the Lagrangian
\begin{equation}
\boxed{
    \begin{gathered}
        L_0 = \langle \bcal,F_\acal\rangle -\frac{1}{2}  \langle \bcal,\Phi(\bcal)\rangle \\
        =
         \kappa(B,F_\omega) +   \tau_I d_\omega e^I +  \kappa(c, Q
        -  \star^{-1}_r {B}) - \frac{t }{2}\tau_I \wedge\tau^I
    \end{gathered}
}
\end{equation}
which gives us 4D gravity. Let us show this explicitly. From now on, similar to section \ref{sec:BFBB}, we omit the inner product $\kappa$ in notation.

\paragraph{Equations of motion}\,\\
The variation of the Lagrangian gives the equation of motion in codimension 0
\begin{equation}
\begin{gathered}
        E_0 = 
    \delta B \wedge (F_\omega - \star^{-1}_r {c})
    + \delta \tau_I \wedge (d_\omega e^I - t \tau^I)
    + \delta c\wedge (Q -\star^{-1}_r {B})\\
    - (
    d_\omega B + \tau\wedge e -[c,S]
    )\wedge\delta\omega
    -(d_\omega \tau_I + c_{IJ}\wedge e^J)\wedge\delta e^I - d_\omega c\wedge\delta S
\end{gathered}
\end{equation}
which can be reduced to ECH gravity as follows. Solve the first three equations for $B,\tau,c$, by assuming invertibility. Then, reinserted into the other three equations, we have
\begin{equation}
    E_0 \hat{=} 
    - 
   \frac{1}{2} d_\omega (\star_\beta e^2 )\wedge\delta\omega
    - ( \star_\beta F_\omega)_{IJ}\wedge e^J\wedge \delta e^I 
\end{equation}
where $\boxed{\beta := r + \frac{1}{t}}$. These are simply the equations of motion of ECH gravity. The solution set is therefore exactly the same. Note in particular that the equation $d_\omega c = 0$ is redundant\footnote{`Redundant' in the sense of redundant systems of constraints~\cite{henneauxQuantizationGaugeSystems1992}.} and implied by the Bianchi identity on $F_\omega$. As for the counting of degrees of freedom, we will have to go to the phase space level. This is where the inclusion of $S,c$ will come in handy. We can then give the effective Lagrangian of the theory exactly, as
\begin{equation}
        L_{0,eff} \hat{=} L_{ECH} + d(\frac{1}{2t}e d_\omega e + \star_r S F_\omega) = L_{ECH} + d CS_3^{\Phi^{-1}}(\acal)
    \end{equation}
which differs from ECH gravity precisely by a boundary Chern-Simons term.\\
We note here already two important things: first, the inclusion of $S$ has relaxed the usual simplicity constraint $B= \frac{1}{2}\star_\beta e^2$ into
\begin{equation}
   B=\star_\beta Q = \star_\beta( d_\omega S + \frac{1}{2}e^2).
\end{equation}
This, as we will see shortly, does not modify the codimension 1 bulk phase space at all compared to ECH gravity. Furthermore, because the field $c$ acts as a Lagrange multiplier for this constraint, which \textit{linearly} identifies $B$ with $e\wedge e$, there is \textit{no} ambiguity of sectors, as there would be in Plebanski-type formulations, where a constraint is made on $B^2$. Second, the field $\tau$ is not just eliminated from the theory, it is actually set to \textit{vanish}. That is, if we solve for $B,\tau,c$, then one of the remaining equations is the Gauss constraint $d_\omega e^2 \approx 0$. In the sector of the field space where the tetrad $e$ (conversely, the metric) is invertible, the crucial implication
\begin{equation}
    d_\omega e^2 = d_\omega e \wedge e =0 \implies d_\omega e = 0
\end{equation}
holds on the spacetime level and \textit{independently} sets torsion to zero. This means, by the other equation of motion, that $\tau \approx \frac{1}{t}d_\omega e \approx 0$. This will be important in what follows, as it means that $\tau$ is \textit{not an interesting observable}. It does not give rise to a nonzero corner charge, and we will need to drop it from the phase space to obtain a well-behaved Poisson algebra.

\paragraph{Phase space}\,\\
We induce from the Lagrangian the symplectic potential current
\begin{equation}
    \theta_1 = B\wedge\delta\omega+\tau_I\wedge\delta e^I + c\wedge\delta S
\end{equation}
which is precisely that of a BF theory. On a codimension 1 slice, we can pull back the equations of motion into constraints which we then impose. Here, for now, we simply show the same reduction as above for the equations of motion. Inserting the expressions for the $\bcal$ fields gives 
\begin{equation}
    \theta_1 \hat{=} \star_r ( d_\omega S + \frac{1}{2}e^2 )\wedge\delta\omega+\frac{1}{t} d_\omega e_I\wedge\delta e^I + (\star_r F_\omega)\wedge\delta S.
\end{equation}
Here, we can see that the $S$-terms combine to give $d( S_{r}\wedge\delta\omega ) + \delta(F_\omega\wedge S_{r}) $, so it does not contribute to the bulk symplectic form. The remainder can be recombined\footnote{Using $ d_\omega e\delta e = \frac{1}{2}e^2\delta\omega + d(\frac{1}{2}e\delta e) + \delta ( \frac{1}{2}e d_\omega e) $} into
\begin{equation}
    \theta_1 \hat{=} \frac{1}{2}e^2_\beta \delta\omega + d( \frac{1}{2t}e_I\wedge\delta e^I + \star_r S\wedge\delta\omega ) + \delta( F_\omega\wedge \star_r S + \frac{1}{2t}e_I\wedge d_\omega e^I )
\end{equation}
which, up to exact and codimension 2 pieces, is the ECH symplectic potential. We will soon see the codimension 2 piece reappear once we perform the corner algebra logic. Again, then, if we impose the remaining equations of motion, and deal with the degeneracy of the ECH symplectic structure\footnote{What is meant is that transformations $\omega\mapsto\omega + \bar\Delta$ with $\star_\beta^{IJ}\wedge e_J = 0$ are in the kernel of the symplectic form, leading to \textit{reduced connections} parametrizing the phase space. These are restricted such that the Gauss constraint still implies torsion freeness on the slice $\s$. } as is typical~\cite{cattaneoReducedPhaseSpace2019,langenscheidtNewEdgeModes2025}, then the on-shell set is the same as in ECH, and the symplectic structure is the same, so the theories are equivalent in codimension 1. For more rigorous results on the equivalence of the theories, see~\cite{cattaneoGravityTorsionDeformed2024}.\\
One final remark about the phase space regards the tetrad $e$. We assume that it is nondegenerate, at least after we solve for $B,\tau,c$. This implies in particular that the field $\tau\approx 0$ at the level of the fully on-shell phase space, and so we may set it to zero at any given point. Only on the level of nondegenerate tetrads do we have a full correspondence to the ECH phase space.

\subsection{Codimension $1\rightarrow 2$: Second class constraints }
We first note the generic phenomenon: when $\Phi$ is no longer equivariant, the constraint system of BF-BB theory is second-class. While the Yang-Mills generators $J_\alpha$ are unaffected, the remaining brackets receive modifications:
\begin{equation}
\begin{gathered}
     \{J_\alpha^{BF-BB},K_\mu^{BF-BB}\}=K_{[\alpha,\mu]}^{BF-BB} + \oint_{\p\s} d\alpha\wedge\mu + \int_\s \bcal \, \Xi(\mu,\alpha) 
    \\
    \{K_\mu^{BF-BB},K_{\bar\mu}^{BF-BB}\}=  \oint_{\p\s} \mu\wedge\Phi(\bar\mu) + \int_\s \acal( \Xi(\mu,\bar\mu)-\Xi(\bar\mu,\mu) ).
\end{gathered}
\end{equation}
The bulk second class constraint matrix is here given by the constant  $\Xi(-,-)$ \eqref{XiBulkMatrix}, multiplied by the fields $\bcal,\acal$. This appears to be a problem for defining the corner algebra due to the second-class pieces. However, we study this case in generality in appendix \ref{App:SecondClassSchematic}, and because of two properties of this algebra, the corner logic still applies. The first of these properties is that the Poisson bracket of a charge with a constraint (vanishing parameter on boundary) gives only constraints. This specifically refers to the structure constants in $\{J,J\}=J, \{J,K\} = K + \cdots$. The second property is that the second-class matrix is ultralocal. The results of the appendix then let us see that on-shell, the algebra, under the \textit{Dirac bracket}, still localises on corners, with the same result as the kinematical bracket, but the discarding the bulk central terms. We can therefore extract the corner algebra in the same way as before. In any case, we will follow the strategy of reducing the BF-BB corner phase space to the gravity one.\\
We can split the BF-BB generators for Yang-Mills and Kalb-Ramond transformations into sets of 3 each, associated respectively with $\hfrak,\pfrak,\hfrak^\ast$. We call the first set of these (we use the names of~\cite{cattaneoGravityTorsionDeformed2024})
\begin{equation}
    H_\alpha = \int_\s B\wedge d_\omega \alpha -(\tau\wedge e)\alpha + [c,S]\alpha 
\end{equation}
\begin{equation}
    I_\phi = \int_\s \tau_I\wedge d_\omega\phi^I - c\wedge (\phi\wedge e)
\end{equation}
\begin{equation}
    J_\gamma = \int_\s c\wedge d_\omega \gamma
\end{equation}
for the Yang-Mills type transformations, separating into Lorentz transformations, tetrad shifts and Bianchi-transformations, which shift $S$ by a derivative and are associated to the redundant constraint $d_\omega c=0$.\\
The second set, the Kalb-Ramond type generators, is given by
\begin{equation}
    M_\mu = -\int_\s (F_\omega- \star^{-1}_r {c})\wedge\mu + \oint_{\p\s} \omega\wedge\mu
\end{equation}
\begin{equation}
    L_\rho = -\int_\s e_I\wedge d_\omega \rho^I - t \tau_I\wedge \rho^I
\end{equation}
\begin{equation}
    K_{\nu} = -\int_\s S\wedge d_\omega \nu + (\frac{1}{2}e^2 - \star^{-1}_r {B})\wedge\nu
\end{equation}
and separates into the `true Kalb-Ramond' $B$-shift, $\tau$-shift and $c$-shift transformations, the latter of which is associated with a modified simplicity constraint $B=\star_r(\frac{1}{2}e^2+d_\omega S)$. \\
We can see that this set of generators has as boundary value precisely the fundamental fields, as we wanted. They also all vanish in the bulk on-shell, as they are constructed from the equations of motion. We report their associated gauge transformations and algebra in appendix \ref{App:BulkAlgebra}. \\
Among the constraints, $K,L,I$ are a second-class subset. %In appendix \ref{App:SolveSecondClass}, we will investigate the bulk phase space by solving the second-class part of the constraints and analysing the rest, while in the main text w
We simply proceed along the lines of the BF-BB result to see where it takes us.\\
Summarising, we have presented 4D gravity in a parametrisation in which there is a corner charge for each field of the theory, at the cost of the presence of second-class constraints. The centerpiece of this is the Maxwell algebra $\gfrak=\hfrak\ltimes(\pfrak\tilde\oplus \hfrak^\ast)$, which contains dual Lorentz transformations. A non-equivariant Lie algebra map $\Phi$ creates a difference to the fully topological BF-BB case we studied before.  

\subsection{Codimension $2\rightarrow 3$: Failure of Jacobi, Chern-Simons data}

We begin by naively splitting up the BF-BB corner algebra into the pieces of $\gfrak$. The result is a large algebra which we report in appendix \ref{App:CornerAlg}. The presence of $\Phi$ makes itself known in the the brackets of $\{\omega,S\}$ and $\{e,e\}$. This result, by itself, is not without issue, as the Jacobi identity fails in exactly one case, between $\tau,e,S$: 
\begin{equation}
    \begin{gathered}
        \{\{\tau_I \phi^I,e^I\wedge\rho_I\},S\wedge\nu\} + \text{cyc.}
        = (\star^{-1}_r  \nu + t \nu )\wedge (\rho\wedge\phi) \neq 0
    \end{gathered}
\end{equation}
This is the first and most severe pathology we encounter. Since all we eventually care about are the fields $e,\omega$, and actually $\tau=0$ on-shell, there is a natural, but manual resolution here: \textit{we remove $\tau$ from the corner algebra}. The remainder, involving $B,c,\omega,e,S$ is then a Poisson algebra. \textit{Any expression involving $\tau$ will then have $\tau$ replaced by zero, in accordance to the equations of motion}. \\
This is sensible because $\tau$, in the gravitational model we consider, never presents a nontrivial observable: It is always set to zero when we consider nondegenerate tetrads $e$. Physically, it is devoid of content, and therefore even if we included it by modifying the Poisson algebra another way, we would just end up setting it to zero in the end. The main reason is that the constraints $ t \tau - d_\omega e\approx 0 \approx d_\omega  e$ are independently true, one due to being a bulk equation of motion, and the other due to being a consequence of the equations of motion and nondegeneracy. As nondegeneracy is an open condition, this situation describes all but a small subset of field configurations, which are also not relevant for gravity, where metrics are nondegenerate\footnote{We do not consider topology change for this work.}.\\

We see several interesting features. In particular, $(\omega,S)$ form a canonical pair, as well as the two components of $e$ with each other. The latter are in a Chern-Simons-like form for this reason. We can also see the appearance of central, derivative terms between the bulk pairs $(B,\omega),(c,S)$. 
%These are in principle ambiguous and we make the choice we presented, in which global (i.e. $d\alpha=0$) Yang-Mills type transformations receive no central term. \\
In principle, we could stop here and only consider the Poisson algebra. However, it will be useful to consider associated symplectic descriptions. 

We cannot peruse the BF-BB result immediately, as there is an issue: the term $-\frac{1}{2}\psi\Phi(\psi)$ in the corner symplectic form is not closed when \eqref{eq:TopProperty} does not hold. This is dual to the statement before that the corner algebra is not Poisson if one includes $\tau$.\textit{ Hence, to remove $\tau$ from the Poisson algebra, we also remove its conjugate, given by the $\chi_\pfrak$ frame, so we set $\chi_\pfrak=0$}. This then makes the symplectic form closed, and we can proceed as before.

\paragraph{Corner constraints} \label{sec:Gravity:CornerConstraints}\,\\
We now want to impose the same constraints as in the bulk, i.e. the Kalb-Ramond constraints, which can be pulled back onto the corner. Their differentiable equivalents read, on the punctured corner $S$,
\begin{equation}
    \bar M_\gamma = -\int_S (F_\omega- \star^{-1}_r {c})\, \gamma + \oint_{\p\s} (\omega- \star^{-1}_r  q_{\hfrak})\, \gamma
\end{equation}
\begin{equation}
    \bar K_{\alpha} = -\int_S S\wedge d_\omega \alpha + (\frac{1}{2}e^2 - \star^{-1}_r {B})\,\alpha -\oint_{\p S} \star^{-1}_r  q_{\hfrak^\ast}\alpha.
\end{equation}
This also technically includes the constraint $d_\omega e = t \tau$, in which we however have the appearance of $\tau$, which we set to zero before doing any corner phase space calculations, in compliance with our general rule:
\begin{equation}    
    \bar L_\phi = -\int_S e_I\wedge d_\omega \phi^I.
\end{equation}
This constraint is on the corner fields $e,\omega$ themselves, and not on the same status as the first two constraints. To get to the geometrical phase space, one only needs to eliminate $B,\tau,c$, and $\bar L_\phi$ is unnecessary for this.  

The constraints $\bar M$ are fully first-class, while already $\bar L, \bar K$ by themselves are second class. $\bar L,\bar K$ are also second class with each other.
%while $\bar K$ are second class.
This is the well-known statement that the simplicity constraints on the corner are second class.
As in codimension 1, this leads to an in-principle problem with defining the puncture algebra, which here would be defining brackets between $\tilde\omega = \omega - \star^{-1}_r  q_\hfrak, \tilde S = S-\star^{-1}_r  q_{\hfrak^\ast} $ and $\tilde e = e$, in analogy to the algebra of $\acal^{op}$ in the topological case.\\
Specialising directly the result from there gives the puncture algebra
\begin{equation}
    \begin{aligned}
        \{ \tilde S\alpha ,\tilde S \bar\alpha \}_{P} &= \tilde S \star_r^{-1}[\bar\alpha,\alpha] \\
        \{\tilde S\alpha ,\tilde \omega \gamma \}_{P} &=  \tilde \omega \star_r^{-1} [\alpha,\gamma] + \gamma \star_r^{-1} d\alpha \\
        \{\tilde S \alpha ,\tilde e_I \phi^I \}_{P} &= \tilde e_I  \frac{k_\pfrak}{k_\hfrak}t   \alpha^{IJ}\phi_J \\
        \{\tilde e_I \phi^I ,\tilde e_J \bar \phi^J \}_{P} &= \tilde \omega k_\hfrak  \star_r^{-1} (\phi\wedge\bar\phi) + t \bar \phi_I d \phi^I 
    \end{aligned}
\end{equation}
This, once again, is not an algebra for $S,e,\omega$.
However, all we care about at this stage is the form of the corner phase space on-shell of $\bar M, \bar K$. This can be much simpler, because we know how to explicitly solve the constraints in question. We study this reduction of the corner symplectic form in detail in appendix \ref{App:CornerSympForm}. The result is that the corner phase space is parametrised by $\omega,S$ and $e$ as well as some of the frames, and their Poisson relations are unchanged from the kinematical corner symplectic structure.  \\
%The nontrivial difference to the simple BF-BB setting is that the reduction is not uniform - the kernel of the on-shell symplectic form is defined through partial differential equations. As a result, it is not a priori clear to which degree one can remove the gauge frames $\chi_\hfrak,\chi_\pfrak$ from the phase space. \\
One finds that the symplectic form of the corner takes the shape 
\begin{equation}\label{Eq:SampleSympForm}
    \boxed{\Omega_S = \int_S \delta S\wedge \star_r\delta\omega + \frac{1}{2t }\delta e_I\wedge\delta e^I +\delta(\frac{1}{2}(\star_\beta e^2) \chi_\hfrak) }
\end{equation}
in which we see the conjugation of $(S,\omega),(e,e)$ as one would have it in BF-BB theory. In particular, we see the same conjugation of $(e,e)$ that has appeared in the literature previously~\cite{freidelBubbleNetworksFramed2019,freidelEdgeModesGravity2021}, implying that the corner metric components $q_{ab}=e_a\cdot e_b$ form an $SL(2;\rbb)$ Poisson algebra. \textit{However, in addition, we included, to the best of our knowledge, for the first time the spin connection $\omega$ on the corner into the Poisson algebra in a way that is consistent with the symmetries and that gives a bulk meaning to its conjugate variable $S$.} The only constraint that is left to impose on this phase space is the torsion constraint $d_\omega e =0$. \\ 
Another result, mentioned already in section \ref{sec:BFBB:Puncture} is that the corner charges, $B,c$, are not the generators of gauge transformations on the corner fields. We mean that from the codimension 1 bulk perspective, we would expect that the generator of Lorentz transformations with parameter $\alpha$ is the ECH corner charge, given by
\begin{equation}
    \int_S \frac{1}{2}(\star_\beta e^2) \alpha
\end{equation}
But rather, this is actually the generator of $\chi_\hfrak$-frame rotations from the corner perspective, and instead the Lorentz generator is
\begin{equation}
   \tilde K_\alpha:= \int_S \star_r S\wedge d_\omega \alpha - \frac{1}{2t}e^2 \alpha + \frac{1}{2}(\star_\beta e^2) \alpha. = -\oint_{\p S}  (\star_r S) \alpha +  \int_S  (\star_r Q)  \alpha .
\end{equation}
This, clearly, contains the Lorentz charge one would get from the bulk, and the charge affects the shift $I_{X_\alpha}\chi_{\hfrak}=\alpha $. These pieces combine to give the dual Lorentz curvature $Q$.
Similarly, the generator of dual Lorentz transformations is 
\begin{equation}
    \tilde M_\gamma = \int_S (\star_r F_\omega) \gamma - \oint_{\p S} (\star_r\omega) \gamma ,
\end{equation}
where again the Lorentz curvature $F_\omega$ appears. There is once again a swapping between the Lorentz and dual Lorentz pieces here.
The codimension 1 bulk has no direct corresponding corner charge; however, the Bianchi identity $d_\omega F_\omega=0$ allows one to extend this generator trivially into the bulk.
Finally, the charges for translational $\rbb^{1,3}$-Yang-Mills transformations is the most different: the original charge $\tau$ vanishes on-shell, or is not even considered on the corner. Instead, the translational Yang-Mills transformations $I_\phi$ are not even integrable/symplectic on the corner phase space: the contraction with $\Omega_S$ above gives 
\begin{equation}
    \delta(-I_{I_\phi}\Omega_S) = \int_S \phi_I (\star_\beta \delta\omega)^{IJ}\wedge\delta e_J \neq 0.
\end{equation}
Due to the pathologies involving $\tau$ and $\chi_\pfrak$, the situation here is slightly more complicated. Let us therefore first make a few insightful remarks about the first two generators.\\
We recognise that the first two of these tilde-generators are the same as the barred generators $\bar M,\bar K$ if the $\bcal$-fields vanish. If we once again take the BF-BB perspective, this makes sense if we view these $\bcal$ fields as \textit{sources} for the constraints of a Chern-Simons-like theory living in codimension $2$, on $S$. In the sourceless case, then, the constraints (the barred generators) are identical to the generators of gauge transformations on the corner fields. If sources are present, then these sources come with their own gauge frames encoded in $\chi$, and we can have separate generators which rotate said frames, coming from the constraints relating the bulk frames with the corner fields. \\
This gives us an alternative recipe for what the `correct' Yang-Mills $\rbb^{1,3}$ generator is: if we copy the structure of $\tilde K_\alpha$, which involves $S$ on punctures, and a curvature term $Q$, 
%in the rough form
%\begin{equation}
%   - \oint_{\p S}\langle \mathcal{X},\Phi^{-1}\acal\rangle + \int_{ S}\langle \mathcal{X},\Phi^{-1}F_\acal\rangle - \langle \mathcal{X},C\rangle 
%\end{equation}
%where $\mathcal{C}$ contains the corner charges coming from the codimension 1 bulk, 
then a sensible translational pendent is
\begin{equation}
    \tilde{L}_\phi :=  \int_{ S} \frac{1}{t} e_I\wedge d_\omega\phi^I.
\end{equation}
This is given by the translational curvature $d_\omega e$, but generates not the translational Yang-Mills gauge transformations; instead, 
\begin{equation}
    \{ \tilde L_\phi, - \} = d_\omega\phi  \frac{\delta}{\delta e} - \frac{1}{t}\star_r^{-1} (e\wedge\phi ) \frac{\delta}{\delta S},
\end{equation}
which are redundancies on-shell of $d_\omega e\approx 0$.
We note that the nonintegrability of the transformations can be cancelled if one does not look at the corner in isolation, but as \textit{coupled} to a codimension 1 ECH bulk $\Omega_\s$. However, in this case, the combined charge becomes
\begin{equation}
    P_\phi = -\int_\s (\phi\wedge e)\star_\beta F_\omega -\oint_{\p S} \frac{1}{t} e_I \wedge d_\omega \phi^I
\end{equation}
and so it vanishes on-shell unless one changes the setup carefully to allow punctures to appear in the expression. We want to avoid such specifics here, and particularly see the corner in isolation, to allow for reconstruction of the codimension 1 bulk from it.

\paragraph{Chern-Simons-like corner phase space}\,\\
The upshot of this bulk-to-boundary induction is that we will now consider as our phase space of interest the symplectic form
%, without the $\chi$-frames that remain, as we now know they pertain moreso to the bulk 
(again, $r+\frac{1}{t}=\beta$):
\begin{equation}
    \Omega_S = \int_S  \delta S\wedge \star_r\delta\omega + \frac{1}{2t }\delta e_I\wedge\delta e^I +\delta(\frac{1}{2}(\star_\beta e^2)\chi_\hfrak)
\end{equation}
with the tilde-generators $\tilde K, \tilde M, \tilde L$ as generators of gauge transformations. They satisfy the algebra 
\begin{equation}\label{eq:CornerAlgGrav4D}
    \boxed{\begin{aligned}
        \{\tilde K_\alpha, \tilde M_{\gamma}\} &= \tilde M_{[\alpha,\gamma]}-\oint_{\p S} \gamma \star_r d\alpha \\
        \{\tilde K_\alpha, \tilde L_{\phi}\} &=\tilde L_{\alpha\cdot\phi}\\
        \{\tilde K_\alpha, \tilde K_{\bar\alpha}\} &= \tilde K_{[\alpha,\bar\alpha]}\\
        \{\tilde L_\phi, \tilde L_{\bar\phi}\} &= \tilde M_{\frac{1}{t} \star^{-1}_r (\bar\phi\wedge\phi)} + \oint_{\p S} \frac{1}{t} \phi_I d\bar\phi^I
    \end{aligned}}
\end{equation}
Note that the structure constants in the first and last lines can easily be reabsorbed into the definition of $\tilde M_\gamma$, by switching to the alternative definition
\begin{equation}
    \tilde M_\gamma' := -\int_S F_\omega \gamma + \oint_{\p S} \omega \gamma
\end{equation}
which changes the Poisson brackets to 
\begin{equation}
    \begin{aligned}
        \{\tilde K_\alpha, \tilde M_{\gamma}'\} &= \tilde M_{[\alpha,\gamma]}-\oint_{\p S} \gamma d\alpha \\
        \{\tilde L_\phi, \tilde L_{\bar\phi}\} &= \tilde M_{\frac{1}{t}  (\phi\wedge \bar\phi)}' + \oint_{\p S} \frac{1}{t} \phi_I d\bar\phi^I
    \end{aligned}
\end{equation}
Of these, $\tilde L$ always becomes a constraint on-shell in vacuum due to torsion freeness $d_\omega e \approx 0$. The other two, in principle, are not constraints as curvature and dual curvature can be nonzero. In~\cite{ freidelLoopGravityString2017,freidelGravitationalEdgeModes2019}, this setting was strengthened by enforcing also $\tilde M$ to be a constraint, which makes the set $\tilde L,\tilde M$ first-class. In codimension 3, we will make use of this idea.\\
Summarising, we have specialised the result of the BF-BB case to the Maxwell algebra, and found that one needs to remove certain pieces in the algebra. However, after this, a reduction to the geometric phase space is explicitly possible, and yields a result almost identical to BF-BB theory. We find that the thus informed corner phase space carries a Chern-Simons structure, supplemented by gauge frames which have the bulk gravitational charges as their conjugate partners. The main difference to Chern-Simons theory is that no flatness is imposed a priori.

\subsection{Codimension $ 3$: Maxwell current algebra}

We can see that when $F_\omega = 0$, the constraints in $\tilde L$ are first class, and otherwise second class. In principle, the corner phase space here carries enough data to account for generic gravitational states, and it is an enticing idea to take full control of it in order to construct gravitational states, or boundary dynamics that gives rise to them.\\
However, in this work we will take only a preliminary step in this direction by considering the special subset of phase space where $\tilde M\approx 0$, i.e spacetime-flat corners. This is a practically very feasible sector in which not only the constraints become first-class but also the equations become essentially solvable. For simplicity we will also make $\tilde K$ into constraints, which does not restrict the gravitational flux $\frac{1}{2}e^2$ because we relaxed the simplicity constraint to include $S$. This means that we really consider the subset $F_\acal=0$, which is in total analogy to the BF-BB case. This allows us to import most of the results from there, and make the strongest possible analogy to previous works. \\
Practically speaking, this restriction really means that we have reduced ourselves to the phase space of a Chern-Simons theory with gauge algebra $\gfrak=\mathfrak{h}\ltimes (\mathfrak{p}\tilde\oplus \mathfrak{h}^\ast)$ and a specific inner product. Then, if considered interesting, curvature in the system can be reintroduced by defects, e.g. by insertions of Wilson line operators. Thus, we are not necessarily restricting ourselves to \textit{only} flat states of the corner.  \\
We can explicitly solve the equations defining the phase space in this setting, locally in the form
\begin{equation}
    \omega = g^{-1}dg\quad e = g^{-1}\cdot dX\quad S = g^{-1}( d\gamma - \frac{1}{2}X\wedge dX )g
\end{equation}
which have degrees of freedom in $g,X,\gamma$, up to their constant zero-modes. Let us forget about the $\psi$-frames for this discussion. The corner symplectic form of BF-BB theory, when $F_\acal=0$, localises in codimension 3, due to $\delta\acal=d_\acal\zeta$, in the form
\begin{equation}
\begin{gathered}
        \Omega_S = \frac{1}{2} \oint_{\p S} \langle\zeta, \Phi^{-1} d_\acal \zeta\rangle \\
        = \frac{1}{2}\oint_{\p S}
    \zeta_{\hfrak^\ast} \star_r (d_\omega \zeta_\hfrak)
    + \zeta_{\hfrak} \star_r (d_\omega \zeta_{\hfrak^\ast} + [S,\zeta_\hfrak]+e\wedge\zeta_\pfrak)
    +\frac{1}{t} \zeta_\pfrak^I (d_\omega\zeta_\pfrak - \zeta_\hfrak\cdot e )_I
\end{gathered}
\end{equation}
where $\zeta=G^{-1}\delta G$, $G=(g,X,\gamma)$. This is a non-exact symplectic form and is well-known from WZW theory\cite{murayamaExplicitQuantizationChernSimons1990,gawedzkiCanonicalQuantizationBoundary2004}. \\
We can see that in this flat sector, we have the degrees of freedom of a group element in $\exp(\gfrak)^{\p S}$ which is a loop group on each puncture. The above symplectic form is then one \textit{on} this group, as opposed to, say, its cotangent bundle. This means the puncture phase space has the structure of a Poisson-Lie group.\\

This, just like in the BF-BB case, corresponds to a puncture algebra which is a Kac-Moody algebra for the currents
\begin{equation}
    \boxed{J_\hfrak = -\star_r S, J_\pfrak = -\frac{\lambda}{t} e, J_{\hfrak^\ast} = \frac{\lambda^2}{t} \omega \qquad 0\neq \lambda\in \rbb}
\end{equation}
which then fulfil
\begin{equation}\label{eq:PunctureKM}
    \begin{aligned}
        \{ J_\hfrak\alpha , J_\hfrak \bar\alpha \}_{P} &= J_\hfrak [\alpha,\bar\alpha] \\
        \{ J_\hfrak\alpha ,  J_{\hfrak^\ast} \gamma \}_{P} &=  J_{\hfrak^\ast} [\alpha,\gamma] + \frac{\lambda^2}{t}\gamma d\alpha \\
        \{ J_\hfrak \alpha , J_\pfrak \phi \}_{P} &=  J_\pfrak\cdot \alpha\cdot\phi \\
        \{ J_\pfrak \phi , J_\pfrak \bar \phi \}_{P} &= J_{\hfrak^\ast} (\phi\wedge\bar\phi) + \frac{\lambda^2}{t} \phi_I d\bar \phi^I 
    \end{aligned}
\end{equation}
where in particular $\omega$ commutes with itself. This is a Kac-Moody algebra for the Maxwell algebra $\gfrak$, equipped with the inner product given by
\begin{equation}
    \boxed{g((X,a,U),(Y,b,V)) := \frac{\lambda^2}{t}(a^I b^J \eta_{IJ} +  \kappa(U,Y)+ \kappa(X,V))}
\end{equation}
which is notably not the inner product we used in codimension 1 and 2. The reason is the definition of $J_{\hfrak^\ast}$ we chose, which ensures that the structure constants in the $\{ J_\pfrak  , J_\pfrak  \}_{P} = J_{\hfrak^\ast}$ bracket are those of the Maxwell algebra.
We can see that the role of the fields has swapped from the intuitive one in the bulk. In the bulk, $\omega$ is the connection for the Lorentz part, while $S$ is for the dual Lorentz part. On the puncture, this is opposite: a Poisson bracket of two translational pieces $e$ gives a connection $\omega$, while $S$ acts as a Lorentz generator would.\\
A particularly interesting application of this algebra and inner product is the construction of the associated quadratic invariant of the codimension 2 components $J_a$ along the surface $S$ (here as an example for $\lambda^2 = t$)
\begin{equation}
    \boxed{g(J_a, J_b) = \frac{1}{t}e_a\cdot e_b - \kappa(S_a ,\star_r \omega_b)-\kappa(S_b ,\star_r \omega_a)}
\end{equation}
which starts with the corner metric $q_{ab}=e_a\cdot e_b$ and is used for example in the Sugawara construction of a stress-energy tensor for the fields $\omega,e,S$, in the form $T_{ab}\propto g(J_a, J_b)$. Note that this first term is consistent with the literature~\cite{freidelLoopGravityString2017} for $\frac{1}{t}=\beta, r=0$. (compare eqn. (77) of the reference). \\
Summarising, we have taken the BF-BB inspired corner phase space for 4D gravity, and reduced it to puncture phase spaces by employing additional constraints as boundary conditions. These effectively make the corner phase space into a Chern-Simons phase space, therefore yielding a Kac-Moody algebra of currents based on the Maxwell algebra.

\section{Discussion + Outlook}\label{sec:Outlook}

The present work had two overall goals. \\
First, we wanted to provide a new angle on corner data by focusing on \textit{Poisson algebras} and \textit{phase spaces} rather than the \textit{symmetries} that act on them. Our overall observation is that using only the kinematical structure of the algebra of gauge generators of the codimension 1 bulk, one can infer a Poisson structure for fields on the codimension 2 boundary. This gives rise to a weak kind of `universality' of the boundary dynamics, where the low energy \textit{kinematics} are fixed by the symmetries coming from the codimension 1 bulk. Several different choices of dynamics can then be supplanted, all on the same kinds of initial data.\\
Second, we argued that such a Poisson algebra in codimension 2 can also make sense for nontopological theories, to some extent. If the theory of interest can be cast in the same kinematics as a BF-BB theory, then, because only kinematics is used, one should be able to use analogous structures also in the theory of interest. For us, this was 4D gravity.\\
Finally, in the case we studied, there were residual constraints in codimension 2 which in principle give rise to a further, codimension 3, Poisson algebra. This data is interesting in more complex setups where different boundary conditions are glued together.\\

A first point of discussion goes to existing work. Perhaps the most important point to mention is the analogy to the Barnich-Troessaert bracket\cite{Barnich:2011mi}. It presents a bracket operation between \textit{on-shell charges} by action of gauge transformations on the charge expressions, and may include additional central terms. Our Poisson structure is highly analogous: it is `read off' from the Poisson brackets of bulk generators, which by definition are given as actions of gauge transformations. In this way, one may see our Poisson structure as a `Barnich-Troessart Poisson structure' on the corner. The principal difference is that the usual Barnich-Troessaert bracket is defined \textit{on charges only}, and so it is not clear whether it comes from a Poisson structure on the space of corner fields. In our setup, due to the theory class we work in, every field \textit{is} a charge, and so the bracket directly acts on all fields of interest. In other works~\cite{Freidel:2021cjp}, this same bracket has been extended to be more like a Poisson bracket already, and we took a parallel route here where the bracket really can be seen as a bracket on a corner phase space of fields.\\
Next, a strong set of analogies can be made between our work and that of Zucchini~\cite{zucchini_4-d_2021}, who studied 2-Chern-Simons theory, which is analogous to the topological BF-BB theories we considered. Our setup is in principle less stringent by not requiring the Pfeiffer identity, but also is a restriction in the sense that all else being equal, we considered only 2-group structures of the type of INN$(\gfrak)$. There are interesting comparisons with Zucchini's work. For example, the way that the higher WZNW functional (\cite{zucchini_4-d_2021}, 4.8.12) does not depend on the gauge frame $\phi$, but on $q$ ($\phi \hat{=} u, q\hat{=} U$), mirrors how the frame $\phi$ drops out of the phase space on $S$ in the topological case. Therefore, if one were to study gravity in a similar manner as the Chern-Simons/WZNW duality, we would expect from current work that the corresponding boundary theory \textit{does} depend on $\phi$, as the main deviation from the topological case.\\
Finally, a series of works~\cite{haggard_sl2c_2015,haggard_four-dimensional_2016,haggard_sl2c_nodate,haggard_encoding_2016,han_complex_2025-1} aiming at the quantisation of homogeneously curved spacetimes in 4D gravity used very similar structures to what we encountered here. This is of course not surprising: within our setup, homogeneous geometries correspond to a slight switch of gauge group for the BF-BB theory, and the removal of the BB term (different formulations are possible). Then, our kinematical analysis tells us that codimension 2 surfaces are naturally equipped with Chern-Simons phase spaces, which is precisely what was used as a starting point for building transition amplitudes in these works. These works also emphasized that extensions to non-homogeneous geometries are included by means of Wilson graph insertions on the corner phase spaces. Therefore, again, the kinematics are rich enough to allow for Hilbert spaces with excitations, and it is thus reasonable that we can apply the same logic of corner phase spaces also to full 4D gravity, as we did in the current paper.\\

Let us return to the topics of the present work.
The structure of 4D gravity at interfaces has been the topic of much research already, but almost exclusively from a symmetry perspective, where the focus lies on diffeomorphisms and making sense of the "would-be-gauge" degrees of freedom that they are associated to on boundaries. Usually, one tries to understand the representations of the boundary symmetry groups, for example at asymptotic null infinity. Such a categorisation of systems applies to gravity beyond specific dynamics, and therefore captures universal infrared properties. Then, the quantisation of these representations can inform us of relevant low-energy quantum gravity effects determined purely by symmetries. \\
However, it is not a priori clear which representations to pick. For a constructive approach to quantum gravity, this is an issue, as it creates two open, difficult questions: (1), one must fix the set of representations one wishes to consider in the system (the `spectral measure'), which must be singled out by some set of unknown criteria, and (2), the representation theory of the corner/boundary symmetry group may not admit interesting observables directly. For example, a generic operator in the corner symmetry group associated to Einstein-Hilbert gravity is the metric determinant~\cite{freidelEdgeModesGravity2021} $|q|$. The representation spaces then all carry such an operator, and in good cases we can take its square root $\sqrt{|q|}$ and talk about areas as geometric observables. On the other hand, the rest of the corner metric $q_{ab}$ is not necessarily an operator in these Hilbert spaces. \\
Our approach attempts to bridge this problem: rather than a corner symmetry algebra, it produces a full corner \textit{Poisson} algebra, which in turn carries one specific, reducible, representation of the corner symmetry group. Using the Poisson algebra as a starting point for quantisation therefore naturally picks out one spectral measure and resolves issue (1). By coming directly from 4D gravity, and capturing all the elementary fields, it also in principle allows us to resolve issue (2). It stands in contrast to simply working with the bulk codimension 1 Poisson algebra, which also contains all the fields, but does not localise on the corner.\\
Simultaneously, the corner Poisson algebra provides us with the right ingredients to work holographically: if we see the bulk fields $\bcal,\acal$ as being radially evolved versions of some composite operators $\mathcal{O}_{\bcal,\acal}[\phi]$ in a boundary theory of fields $\phi$ which presents the holographic dual, then the corner Poisson brackets provide us with the brackets of the composites $\mathcal{O}_{\bcal,\acal}[\phi]$ in the boundary theory. This is analogous to the situation in CS/WZW. This set of composites, and their dynamics, then form a sort of `hydrodynamics' of the boundary dual in the sense of encompassing generic, low energy observables that the bulk theory cares about.
One can then think of quantising the corner phase space as an intermediate step towards studying the low-energy region of a holographic dual. It provides a setup in which one can radially evolve the boundary fields in order to define the bulk.\\

Our aim was to be as systematic as possible while also keeping practical concerns in mind. Using the analogy to BF-BB theories, we determined a sensible candidate for the corner Poisson brackets which follows from corner charge algebras alone. Our corner algebra logic, in a sense, was therefore to switch to a presentation of the theory where the symmetry algebra and the Poisson algebra are identical, and make inferences from there.\\
Our results, importantly, do not require or mention boundary conditions in any way, but are of course still dependent on the specific presentation. In the Maxwell-algebra based BF-BB presentation we chose, we found that codimension 2 surfaces should naturally be assigned phase spaces with symplectic form
\begin{equation}
    \Omega_S = \int_S \delta S \wedge\star_r \delta\omega + \frac{1}{2t}\delta e \wedge \delta e + \delta(\frac{1}{2}(\star_\beta e)^2 \chi_\hfrak(\phi))
\end{equation}
which closely resembles the one of a Chern-Simons theory. This matches expectations from previous studies, and naturally hints at the presence of noncommuting geometric observables in gravity. Similarly, in codimension 3, by analogy to the CS/WZW correspondence, we argued that at least in the partially flat sector of the corner phase space, there is also a natural current Kac-Moody algebra \eqref{eq:PunctureKM} present.\\

Let us now mention a few caveats.
First, the general corner algebra logic and its dependence on the embedding into BF-BB theory. While generally speaking, the gauge group of a theory depends on the specific presentation made, this is especially visible in our study. While the result is sensible and indeed deviates from the purely topological result in ways that transparently relate $\Omega_S$ to gravitational corner symmetry charges such as $\frac{1}{2}(\star_\beta e^2)$, we found it difficult to anticipate the symplectic form on general grounds. A more detailed understanding of the procedure and the generic differences between the topological and non-topological cases is necessary. In particular, the question of how to sensibly work with the second-class constraint systems of the bulk needs to be understood more thoroughly. A possible avenue for this is making more strict connections to BV-BFV methodology, which produces similar but seemingly distinct Poisson bivectors in codimension 2~\cite{canepaCornerStructureFourDimensional2024}.\\
Second, the pathology of $\tau, \chi_\pfrak$ is intriguing. It is known that the codimension 1 bulk of 4D gravity carries a translational corner charge $p^I=-(\star_\beta\omega)^{IJ}_\wedge e_J$, which might have appeared in the same place in the symplectic form as the Lorentz one $\frac{1}{2}(\star_\beta e^2)$. However, we could not derive this here, as we had to remove $\tau$ from the Poisson algebra to ensure the Jacobi identity holds. Whether a manual modification, or a derivation from a more careful analysis can yield a corner Poisson algebra including $\tau$, is an interesting question.\\
Furthermore, we did not impose one final geometric constraint on the proposed corner phase space of gravity yet, the torsion constraint $d_\omega e =0$. As can be seen from ~\eqref{eq:CornerAlgGrav4D}, this is a second-class constraint whenever $F_\omega$ is nonzero, so one has to make a more involved phase space reduction. In particular, the constraints are not as easy to solve as the somewhat trivial algebraic constraints we dealt with here. Still, one can make a few guesses on the final result. The first is the degree of freedom count: $\phi$ carries $6$ degrees of freedom, $S,\omega$ $2\times2\times6$, and $e$ $2\times 4$. The torsion constraint removes $4$ degrees of freedom from either $\omega,e$ and so leaves us with those of $\phi,S$ and $16$ degrees of freedom in $e,\omega$ on the corner. \\
Beyond this, we note that the Maxwell algebra, being the symmetry algebra of a particle in an electromagnetic field, allows for some analogies through its representation theory. In particular, seeing $d_\omega e$ as the momentum in the Maxwell algebra, which in the particle scenario is the canonical momentum $p-e A$, one can interpret the representations of the Maxwell algebra that appear in the physical corner phase space as being the same ones as the particle in a restricted \textit{guiding center} approximation\cite{jacobson_guiding-centre_2025}. Alternatively, we may try to skip to the quantum level and attempt reducing irreducible unitary representations of the Maxwell algebra\cite{Schrader:1972zd}. 
%For a generic $F_\omega^{IJ}$, we may use Lorentz transformations to put it into a standard form. Then, the space $\rbb^{1,3}=V\oplus W$ splits such that $a_I F_\omega^{IJ} b_J=0\,\forall\, a,b\in V$. The representation space is then $L^2(W)$, and $V$ as a part of the translations in the Maxwell group acts only as a phase factor on it, while $W$ acts through translations. Requiring invariance 
The second class nature of the constraint is reflected here: there is no normaliseable state in any nontrivial irreducible representation that satisfies $P_I=0$. A good analogy might be to look for a zero-energy, zero-momentum state within the quantum harmonic oscillator, which is again not possible. Therefore, one might need to accept that the constraints, when imposed on the quantum level, allow for a very slight torsion defect. Still, let us mention one final intriguing point: The Cartan elements of the Maxwell algebra are strictly speaking of the form $(X,0,u)$, with the $\rbb^{1,3}$ part vanishing. By standard folklore, curvature defects like for $F_\acal$ are labeled by Cartan elements. The standard kinds of curvature defects within the Maxwell algebra are then exactly in line with the physical torsion constraints.\\

Let us now turn to some future directions. 
We begin with the most simple extension of what we did: the inclusion of a cosmological constant $\Lambda$. It is known that there is a 1-parameter family of deformations~\cite{gomisDeformationsMaxwellAlgebra2009} of the Maxwell algebra, corresponding to the inclusion of a cosmological constant. These, in fact, are much simpler in structure than the Maxwell algebra, being semisimple and isomorphic to
\begin{equation}
    \gfrak_{\Lambda>0}\cong \mathfrak{so}(1,4)\oplus\mathfrak{so}(1,3)\qquad \gfrak_{\Lambda<0}\cong \mathfrak{so}(2,3)\oplus\mathfrak{so}(1,3)
\end{equation}
which makes certain aspects of the representation theory simpler. We focused on the $\Lambda=0$ in this work to emphasize the differences to the naively expected Poincare algebra $\mathfrak{so}(1,3)\ltimes \rbb^{1,3}$, but there are no obstructions to performing our analysis with $\Lambda\neq 0$.\\
Beyond this, the main two avenues for applying the corner Poisson algebra are for informing quantum gravity model building and for preparing physical states. \\

The former, loosely speaking, asks how to write down concrete partition functions and Hilbert spaces for quantum gravity, and what data (e.g. what kind of representation data) they carry. This is affected by our results by providing a classical continuum phase space to which usual techniques of quantisation can be applied, which then can be compared to existing ideas. For example, a common idea is that the representation data of 4D gravity should be organised in terms of representations of 2-groups~\cite{girelliDiscretization4dPoincare2022,mikovicHamiltonianAnalysisBFCG2022,Korepanov:2002tp,Yetter:1993dh}. While the continuum picture does not immediately admit this interpretation, the presence of extended, independent connection $\omega$ and tetrad data $e$ bear certain similarities to models built on 2-group structures. When restricting attention further to discrete approximations to partition functions in the sense of state sum models~\cite{Asante:2019lki,Baratin:2006gy}, there are certain advantages to the presentation of gravity we worked with. The case in point is the $\bcal\bcal$ term, which generically is not easily put on a lattice (more generally, a cellular complex). Conventionally, $\bcal$ is put as variables on the 2-skeleton of the 4D cellular complex which cellulates a given spacetime, and then $\bcal\wedge\bcal$ cannot be discretised. However, the specific pairing $\hfrak\ltimes\hfrak^\ast$ makes this easier: If one places $B_f$ on the 2-skeleton, and $c_{\tilde f}$ on the dual 2-skeleton, then the BB term can be written appropriately as $\sum_{\langle f,\tilde f\rangle} \kappa(B_f, \star_r^{-1} c_{\tilde{f}}) $. Related to this, the relaxed simplicity constraint $B_f = (e\wedge e)_f + (d_\omega S)_f$ can possibly modify some aspects or issues found when implementing simplicity within state sums~\cite{Dittrich:2010ey}. \\

The latter avenue is about trying to find solutions to the gravitational constraints by constructing them from boundary data.
The perspective here is to prepare wavefunctionals for the codimension 1 bulk $\s$ as Euclidean partition functions of a theory which has Hilbert spaces in codimension 2, i.e. 
\begin{equation}
    \Psi_{\s}[\acal] = Z_{\s}[\acal;\phi] ,\phi \in \hbb_{\p\s}. 
\end{equation}
where the bulk fields $\acal$ are seen as sources, and probe the response of the theory on $\p\s$ which also provides the boundary condition for $Z_{\p\s}$. The gravitational constraints then are equivalent to Ward identities for $Z_{\p\s}$, i.e. they constrain what possible dynamics we may put on the Hilbert space $\hbb_{\p\s}$. It should be clear how our current work fits into this picture: we have from the corner Poisson algebra a candidate for how to construct the corner Hilbert space $\hbb_{\p\s}$. Any Euclidean dynamics we put on it, in particular the Euclidean evolution of $F_\omega$, will specify a state on $\s$ if it complies with the constraints.\\
The most direct way in which this can be done is to prepare the `vacuum', meaning the flat state which classically carries $F_\omega=Q=d_\omega e = 0$. This is of course, as is well known~\cite{horowitzExactlySolubleDiffeomorphism1989}, described by a Chern-Simons theory on $\s$, and this perfectly matches with the Poisson relations we found. In turn, the boundary conditions of the partition function are fixed by gauge transformations in $\gfrak$ only, as the flat state only carries gauge information. These boundary conditions are then well-described by WZW models, which exactly describe mappings from $\p\s$ into the gauge group, i.e. they capture boundary gauge transformations.\\
Inspired by this, we can also introduce defects. In the continuum, this could be done naturally along a Heegaard decomposition of $\s$, leading to an analogous picture to that of Poincare networks~\cite{dittrich_3_2017,freidelGravitationalEdgeModes2019}. These same networks then also inform Hilbert spaces for use in lattice gauge theory-like approaches to build gravity Hilbert spaces~\cite{freidelContinuousFormulationLoop2013}. \\
Staying in the discrete, one might also directly try to construct Hilbert spaces for the corner Poisson algebra, for example by performing a naive smearing of the variables and working on cellulations of $\s,\p\s$, or by employing some version of the Fock-Rosly construction~\cite{Alekseev:1994pa,Alekseev:1994au}, which is adapted to the kind of Chern-Simons-like phase space we found. This in particular would require an understanding of the relevant quantum group deformations in this setting, of which there may be multiple, which can lead to inequivalent theories~\cite{Gaiotto:2024osr}. This is particularly interesting in relation to 3D gravity, where the quantum group structure is made clear by presence of a Drinfel'd double $\mathfrak{so}(1,2)\ltimes \mathfrak{so}(1,2)^\ast$. The 4D case then requires more careful study as we instead find a nontrivial quadratically extended Lie algebra, whose quantum deformations may be more complicated. \\
Lastly, the corner Poisson algebra we found in this work is an interesting subject of study on its own; one should study its decomposition in representations of the corner symmetry algebras previously found in gravity, in particular the universal corner symmetry~\cite{Ciambelli:2021vnn,Ciambelli:2022cfr,Neri:2025fsh} algebra of diffeomorphisms.\\
We will investigate these points in future publications.

\paragraph*{Acknowledgements}
The author thanks G.Neri, F.Girelli, L.Freidel, B.Dittrich, R.Silva, and especially D.Oriti for useful discussions, thanks C.Pollack for stressing to verify the Jacobi identity, and Q.Pan, A.Riello for pointing out relevant literature. The author also thanks the anonymous referee for giving suggestions that greatly improved the presentation of the paper. \\
The author gratefully acknowledges the support of the Bruce H. Mitchell Foundation, whose funding helped make this research possible.\\
The author would also like to thank the Perimeter institute for hospitality. Research at the Perimeter Institute is supported in part by the Government of Canada through NSERC and by the Province of Ontario through MEDT.

\appendix
\section{Reduction in the topological case}\label{App:TopReduction}
Here, we go step-by-step through the symplectic reduction of the corner phase space by the constraints coming from the bulk, for the topological case where (\ref{eq:TopProperty}) holds. We start from the symplectic form
\begin{equation}
    \begin{gathered}
        \Omega_S =\int_{S}
        \delta\bcal\chi - \bcal\frac{1}{2}[\chi,\chi] 
         + (\delta\acal-d_\acal \chi)\wedge\psi 
         - \frac{1}{2}\psi\wedge\Phi(\psi)
        -\oint_{\p S} \psi \chi + q \frac{1}{2}[\chi,\chi]
    \end{gathered}
\end{equation}
and its contraction with a generic field space vector field $V$,
\begin{equation}
    \begin{aligned}
        -I_V \Omega_S =&
        \int_S \delta\bcal\,  I_V\chi + \delta\acal\wedge I_V \psi 
        -(V[\bcal]-[\bcal,I_V\chi]-d_\acal I_V\psi)\, \chi \\&- (V[\acal]-d_\acal I_V\chi - \Phi(I_V\psi) )\wedge\psi  
        - \oint_{\p S} \delta q\,  I_V\chi.
    \end{aligned}
\end{equation}
Now, $\delta \bcal = \delta(\Phi^{-1}F_\acal) = \Phi^{-1}d_\acal\delta\acal$ modifies the first two terms in the contraction to
\begin{equation}
    (\Phi^{-1}d_\acal\delta\acal)\,  I_V\chi + \delta\acal\wedge I_V \psi 
    =
      \delta\acal\wedge (I_V \psi + \Phi^{-1} d_\acal I_V\chi )
      + d( \delta\acal \, \Phi^{-1}I_V\chi )
\end{equation}
Meanwhile the coefficient of $\chi$ becomes:
\begin{equation}
    \begin{aligned}
        V[\bcal]-[\bcal,I_V\chi]-d_\acal I_V\psi
    &=
    \Phi^{-1}d_\acal V[\acal]  -\Phi^{-1}[F_\acal,I_V\chi]-d_\acal I_V\psi\\
    &=
    \Phi^{-1}d_\acal( V[\acal] - d_\acal I_V\chi - \Phi (I_V\psi) )
    \end{aligned}
\end{equation}
where we see the coefficient of $\psi$ in the bulk appear. There is, importantly, no additional kernel equation coming from the coefficient of $\chi$. Now look for a nontrivial kernel $-I_V\Omega_S=0$. This requires
\begin{equation}
\begin{gathered}
    I_V \psi + \Phi^{-1} d_\acal I_V\chi =0\\
    V[\acal] - d_\acal I_V\chi - \Phi (I_V\psi)=0 \Longleftrightarrow V[\acal]=0
\end{gathered}
\end{equation}
and on the boundary $\p S$:
\begin{equation}
    \delta (\acal - \Phi q) \, \Phi^{-1}I_V\chi = 0.
\end{equation}
This can be easily solved nontrivially with free data being any $I_V\chi$ which vanishes on $\p S$. Therefore, there is a large kernel
\begin{equation}
    V_\alpha[\phi] = \phi\alpha\quad V_\alpha
    [q] = [q,\alpha]- \Phi^{-1}d_\acal \alpha
\end{equation}
in the symplectic form which we need to remove by a gauge fixing condition. The sought object is a function $f(\acal,\phi,q)$ which transforms nontrivially under all kernel vector fields. Since $\acal$ does not transform under those, and $q$ only transforms if $\phi$ does, the dependence on $\phi$ is all we care about.\\
Second, we want such a function that does not restrict $\phi$ on $\p S$. For this, note that in the interior of $S$, only $d_\acal I_V\chi$ appears, so there appears to be an obvious differential gauge fixing. If we impose $d_\acal \chi = 0$ with $I_V\chi \stackrel{\p S}{=} 0$, then the kernel trivialises. This entails
\begin{equation}
    d_\acal \chi = \delta\acal -\phi^{-1} \delta \acal_\phi \, \phi = 0,  \quad
    \acal^\phi:=  \phi \acal \phi^{-1} - d\phi \phi^{-1}.
\end{equation}
What this means practically is that we can impose an arbitrary standard gauge fixing condition from Yang-Mills theory on $\acal^\phi$ and it will be a good gauge fixing for this phase space. \\ 
A simple choice which fixes the entire kernel is to just set $d \phi \phi^{-1}=0 $, which implies that $d(I_{V_\alpha}\chi)= d\alpha=0$, and so only constant $\alpha$ are allowed. But they must vanish on $\p S$, so $\alpha=0$. We use this in the main text.\footnote{A second choice is a Laplace-type equation with respect to some auxiliary 2D metric $q$ on $S$, for example
\begin{equation}
    \text{div}_{q}(\acal^\phi):=\ast_q^{-1} d \ast_q \acal^\phi   = 0, \quad i_{n_{\p S}}\acal^\phi = 0.
\end{equation}
This option has some residual gauge freedom, namely the parameters satisfying $d_\acal I_V\chi=0$ subject to $I_V\chi \stackrel{\p S}{=} 0$.}\\
With the redundancy freedom fully fixed, we have $d\chi = 0$, so the symplectic form reduces to pieces involving the remaining local fields $\acal,q$ and a global contribution from $\chi$:
\begin{equation}
    \begin{gathered}
        \Omega_S =\int_{S}
         \delta(\phi\acal\phi^{-1})\wedge\delta(\phi q\phi^{-1}) 
         - \frac{1}{2}\delta(\phi q\phi^{-1})\wedge\Phi \delta(\phi q\phi^{-1})
        + \Omega_S^\chi
    \end{gathered}
\end{equation}
with
\begin{equation}
    \Omega_S^\chi = \delta \left(  ( \int_S \Phi^{-1}F_\acal -\oint_{\p S}  q) \;\chi \right).
\end{equation}
We can recognise $\Omega_S^\chi$ as the variation of the differentiable smearing of $B$ against $\chi$. In the remaining terms, $\phi$ is a spacetime constant, so all the terms involving $\chi$ may in principle be grouped into global contributions as well. If we also recoordinatise to $\bar\acal
=\phi\acal\phi^{-1},\bar q=\phi q\phi^{-1}$, then the data involving $\phi$ neatly splits from the rest, 
\begin{equation}
    \begin{gathered}
        \Omega_S =\int_{S}
         \delta \bar\acal\wedge\delta \bar q 
         - \frac{1}{2}\delta \bar q\wedge\Phi \delta \bar q + 
         \delta \left(  ( \int_S \Phi^{-1}F_{\bar\acal} -\oint_{\p S}  \bar q) \;\delta\phi \phi^{-1} \right).
    \end{gathered}
\end{equation} 
This shows that the constraint surface is coisotropic, and that the reduction can be performed easily on the level of symplectic forms. In particular, $\bcal$ completely becomes a function of the other fields. This is characteristic of the topological case, where the constraint $F_\acal=\Phi(\bcal)$ is first class.\\
A final technical remark is in order: $q$ was not a Dirac observable on the original phase space, but persists into the reduced, gauge-fixed phase space. This is because what actually persists is not $q$, but rather a gauge invariant combination which depends on the constant value $\phi_0$ we gauge fix $\phi$ to:
\begin{equation}\label{eq:qDiracObs}
    \bar q (\acal,q,\phi;\phi_0) := \phi q\phi^{-1} + \Phi^{-1}(
    \phi_0 \acal \phi_0^{-1} - \phi \acal\phi^{-1} + d\phi \phi^{-1}
    )\stackrel{\phi \equiv \phi_0}{ =} \phi_0 q\phi^{-1}_0
\end{equation}
This is a Dirac observable with respect to the constraints including the gauge fixing $\phi\equiv \phi_0$. There is such a function for every choice of $\phi_0$. The reduced phase space is thus more invariantly coordinatized by $(\acal,\bar q, \phi_0)$.

\section{Extended 3D gravity phase space}\label{app:3D}
As stated in the introduction, there is a clean derivation of the corner Poisson algebra of 3D gravity~\eqref{eq:3DGravPoissonAlg} from the algebra of gauge generators in codimension 1. No further constraints arise, so no real issues remain in principle. If one insists on a symplectic description, however, one will again introduce frames $\phi,q$ that also arise in an extended phase space description, where $q$ is now a 0-form. The relevant corner symplectic form is that of the dressed fields $e,\omega$ and the gauge frames,
\begin{equation}
    \Omega_S = \int_S (\delta\star e - d_\omega\delta q) \chi - \star e\frac{1}{2}[\chi,\chi] + \delta\omega\delta q.
\end{equation}
This is presented to already include a possible puncture modification. Of course it looks analogous to the symplectic form we use in the main text, but the $\delta q\delta q$ term is absent (and impossible for dimensional reasons). The smeared fields $\star e,\omega$ now again fulfil the $\gfrak\ltimes\gfrak^\ast$ current algebra we mentioned in the introduction, and generate shifts of the gauge frames $\phi,q$ respectively. Bulk fields $\star e_b, \omega_b$ will be related to the dressed corner fields by
\begin{equation}
    \omega= \phi^{-1}\omega_b \phi + \phi^{-1}d\phi \quad \star e = \phi^{-1}\star e_b \phi + d_\omega q.
\end{equation}
Then, an extended phase space of $(e_b,\omega_b;e,\omega,\phi,q)$ with symplectic form $\Omega_\s^e = \Omega_\s + \Omega_{\p \s} $ will have all bulk gauge transformations (incl. the change of gauge frames) be pure redundancy, so they can be gauge fixed at will. In such a scenario, we can either treat 3D gravity as a strict BF theory, and gauge fix $e_b=0=\omega_b$, in which case the symplectic form reduces to
\begin{equation}
    \Omega_S \stackrel{BF}{=} -\int_S   \chi\, d_{\phi^{-1}d\phi}\delta q  + \oint_{\p S} \chi \delta q - q \frac{1}{2}[\chi,\chi] 
\end{equation}
which describes the phase space of 2 chiral scalars, one group-valued, one Lie algebra-valued. See e.g. ~\cite{geillerExtendedActionsDynamics2020}, Eqs. (4.79-4.83) for comparison. However, we can also be more careful and remember that $e_b$ must be nondegenerate in gravity. Then, we cannot set it to zero by a gauge transformation, and the same goes $\omega_b$. Still, if we set $\delta e_b=0$ and $ \omega_b = \omega_b(e_b)$ satisfying the torsion constraint at the surface $S=\p\s$, then we still get a reduction of the corner symplectic form to
\begin{equation}
    \Omega_{\p\s} \stackrel{GR}{=} \int_{\p \s} \star e_b \frac{1}{2}\phi [\chi,\chi]\phi^{-1} + \omega_b(e_b) \, \phi[\chi,\delta q + [q,\chi]]\phi^{-1}  -\chi\, d_{\phi^{-1}d\phi}\delta q.
\end{equation}
This shows that the Poisson relations on the phase space that lives on the corner is different between 3D BF theory and 3D gravity with nondegeneracy. This highlights that using bulk gauge fixings must also be handled with care in the 4D case discussed in section~\ref{sec:Gravity}, if one is interested in the pure corner data.  

\section{Abelian examples}\label{sec:Examples}

Here we discuss a full example with bulk-boundary dynamical interactions and codimension 3 strata. For this, we restrict to gauge group $G=\rbb$, and our homomorphism is $\Phi = k \,\text{id}$ with $k \neq 0$. We will consider a big ambient spacetime $M=\s \times \rbb$, with closed slice $\s$, in which we consider two smaller, overlapping spacetimes $M_L=\s_L\times\rbb, M_R=\s_R\times\rbb$, with overlapping (codimension 1)\footnote{Codimension $k$ will as usual always refer to the ambient spacetime, e.g. codimension $3$ is always $1$D.} Cauchy slices $\s_L\cap \s_R \neq \emptyset$. We let the slices $\s_L,\s_R$ have boundaries each, and the fact that they overlap creates a codimension 3 locus $\ccal:=\p\s_L\cap \p\s_R$. Then we let $S_L:= \p\s_L\setminus \s_R,S_R:= \p\s_R\setminus \s_L$ be the (codimension 2) parts of the boundaries $\p\s_L,\p\s_R$ which are not in the overlap, and identify $\ccal = \p S_L$ as the (codimension 3) boundary of $S_L$.\\
We will study the reduction of BF-BB theory to the two smaller spacetimes, and impose boundary conditions separate on the boundaries of the two smaller spacetimes. On $S_L\times\rbb$, we will impose the electromagnetic boundary conditions, on $S_R\times\rbb$, we will impose topological boundary conditions. This will induce nontrivial effects on the evolution of the codimension 3 puncture, $\ccal\times\rbb$. To see them, we then further reduce the theory on $(S_L\cup S_R)\times \rbb$ to $S_L\times\rbb$.\\
The two boundary conditions we impose are as follows, fixing the timelike components of the fields:
\begin{enumerate}
    \item \textit{Electromagnetic}: $\bcal_t = m\ast_2 \acal$, $\acal_t = \frac{1}{m} \ast_2 \bcal$
    \item \textit{Topological}:  $\bcal_t = 0 = \acal_t$.
\end{enumerate}
Electromagnetic boundary conditions correspond to the spacetime condition $\bcal = m \ast_3 \acal$, and make reference to a background 3-metric on $\p \s_L\times\rbb$ through the Hodge dual $\ast_3$. These may be considered the `standard' boundary conditions in BF theories, and are studied in many situations~\cite{maggiore_chiral_1992,maldacena_d-brane_2001,amoretti_three-dimensional_2012,fliss_entanglement_2023,fliss_entanglement_2025,antinucci_holographic_2024,joungExtractingEdgeModes2026}, and are naturally produced under RG flow of the system. They lead to an effective Maxwell-Chern-Simons theory on the boundary segment in terms of the field $q$. In contrast, topological boundary conditions only make use of a global vector field (which here we use as a time) to specify which component vanishes. These conditions are in a sense the `minimal' ones one can impose, as they do not require the choice of a volume form to dualise $\bcal$, instead opting for a `time vector field' $\zeta$ as the structure used to define the boundary condition. They have been studied in pure BF theory both in 3D~\cite{freidelQuantumGravityDisk2021} and higher dimensions~\cite{maldacena_d-brane_2001}, and lead to an effective Chern-Simons theory for $q$.
$\ast_2$ refers to the spatial Hodge dual on constant time cuts of $\p \s_L\times\rbb$.\\

A priori, it is not clear what data we should have on the puncture, kinematically speaking. We thus begin from the kinematics on the Cauchy slices again. To start though, we need the dynamics of the bulk, given by 
\begin{equation}
    L_0 = \bcal\wedge d\acal- \frac{k}{2}\bcal\wedge\bcal.
\end{equation}
We will use that the equations of motion, $d\bcal=0=d\acal-k\bcal$, pull back onto the Cauchy slice $\s$. The overall phase space is equipped with 
\begin{equation}
    \Theta_\s = \int_\s \bcal\wedge\delta\acal = \Theta_{\s_L\cup\s_R} + \Theta_{\s\setminus( \s_L\cup\s_R)}.
\end{equation}
We first localise the theory by freezing the dynamics outside the subregion $\s_L\cup\s_R$ of interest. We do so by writing the fields outside as a gauge fixed and a gauge contribution\footnote{This is known as an intrinsic reference frame split in the corner literature\cite{Carrozza:2021gju,araujo-regadoSoftEdgesMany2025}.},
\begin{equation}
    \bcal = \bcal_{gf}- dq \quad \acal= \acal_{gf} - k q - d\lambda
\end{equation}
By inserting this into the symplectic potential for the outside region, we achieve
\begin{equation}
    \Theta_{\s\setminus( \s_L\cup\s_R)} \approx \Theta_{\s\setminus( \s_L\cup\s_R)}^{gf} -  \int_{\s\setminus( \s_L\cup\s_R)}\delta(q\wedge d\acal-\frac{k}{2}q\wedge dq ) - \Theta_{\p(\s_L\cup\s_R)} 
\end{equation}
where we use the constraints outside the region of interest. The first term accounts for gauge-fixed degrees of freedom of the outside, and does not imprint nontrivially on the region $\s_L\cup\s_R$. The second is just an exact contribution and does not affect our region either. The last term localises on the codimension 2 boundary $ \p(\s_L\cup\s_R)$, and is given by
\begin{equation}
    \Theta_{\p(\s_L\cup\s_R)} = \oint_{\p(\s_L\cup\s_R)} \bcal_{gf}\delta\lambda+ q\wedge\delta (\acal_{gf}-d\lambda) - \frac{k}{2}q\wedge \delta q. 
\end{equation}
This is of course the specialisation of the general formula for the symplectic potential on the corner. We identify the gauge fixed fields $A_{gf}$ of $\s\setminus( \s_L\cup\s_R)$ as the analogue of the dressed fields - on $\s\setminus( \s_L\cup\s_R)$, they are dressed by an \textit{intrinsic/internal} reference frame $q,\lambda$, which technically is part of the fields\footnote{I.e. it is determined by a Hodge decomposition of the fields.}. However, those same reference frames are then \textit{external} to the region $ \s_L\cup\s_R$ and can serve for a dressing of the fields on $\s_L\cup\s_R$. The piece $\Theta_{\p(\s_L\cup\s_R)}$ now imprints on this region with opposite orientation, giving the extended potential
\begin{equation}
    \Theta_{\s_L\cup\s_R} = \int_{\s_L\cup\s_R} \bcal\wedge\delta\acal + \Theta_{\p(\s_L\cup\s_R)}.
\end{equation}
This extended phase space of the subregion accounts for the gauge imprint of the environment. \\
We now switch notation back to what we are used to from the main text: the fields in the slice $\s_L\cup\s_R$ will be called $\acal_b,\bcal_b$ referring to the bulk, and the boundary fields $\acal_{gf},\bcal_{gf}$ will be renamed $\acal,\bcal$. Again, it holds that $\bcal_b + dq = \bcal$, $\acal_b + k q + d\lambda = \acal$ as constraints by continuity.\\
The differentiable codimension 1 gauge generators of the slice are now
\begin{equation}
    J_{\alpha} = \int_{\s_L\cup\s_R}\bcal_b\wedge d\alpha + \oint_{\p(\s_L\cup\s_R)} q\wedge d\alpha - \bcal\alpha  
\end{equation}
\begin{equation}
    K_\mu = -\int_{\s_L\cup\s_R}(d \acal_b-k\bcal_b)\wedge \mu + \oint_{\p(\s_L\cup\s_R)} (\acal_b +k q -\acal)\wedge\mu - \lambda d\mu
\end{equation}
and to define the phase space we set them to zero for \textit{all} parameters $\alpha,\mu$, including those nonvanishing at boundaries. The bulk gauge transformations affect, among the boundary fields, only the frames $\lambda,q$ as $\{J_\alpha,\lambda\}=-\alpha, \{K_\mu, q\}= -\mu$. On the other hand, we have the corner charges
\begin{equation}
    \oint_{\p(\s_L\cup\s_R)}\bcal\alpha\qquad \oint_{\p(\s_L\cup\s_R)} \acal\wedge\mu
\end{equation}
which now generate the usual gauge transformations $\delta_{(\alpha,\mu)}\acal=k\mu +d\alpha, \delta_{(\alpha,\mu)}\bcal=d\mu$ of the bulk, but on the boundary fields. They fulfil the corner Poisson algebra we are interested in. They also, notably, fulfil the same algebra as the pure bulk pieces of $J_\alpha,K_\mu$. The phase space coordinatized by $\bcal,\acal,q,\lambda$ thus keeps track of the imprint of the environment, and its fundamental Poisson brackets are given by the algebra of corner charges of the subregion. \\
Ultimately, we will want to put a Hamiltonian on this phase space, but our goal was to introduce two different boundary conditions on $S_L$,$S_R$ each. For this purpose, we now consider the subregions individually before introducing the constraints. The first step is that on $S_L$ (same for $S_R)$), we need to make the basic variables differentiable. As already stated, this leads to a modification into 
\begin{equation}
\begin{aligned}
        \Theta_{S_L} &= \int_{S_L} \bcal\delta\lambda+ q\wedge\delta (\acal-d\lambda) - \frac{k}{2}q\wedge \delta q - \oint_{\p S_L} q\delta\lambda\\
        &=\int_{S_L} \bcal\delta\lambda+ q\wedge\delta \acal - dq\,\delta \lambda - \frac{k}{2}q\wedge \delta q 
\end{aligned}
\end{equation}
which just partially integrates the $q\wedge d\delta\lambda$ term, and the basic coordinates on the phase space are now
\begin{equation}
    \acal(\mu)=\int_{S_L}\acal\wedge\mu\qquad \bcal(\alpha) = \int_{S_L}\bcal\alpha - \oint_{\p S_L}q\alpha.
\end{equation}
Note that this extension does not include the effects of gauge imprints from the other side (and we have not imposed the leftover constraint at this point anyway). It is purely for differentiability of the basic variables in the presence of $\p S_L$. A similar construction applies to $S_R$.\\
The leftover constraint can be discussed on either side $S_L,S_R$. On both sides, its differentiable form is
\begin{equation}
    \bar\kkcal_\alpha = -\int_{S_L} d\acal\,\alpha - k\bcal\alpha + \oint_{\p S_L} (\acal-k q) \alpha
\end{equation}
and generates $ \{\bar\kkcal_\alpha,\lambda\} = k\alpha, \{\bar\kkcal_\alpha, q\} = -d\alpha $ as a redundancy. As in general, we can gauge fix using $d\lambda=0$, which allows only constant $\alpha$ as residual transformations, which however are not redundancies. This reduces the symplectic form to ($\acal^{op}=\acal-kq$)
\begin{equation}
\begin{aligned}
        \Theta_{S_L} &\approx \int_{S_L}  q\wedge\delta \acal - \frac{k}{2}q\wedge \delta q + \delta(\int_{S_L} \bcal - \int_{S_L} q) \delta\lambda\\
        &=
        \int_{S_L}  \frac{1}{2k}\acal\wedge\delta \acal - \frac{1}{2k}\acal^{op}\wedge\delta \acal^{op}
        +\delta(\frac{1}{k}\oint_{\p S_L} \acal^{op})\delta\lambda
\end{aligned}
\end{equation}
and so again the physical data is encoded in $\acal,\acal^{op}$ and the constant part of $\lambda$. The generators obey the algebra $ \{ \bar\kkcal_\alpha, \bar\kkcal_{\bar\alpha} \}= k \oint_{\p S_L}  d\alpha \, \bar\alpha $, so from the un-extended phase space on $S_L$ point of view, there are corner charges $\oint_{\p S_L} \acal^{op}\alpha$ which generate an $\rbb$-Kac-Moody algebra at level $k$.\\

We now further reduce the theory from $(S_L\cup S_R)\times \rbb$ to $S_L\times\rbb$.
For this we again find an extended phase space on $S_L$. For this purpose, take the potential on $S_R$ and again perform an intrinsic-gauge frame split $q = q_{gf} + d\gamma$, $\lambda = \lambda_{gf} -k \gamma $, and insert this into the potential. On-shell of the constraint, then the potential on $S_R$ splits again as
\begin{equation}
    \Theta_{S_R} = \Theta_{S_R}^{gf} - \delta(\int_{S_R} \gamma d\acal + \frac{k}{2}d\gamma\wedge q  - \oint_{\p S_R} \gamma \acal ) - \Theta_{\ccal}
\end{equation}
with the \textit{puncture potential}
\begin{equation}
    \Theta_{\ccal} = \oint_{\p S_L} \acal^{op}_{gf}\delta\gamma - \frac{k}{2}d \gamma \delta\gamma 
\end{equation}
with $\acal^{op}_{gf}=\acal-k q_{gf}$, which yields the extended phase space on $S_L$ if we add it back to $\Theta_{S_L}$. In that setting, we must see $q_{gf}$ as independent from $q$, and their relation holds on-shell. In this phase space, we speak of the field $\mathfrak{A}^{op}$ on the puncture $\p S_L$, and the extended symplectic potential is
\begin{equation}
    \Theta^e_{S_L} = \int_{S_L} \bcal\delta\lambda+ q\wedge\delta \acal - dq\,\delta \lambda - \frac{k}{2}q\wedge \delta q + \oint_{\p S_L} \mathfrak{A}^{op}\delta\gamma - \frac{k}{2}d \gamma \delta\gamma 
\end{equation}
The extended generator of corner redundancies is then
\begin{equation}
    \bar\kkcal^{e}_\alpha = -\int_{S_L} d\acal\,\alpha - k\bcal\alpha + \oint_{\p S_L} (\acal^{op}-\mathfrak{A}^{op}) \alpha - k\,\gamma d\alpha
\end{equation}
which again we set to zero for \textit{all} $\alpha$, thus defining the on-shell extended phase space with the matching constraint $\acal^{op} + k d\gamma =\mathfrak{A}^{op} $. In this phase space, $\lambda=0$ is, kinematically speaking, a valid gauge fixing condition for the constraint $d\acal-k\bcal=0$, and the effective phase space is therefore described well by
\begin{equation}
    \Theta^e_{S_L} \approx \int_{S_L}  q\wedge\delta \acal - \frac{k}{2}q\wedge \delta q + \oint_{\p S_L} \acal^{op}\delta\gamma + \frac{k}{2}d \gamma \delta\gamma 
\end{equation}
The corresponding corner charge on the extended phase space is instead
\begin{equation}
    \mathfrak{A}^{op}(\alpha) = \oint_{\p S_L} \mathfrak{A}^{op}\alpha \approx \oint_{\p S_L} \acal^{op}\alpha - k\gamma d\alpha
\end{equation}
and fulfils an $\rbb$-Kac-Moody algebra at level $k$, and generates $\{\mathfrak{A}^{op}(\alpha),\gamma\}=\alpha$ on the gauge frame. 
\\

If we now want to induce dynamics, the first piece to consider is the interior of $S_L$ or $S_R$. For electromagnetic boundary conditions, a suitable Hamiltonian density is
\begin{equation}
    h_L = \frac{m}{2} \acal\wedge\ast_2\acal + \frac{1}{2m} \bcal\wedge\ast_2\bcal
\end{equation}
which generates the flow
\begin{equation}
    X_L[q]=m\ast_2\acal,X_L[\lambda]=\frac{1}{m}\ast_2\bcal,X_L[\acal]=k m\ast_2 \acal + \frac{1}{m}d\ast_2\bcal,X_L[\bcal]= m d\ast_2 \acal.
\end{equation}
This can be read as equivalent to a diffeomorphism if we identify $\acal_t = \frac{1}{m}\ast_2\bcal$, $\bcal_t = m\ast_2\acal$, which can be combined into the condition $\bcal = m \ast_3 \acal$. It also exactly preserves the leftover constraint $d\acal=k\bcal$, and the transformations of $q,\lambda$ are just shifts by $\bcal_t,\acal_t$ respectively.\footnote{Note, however, that it does not preserve the gauge fixing condition $d\lambda=0$. Instead, it ties the value of $\lambda$ dynamically to that of $\bcal$, and so if we just specify $\lambda=0$ at an initial time, then the evolution of $\lambda$ is fixed without a gauge condition.}
Demanding that this vector field actually preserves the extended generator $\bar \kkcal_\alpha$, fixes
\begin{equation}
    X_L[\mathfrak{A}^{op}]-k d X_L[\gamma] = \frac{1}{m} d \ast_2 \bcal
\end{equation}
This time evolution is made of field-dependent gauge transformations.
Contracting this vector field with the symplectic form \textit{including the puncture extension} gives a flux term
\begin{equation}
    -I_{X_L}\Omega_{S_L}^e = \delta(\int_{S_L}h_L) + \oint_{\p S_L} \delta\mathfrak{A}^{op} X_L[\gamma]  -\frac{1}{m} \delta (q-d\gamma)\wedge \ast_2 \bcal 
\end{equation}
which shows that either we need to impose boundary conditions to make the evolution on $S_L$ well-defined, or we need to introduce degrees of freedom on $\p S_L$ to compensate the flux term. We will for simplicity choose $X_L[\gamma]=0$.  \\
Contrast this situation with $S_R$, where we want to impose topological boundary conditions $\acal_t=0=\bcal_t$. These can be implemented e.g. by a trivial Hamiltonian $h_R=0$, as a diffeomorphism with these parameters is on-shell just the trivial transformation.

A boundary condition on $\p S_L$ can easily be supplied: We give it as a twisted self-duality on $q-d\gamma=q_{gf}$,
\begin{equation}
    q_{gf} - \sigma \bar\ast_2 q_{gf}+ d\lambda_{gf} \stackrel{\p S_L}{=} 0
\end{equation}
where $\bar\ast_2$ refers to the Hodge dual on $\p S_L\times \rbb$, not that on $S_L$. In the effective Maxwell-Chern-Simons dynamics of the boundary, where $q$ takes on the role of a connection 1-form, this becomes analogous to the usual CS Dirichlet condition that leads to WZW dynamics on the boundary. This boundary condition can be arranged by use of a Lagrangian on $\p S_L\times \rbb$, 
\begin{equation}
    L_2 = -( \int_{\p S_L\times \rbb} \gamma d\acal + \frac{k}{2}d\gamma\wedge q_{gf} + \frac{a}{2}q_{gf}\wedge\bar\ast_2 q_{gf} + q_{gf}\wedge\acal    )
\end{equation}
so then the variation piece in the action on $\p S_L\times \rbb$ becomes
\begin{equation}
    \int_{\p S_L\times \rbb} (\bcal- dq_{gf})\delta\lambda_{gf}+ (\acal- \frac{k}{2}q_{gf}- a \bar\ast_2 q_{gf})\wedge \delta q_{gf}  - d\theta_{3} + \delta L_2
\end{equation}
with symplectic potential imprint on $\p S_L$ 
\begin{equation}
    \theta_{3} =     \mathfrak{A}^{op}\delta\gamma - \frac{k}{2}d \gamma \delta\gamma.
\end{equation}
With this Lagrangian, and using the bulk gauge fixing $\bcal_b=0=\acal_b$, we have
\begin{equation}
    \begin{gathered}
        \bcal- dq_{gf}=\bcal_b +dq-dq_{gf}=0\\
        \acal- \frac{k}{2}q_{gf}- a \bar\ast_2 q_{gf} = kq + d\lambda - \frac{k}{2}q_{gf}\\
        =
        \frac{k}{2} q_{gf}- a \bar\ast_2 q_{gf} + d(\lambda+k\gamma) = k q_{gf}- a \bar\ast_2 q_{gf} + d\lambda_{gf}
    \end{gathered}
\end{equation}
So we impose the twisted self-duality condition with $\sigma = \frac{a}{k}$. Then, these boundary conditions imply on the flux term of the Hamiltonian that
\begin{equation}
    -\frac{1}{m} \delta (q-d\gamma)\wedge \ast_2 \bcal 
    \approx
    -\frac{1}{m} \delta q_{gf}\wedge \ast_2 d q_{gf}\approx -\frac{a}{km} \delta q_{gf}\wedge \ast_2 d\bar\ast_2 q_{gf}.
\end{equation}
When we choose a gauge fixing of the $q\mapsto q + d\gamma$ transformations such that $d\bar\ast_2 q_{gf}=0$, this makes the nonintegrable term in the Hamiltonian disappear. Then, the dynamics of the subregion $S_L$ is well-defined on its own. \\
The Lagrangian $L_2$, by itself, can be reduced to the effective
\begin{equation}
    L_{2,eff} = \frac{a}{2}d\gamma\wedge\bar\ast_2 d\gamma,
\end{equation}
which is just a free scalar on $\p S_L\times\rbb$, which is the simplest abelian analogue of a WZW model.\\

In summary, we studied the simple $\rbb$-BF-BB theory on a pair of overlapping subregions. Each of these was supplanted with boundary Lagrangians, and we constructed a Lagrangian on the overlap of their boundaries procedurally by studying first the kinematics of the subregion $S_L$, and then asking for integrability of the time evolution on $S_L$. The structures we found were those expected of the usual dimensional ladder BF-BB $\rightarrow$ Chern-Simons $\rightarrow$ WZW. It all also illustrates that the kinematics we found can be reduced to standard cases, by supplying Lagrangians or Hamiltonians to them.

\section{Gravity gauge transformations in codimension 1}\label{App:BulkAlgebra}
Recall that our convention is
\begin{equation}
    I_V\Omega + \delta F_V = 0\implies\{F_V,-\} = V[-]
\end{equation}
we thus display the transformations, i.e. the vector fields on phase space, as Poisson brackets. For the parameters, we have 
\begin{equation}
    \begin{gathered}
        \alpha,\gamma\in \Omega^0(M;\hfrak),\phi\in\Omega^0(M;\pfrak)\\
    \mu,\nu\in \Omega^1(M;\hfrak),\rho\in\Omega^1(M;\pfrak).
    \end{gathered}
\end{equation}
\begin{equation}
    \{H_\alpha,-\} = 
    d_\omega\alpha\frac{\delta }{\delta \omega}
    + [B,\alpha] \frac{\delta}{\delta B} 
    + [c,\alpha]\frac{\delta}{\delta c}
    + [S,\alpha] \frac{\delta}{\delta S} 
    -\alpha^I_J e^J \frac{\delta }{\delta e^I} 
    - \alpha^I_J \tau^J \frac{\delta }{\delta \tau^I}    
\end{equation}
\begin{equation}
    \{I_\phi,-\} = 
     (\phi\wedge\tau) \frac{\delta}{\delta B} 
    + (e\wedge\phi) \frac{\delta}{\delta S} 
    +d_\omega\phi \frac{\delta }{\delta e^I} 
    - c^I_J\phi^J \frac{\delta }{\delta \tau^I}    
\end{equation}
\begin{equation}
    \{J_\gamma,-\} = 
     [c,\gamma] \frac{\delta}{\delta B} 
    + d_\omega\gamma \frac{\delta}{\delta S}  
\end{equation}
\begin{equation}
    \{M_\mu ,-\} = 
    d_\omega\mu \frac{\delta}{\delta B} 
    + \star^{-1}_r {\mu} \frac{\delta}{\delta S}     
\end{equation}
\begin{equation}
    \{L_\rho,-\} = 
    - (e\wedge \rho) \frac{\delta}{\delta B} 
    +t \rho^I \frac{\delta }{\delta e^I} 
    + d_\omega \rho^I \frac{\delta }{\delta \tau^I}    
\end{equation}
\begin{equation}
    \{K_\nu,-\} = 
    \star^{-1}_r {\nu} \frac{\delta }{\delta \omega}
    + [S,\nu] \frac{\delta}{\delta B} 
    + d_\omega\nu \frac{\delta}{\delta c}
    - \nu^I_J\wedge e^J \frac{\delta }{\delta \tau^I}    
\end{equation}
This leads to an algebra of generators that looks as follows.
\begin{equation}
    \{H_\alpha, H_{\bar\alpha}\}=H_{[\alpha,\bar\alpha]}, 
\{H_\alpha, J_{\bar\alpha}\}=J_{[\alpha,\bar\alpha]},\dots, 
    \{H_\alpha, I_{\phi}\}=I_{\alpha \cdot \phi},\dots
\end{equation}
In other words, $H_\alpha$ simply acts as a (negative) Lorentz transformation on parameters of the other generators.
\begin{equation}
    \begin{aligned}
        \{I_\phi, I_{\bar\phi} \} &= J_{\phi\wedge\bar\phi} \\
        \{I_\phi, L_\rho \} &= M_{\rho\wedge\phi} - \int_\s (tc+\star^{-1}_r  c)\wedge(\rho\wedge\phi) - \oint_{\p S}\phi_I d\rho^I \\
        \{I_\phi, K_\nu \} &= L_{\nu\cdot \phi}-\int_\s \tau_I\wedge (t\nu + \star^{-1}_r  \nu)^I_J \phi^J\\
        \{ J_\gamma, K_\nu \} &= M_{\gamma,\nu} - \oint_{\p S} \gamma d \nu\\
        \{ M_\mu, K_\nu \} &= \oint_{\p S} \mu\wedge\star^{-1}_r  \nu\\
        \{ L_\rho, L_{\bar\rho} \} &= \oint_{\p S} t \rho_I\wedge\bar\rho^I\\
        \{ L_\rho, K_\nu \} &= - \int_\s (t(\rho\wedge e) + \star^{-1}_r  (\rho\wedge e) )\wedge \nu
    \end{aligned}
\end{equation}
In particular, the modified simplicity generators \textit{commute}, $\{ K_\nu, K_{\bar\nu} \} = 0$, as do the \textit{constraints} included in $M,K,L$, where the boundary value of the parameters vanish. Apart from the bulk terms between $(I,L),(I,K),(K,L)$ (which, notably, do not contain derivatives of fields or parameters!), the algebra is first-class and the corner charges weakly or strongly commute with the constraints. This immediately makes $H_\alpha,J_\gamma,M_\mu$ into observables on the reduced phase space, in particular their corner charges $(B,c,\omega)$. \\
For the remaining generators, we would usually need to perform a modification to turn their corner values into proper observables. Conventionally, any function $F$ could be modified into a function $F^\star$ which weakly commutes with the second-class part of the constraints, $I,K,L$:
\begin{equation}
    F^\star = F + I_{\phi_F} + K_{\nu_F}+L_{\rho_F}
\end{equation}
where the additional parameters are required to vanish on the boundary. However, because in this case the second-class matrix between the constraints is ultralocal, it turns out that this is not possible for the charges we have - I.e. the equation for the modification of $I_\phi$ to weakly commute with $L_\rho$ becomes
\begin{equation}
    (t + \star^{-1}_r (-))c \cdot \phi + (t + \star^{-1}_r (-))\nu_\phi\cdot e = 0
\end{equation}
This cannot be satisfied at boundary points with nonvanishing $\phi$ and vanishing $\nu_\phi$ unless $c$ vanishes on the boundary. Therefore, the conventional wisdom breaks in the case of corner charges. 

\section{BF-BB like corner algebra for gravity}\label{App:CornerAlg}
The following is an expansion of the BF-BB corner Poisson algebra in the case of the Maxwell algebra. We only report the nonzero corner brackets.
\begin{equation}\label{eq:GravCornerAlg}
    \begin{gathered}
        \{ B \alpha , B \bar \alpha \}_{S} = B [\alpha,\bar\alpha] \\
        \{B \alpha , \tau_I \phi^I\}_{S} = \tau_I \alpha^I_J \phi^J \\
        \{B \alpha ,c \gamma \}_{S} = c [\alpha,\gamma] \\
        \{B \alpha ,\omega\wedge\mu \}_{S} = d_\omega \alpha\wedge\mu = \omega\wedge[\alpha,\mu] + d\alpha\wedge\mu\\
        \{B \alpha ,e^I\wedge\rho_I \}_{S} = e_I \wedge\alpha^I_J \rho^J\\
        \{B \alpha ,S\wedge\nu \}_{S} = S\wedge[\alpha,\nu]
    \end{gathered}    
\end{equation}
\begin{equation}
    \begin{gathered}
        \{ \tau_I \phi^I ,\tau_J \bar\phi^J \}_{S} = c  (\bar\phi\wedge\phi)\\
        \{ \tau_I \phi^I ,e^I\wedge\rho_I \}_{S} = d_\omega\phi^I\wedge\rho_I = \omega\wedge(\rho\wedge\phi) + d\phi^I\wedge\rho_I \\
        \{ \tau_I \phi^I ,S\wedge\nu \}_{S} = e_I \wedge \nu^I_J\phi^J
    \end{gathered}
\end{equation}
\begin{equation}
    \begin{gathered}
        \{ c\gamma ,S\wedge\nu \}_{S} = d_\omega\gamma\wedge\nu = \omega\wedge[\gamma,\nu] + d\gamma\wedge\nu
    \end{gathered}
\end{equation}
\begin{equation}
    \begin{gathered}
        \{ \omega\wedge\mu ,S\wedge\nu \}_{S} = \star^{-1}_r {\mu}\wedge\nu
    \end{gathered}
\end{equation}
\begin{equation}
    \begin{gathered}
        \{e^I\wedge\rho_I  ,e^I\wedge\bar\rho_I \}_{S} = t \rho^I\wedge\bar\rho_I 
    \end{gathered}
\end{equation}
It is important to note that the bulk-induced corner algebra is antisymmetric only up to codimension 3 terms, in principle. E.g. when derived via transformation properties of the fields,
\begin{equation}
    \begin{gathered}
        \{ \tau_I \phi^I ,e^I\wedge\rho_I \}_{S} = d_\omega\phi^I\wedge\rho_I \\
        -\{  e^I\wedge\rho_I , \tau_I \phi^I \}_{S} = - d_\omega\phi^I\wedge\rho_I = d(-\phi^I\wedge\rho_I) + d_\omega\phi^I\wedge\rho_I
    \end{gathered}
\end{equation}
However, notice that this only pertains to the central cocycles, when derivatives of the transformation parameters are involved, so in $ \{\tau,e\}, \{B,\omega\},\{c,S\} $. Notice that these are precisely the canonical pairs of the codimension 1 bulk. The central terms refer to the codimension 3 puncture data in an essential way (they provide the central charge), so this does not come as a surprise.
We choose to display here the variant that is consistent with the action of the $\bcal$ fields as Yang-Mills type gauge transformations acting on the $\acal$ fields.\\
Much more importantly, though, one needs to verify whether the Jacobi identity holds in the above. In fact, out of 20 identities, exactly one fails:
\begin{equation}
    \label{FailedJacIdent}
    \begin{gathered}
        \{\{\tau_I \phi^I,e^I\wedge\rho_I\},S\wedge\nu\}+\{\{e^I\wedge\rho_I,S\wedge\nu\},\tau_I \phi^I\}+\{\{S\wedge\nu,\tau_I \phi^I\},e^I\wedge\rho_I\}\\
        = (\star^{-1}_r  \nu + t \nu )\wedge (\rho\wedge\phi) \neq 0
    \end{gathered}
\end{equation}
\textit{This means the above is \textit{actually} a Poisson algebra if we remove $\tau^I$ from it, i.e. the subspace generated from $B,c,\omega,e,S$ forms a strict Poisson algebra.} Removing $\tau$ is not a physical restriction, as it is set to zero by the equations of motion.

\section{Gravity corner symplectic form}\label{App:CornerSympForm}
\paragraph{Expanding the BF-BB form}\,\\
Here, we expand the corner symplectic form for BF-BB, with the caveat mentioned in the main text, that we need to fix one of the $\chi$-frames in order to have a closed symplectic form. We also add the puncture extension $-\delta (\langle q,\chi\rangle)$.
\begin{equation}
\begin{gathered}
       \Omega_{S} =\int_{S}
        \langle\delta\bcal,\chi \rangle
        - \langle\bcal,\frac{1}{2}[\chi,\chi] \rangle
         + \langle\delta\acal-d_\acal\chi,\psi \rangle\\
         -
        \frac{1}{2}\langle\psi ,\Phi(\psi )\rangle 
        - \oint_{\p S}
       \langle\psi, \chi\rangle + \langle q, \frac{1}{2} [\chi,\chi]\rangle
\end{gathered}
\end{equation}
The only thing that changes from the BF-BB case is that this symplectic form is no longer exact\footnote{To our knowledge.}: because conjugation by $\phi$ does not pass through $\Phi$, the potential \eqref{eq:codim2BFBBBsymppot} does not give rise to the above form anymore. $\Phi$ also spoils the closedness of the above form, but this will be relatively easy to fix. First, though, we split this into its gravity components using
\begin{equation}
    \bcal = (\frac{1}{k_\hfrak}c,\frac{1}{k_\pfrak}\tau,\frac{1}{k_\hfrak} B)  \quad \acal = (\omega, e,S),
\end{equation}
and an equivalent split of $\chi,\psi$
\begin{equation}
    \chi = (\chi_{\hfrak} , \chi_\pfrak , \chi_{\hfrak^\ast} )\quad \psi = (\frac{1}{k_\hfrak} \psi_{\hfrak} ,\frac{1}{k_\pfrak}\psi_\pfrak  ,\frac{1}{k_\hfrak}\psi_{\hfrak^\ast}  ).
\end{equation}
We do this piece by piece, as the total amounts to 18 terms.
\begin{equation}
    \langle\delta\bcal,\chi\rangle = \delta c \chi_{\hfrak^\ast} + \delta\tau \chi_\pfrak + \delta B \chi_{\hfrak}
\end{equation}
\begin{equation}
    - \langle\bcal,\frac{1}{2}[\chi,\chi] \rangle
    =
    - \frac{1}{2} B [\chi_{\hfrak},\chi_{\hfrak}] - \tau \cdot \chi_{\hfrak}\cdot\chi_{\pfrak} - c [\chi_{\hfrak},\chi_{\hfrak^\ast}] - \frac{1}{2}c (\chi_{\pfrak}\wedge \chi_{\pfrak})
\end{equation}
\begin{equation}
   \langle \delta\acal,\psi \rangle
   = 
   \delta\omega \wedge \psi_{\hfrak^\ast} + \delta e \wedge\psi_\pfrak
   + \delta S\wedge\psi_{\hfrak} 
\end{equation}
\begin{equation}
  - \langle \psi , d_\acal\chi\rangle 
   =
   -\psi_{\hfrak^\ast}  \wedge d_\omega\chi_{\hfrak}
   - 
   \psi_{\hfrak} \wedge (
   d_\omega \chi_{\hfrak^\ast} + [S,\chi_{\hfrak}]+ e\wedge\chi_\pfrak
   )
   - \psi_\pfrak\wedge ( d_\omega \chi_\pfrak - \chi_{\hfrak} \cdot e )
\end{equation}
\begin{equation}
    -\frac{1}{2}\langle \psi,\Phi(\psi)\rangle 
    = -\psi_{\hfrak}\wedge \star^{-1}_r {(\psi_{\hfrak^\ast}) } - \frac{t}{2} \psi_\pfrak\wedge \psi_\pfrak.
\end{equation}
\begin{equation}
    -\langle \psi, \chi\rangle 
    =
    -\psi_{\hfrak} \chi_{\hfrak^\ast}-\psi_{\pfrak} \chi_{\pfrak} - \psi_{\hfrak^\ast} \chi_{\hfrak}
\end{equation}
\begin{equation}
    - \langle q,\frac{1}{2}[\chi,\chi] \rangle
    =
    - \frac{1}{2} q_{\hfrak^\ast} [\chi_{\hfrak},\chi_{\hfrak}] - q_\pfrak \cdot \chi_{\hfrak}\cdot\chi_{\pfrak} - q_\hfrak [\chi_{\hfrak},\chi_{\hfrak^\ast}] - \frac{1}{2}q_\hfrak (\chi_{\pfrak}\wedge \chi_{\pfrak})
\end{equation}

Of special note is the line $-\frac{1}{2}\langle \psi,\Phi(\psi)\rangle $, which indicates the Poisson pairing between $S,\omega$ and the $e$s. It is this piece which is not closed: we have
\begin{equation}
    \delta(-\frac{1}{2}\langle \psi,\Phi(\psi)\rangle )
    =
    \chi_\pfrak^I (\varepsilon \star^{-1}_r  \psi_\hfrak + t \psi_\hfrak)_{IJ}\wedge \psi_\pfrak^J
\end{equation}
which notably only includes the $\chi,\psi$-frames in our phase space. In the corner algebra \eqref{eq:GravCornerAlg}, we found it necessary to remove $\tau$ in order to work with a true Poisson algebra. On the symplectic level, this removal can be mirrored by fixing the conjugate partner to $\tau$, which is given by $\chi_\pfrak$. And as we see here, if we fix $\chi_\pfrak=0$, the non-closure disappears, and the symplectic form is closed, as it should be. \textit{Therefore, while we keep $\chi_\pfrak$ in the expansions for the moment for completeness, we must always set it to zero for the discussion of the physical phase space.}\\
The contraction of the symplectic form with a phase space vector field can be written as
\begin{equation}
\begin{aligned}
    -I_V \Omega_S = &\int_S 
     \delta c I_V\chi_{\hfrak^\ast} + \delta\tau I_V\chi_\pfrak + \delta B I_V\chi_{\hfrak}\\
    &+ \delta\omega \wedge I_V\psi_{\hfrak^\ast} + \delta e \wedge I_V\psi_\pfrak
   + \delta S\wedge I_V\psi_{\hfrak} \\
   &- X_{\hfrak}\chi_\hfrak - X_\pfrak \chi_\pfrak - X_{\hfrak^\ast}\chi_{\hfrak^\ast}\\
   &- Y_{\hfrak}\wedge\psi_\hfrak - Y_\pfrak \wedge\psi_\pfrak - Y_{\hfrak^\ast}\wedge\psi_{\hfrak^\ast}\\
   &- \oint_{\p S} q_V
\end{aligned}
\end{equation}
with
\begin{equation}
    \begin{aligned}
        X_{\hfrak} &= V[B] - [B,I_V\chi_{\hfrak}] - (I_V\chi_\pfrak\wedge \tau   ) - [c,I_V\chi_{\hfrak^\ast}] - d_\omega I_V\psi_{\hfrak^\ast} - [I_V\psi_{\hfrak},S] + (I_V\psi_\pfrak\wedge e) \\
         X_\pfrak &= V[\tau] + I_V\chi_{\hfrak}\cdot\tau + c\cdot I_V\chi_{\pfrak} + I_V\psi_{\hfrak}\cdot\wedge e - d_\omega I_V\psi_{\pfrak}\\
         X_{\hfrak^\ast} &= V[c] - [c,I_V\chi_{\hfrak}] - d_\omega I_V\psi_{\hfrak} \\
         Y_{\hfrak} &= V[S] - d_\omega I_V\chi_{\hfrak^\ast} - [S,I_V\chi_{\hfrak}]- e\wedge I_V\chi_\pfrak - \star^{-1}_r {(I_V\psi_{\hfrak^\ast}) } \\
         Y_\pfrak  &= V[e] -  d_\omega I_V\chi_\pfrak + I_V\chi_{\hfrak} \cdot e - t I_V \psi_\pfrak  \\
         Y_{\hfrak^\ast} &= V[\omega] -  d_\omega I_V\chi_{\hfrak} - \star^{-1}_r {(I_V\psi_{\hfrak}) }  \\
    \end{aligned}
\end{equation}
and the puncture term is
\begin{equation}
\begin{gathered}
        q_V = \psi_\hfrak I_V\chi_{\hfrak^\ast}  + \psi_{\hfrak^\ast} I_V\chi_\hfrak + \psi_\pfrak I_V\chi_\pfrak 
    - [q_\hfrak,I_V\chi_{\hfrak}]\chi_{\hfrak^\ast}\\
    -( [q_\hfrak,I_V\chi_{\hfrak^\ast}]-[I_V\chi_\hfrak,q_{\hfrak^\ast}]+\varepsilon q_\pfrak\wedge I_V\chi_\pfrak )\chi_\hfrak
    -\chi_\pfrak\cdot ( q_\hfrak \cdot I_V\chi_\pfrak - I_V\chi_\hfrak \cdot q_\pfrak )
\end{gathered}
\end{equation}
and only contains the frames.
Note in particular that for the actual phase space discussion, $\chi_\pfrak$ can be set to zero, removing $X_\pfrak$ altogether from the contraction. As for $\tau$, just like in the codimension 0 bulk, we will set it to zero for the following discussions. 

\paragraph{Reducing the BF-BB form}\,\\
Reducing the BF-BB corner symplectic form by the constraint $\bcal=\Phi^{-1} F_\acal$ leads to a symplectic form with kernel
\begin{equation}
    V[\acal]=0= I_V\psi + \Phi^{-1}d_\acal I_V\chi.
\end{equation}
This is sensible as the constraints are first class, meaning this really is a presymplectic form on a coisotropic subset. By gauge fixing the $\chi-$frame $\phi$ to some value, e.g. setting $\chi=0$ in the interior, we get a symplectic form again,
\begin{equation}
   \Omega_{S} :=\int_{S}
    \delta\acal\wedge\psi  -
        \frac{1}{2}\psi\Phi(\psi) - \delta\oint_{\p S} q\chi .
\end{equation}
In the gravity case, this is less straightforward. Because the constraints are mixed, in principle the constrained subset could be coisotropic or not. We will simply look for a symplectic subspace by gauge fixing again. Once again we can hope for a kernel parametrised by the way the $\chi$-gauge frames transform. The only terms that change form when going on-shell are
\begin{equation}
\begin{aligned}
        \langle\delta\bcal,\chi\rangle +\langle \delta\acal,\psi \rangle &\approx 
     \delta S\wedge ( \psi_{\hfrak} +(d_\omega \chi_{\hfrak})_{r})\\
    &+ \delta e \wedge(\psi_\pfrak + \star_r (\chi_\hfrak )\cdot e 
    %+ \frac{1}{t} d_\omega \chi_\pfrak
    )\\
    &+\delta\omega \wedge( \psi_{\hfrak^\ast} + \star_r(d_\omega \chi_{\hfrak^\ast}) 
    %+ \frac{1}{t} (\chi_\pfrak\wedge e) 
    + \star_r[S,\chi_\hfrak] )\\
    &+d( \delta S \star_r \chi_\hfrak + \delta \omega \star_r\chi_{\hfrak^\ast} 
    %+ \frac{1}{t}\delta e \chi_\pfrak 
    )
\end{aligned}
\end{equation}
because the $\delta B,\delta c$ terms get reexpressed. In this and any other expressions, the removal of $\tau,\chi_\pfrak$ is necessary.
We simply replace the fields $B,c$ by their on-shell values and remove $\tau$, which really only affects the term
\begin{equation}
\begin{aligned}
        - \langle\bcal,\frac{1}{2}[\chi,\chi] \rangle
    \approx
    &- \frac{1}{2} \star_r(d_\omega S+\frac{1}{2}e^2) [\chi_{\hfrak},\chi_{\hfrak}] 
    %- \frac{1}{t}d_\omega e \cdot \chi_{\hfrak}\cdot\chi_{\pfrak} 
    \\
    &- \star_r(F_\omega) [\chi_{\hfrak},\chi_{\hfrak^\ast}] 
    - \frac{1}{2}\star_r(F_\omega) (\chi_{\pfrak}\wedge \chi_{\pfrak})
\end{aligned}
\end{equation}
Then, we need to check for a kernel in this pre-symplectic form, for the moment only on closed corners $\p S=\emptyset$. The first of conditions for a vector field $V$ to lie in the alleged kernel are the coefficients of $\delta\omega,\delta e,\delta S$,
\begin{equation}
    \begin{aligned}
    0=& I_V\psi_{\hfrak} + \star_r(d_\omega I_V\chi_{\hfrak})\\
    0=& I_V\psi_\pfrak + \star_r(I_V\chi_\hfrak )\cdot e 
    %+ \frac{1}{t} d_\omega I_V\chi_\pfrak
    \\
    0=&I_V \psi_{\hfrak^\ast} + \star_r(d_\omega I_V\chi_{\hfrak^\ast}) 
    %+ \frac{1}{t} (I_V\chi_\pfrak\wedge e) 
    + \star_r[S,I_V\chi_\hfrak]
    \end{aligned}
\end{equation}
These determine the transformation of the $\psi$-frames through those of the $\chi$-frames. Next, we have the conditions $Y_{\hfrak},Y_{\pfrak},Y_{\hfrak^\ast}=0$, which determine the transformations of $S,\omega,e$ through the ones of $\psi$. 
They give
\begin{equation}
    \begin{aligned}
       0=Y_{\hfrak} &= V[S] %+ I_V\chi_\pfrak\wedge e %+ \frac{1}{t}  \star^{-1}_r (I_V\chi_\pfrak\wedge e)  
       \\
         0=Y_\pfrak  &= V[e]  + I_V\chi_{\hfrak} \cdot e + t \star_r(I_V\chi_\hfrak )\cdot e  \\
         0=Y_{\hfrak^\ast} &= V[\omega] 
    \end{aligned}
\end{equation}
From this we can see that $\omega,S$'s degrees of freedom are not in the kernel at all. Further, unlike in BF-BB, this is not necessarily true of $e$, which may transform in the kernel in principle due to the choice of homomorphism $\Phi$. 
Finally, there are still the conditions $X_{\hfrak},%X_{\pfrak},
X_{\hfrak^\ast}=0$, in which we again replace the expressions of $B,c%,\tau
$ by their on-shell values, i.e $X_{\hfrak^\ast}$ becomes
\begin{equation}
\begin{aligned}
        0=X_{\hfrak^\ast} &= \star_r V[F_\omega] - \star_r[F_\omega,I_V\chi_{\hfrak}] - d_\omega I_V\psi_{\hfrak} \\
        &=\star_r V[F_\omega] - \star_r[F_\omega,I_V\chi_{\hfrak}] + \star_r[F_\omega, I_V\chi_{\hfrak}] \\
        &= \star_r V[F_\omega] ,
\end{aligned}
\end{equation}
which is consistent with $V[\omega]=0$; and secondly,
\begin{equation}
    \begin{aligned}
      0=  X_{\hfrak} =& V[B] - [B,I_V\chi_{\hfrak}] 
      %- (I_V\chi_\pfrak\wedge \tau   ) 
      - [c,I_V\chi_{\hfrak^\ast}] \\
      &+ \star_r d_\omega ((d_\omega I_V\chi_{\hfrak^\ast}) 
      %+ \frac{1}{t} (I_V\chi_\pfrak\wedge e) 
      + [S,I_V\chi_\hfrak]) \\
      &+ \star_r[d_\omega I_V\chi_{\hfrak},S] 
      - (\star_r(I_V\chi_\hfrak )\cdot e 
      %+ \frac{1}{t} d_\omega I_V\chi_\pfrak
      )\wedge e\\
      %%
      %%
      %=& V[d_\omega S + \frac{1}{2}e^2]\\
      %%
      %%
      =&
      %- d_\omega (I_V\chi_\pfrak\wedge e + \frac{1}{t}  \star^{-1}_r (I_V\chi_\pfrak\wedge e)) + 
      [\frac{1}{2}e^2, t \star_r(I_V\chi_\hfrak )+I_V\chi_{\hfrak} ]
    \end{aligned}
\end{equation}
The final condition has a direct, algebraic solution: Fix first a reference 2-form $w$ on $S$. Then, taking $e^2$ on the 2-surface $S$ determines a normalised internal 0-form bivector $\epsilon_\parallel$, via $e^2 \propto \epsilon_\parallel w$ which gives the 2D internal space spanned by $e$, when pulled back to $S$. There is also an associated normal bivector $\epsilon_\perp=\star\epsilon_\parallel$. Then, the condition $X_\hfrak=0$ is solved by
\begin{equation}
    I_V\chi_\hfrak = (a+b\star) \epsilon_\perp\qquad a,b\in \rbb.
\end{equation}
This means the only $\chi_\hfrak$ changes which are in the kernel are those field-dependent ones that rotate in the planes determined by $e^2$. Similarly, the equation from $Y_\pfrak$ shows that to be in the kernel of $\Omega_S$, this change must be accompanied by an associated rotation of $e$ in the same planes. Of these, only the ones in the normal plane $\epsilon_\perp$ have a nontrivial effect on $e$. In turn, the $\psi$-frames must all be affected in kernel transformations. The meaning of this kernel is essentially that when the constraints are imposed, a part of them is first class, and on-shell one should think of the $\chi_\hfrak$ frame as being a part of the fields $e$; therefore, transforming one is the same as transforming the other, and a redundancy appears in the description. \\
In contrast, $\omega,S$ are unaffected by the kernel, and there are no restrictions on $I_V\chi_{\hfrak^\ast}$, meaning we can gauge fix this frame arbitrarily, and set $\chi_{\hfrak^\ast}=0$. This makes it part of the derivative shift degrees of freedom in $S$, which are generated on the corner by $\bar K$. The kernel thus allows us to slightly shift $\chi_\hfrak$ and arbitrarily shift $\chi_{\hfrak^\ast}$.\\
\textit{However, including the puncture pieces on $\p S$ gives one additional condition on kernel vector fields: It requires that $I_V\chi_{\hfrak^\ast}=0$ on $\p S$}. Hence, strictly, we can only remove $\chi_{\hfrak^\ast}$ in the interior of $S$.
Removing $\chi_{\hfrak^\ast}$ in the cavalier way of only removing the bulk terms gives then the partially reduced symplectic form
\begin{equation}
    \begin{aligned}
        \Omega_S \approx& \int_S
     \delta S\wedge ( \psi_{\hfrak} + \star_r(d_\omega \chi_{\hfrak}))
    + \delta e \wedge(\psi_\pfrak + \star_r (\chi_\hfrak ) \cdot e 
    %+ \frac{1}{t} d_\omega \chi_\pfrak
    )
    +\delta\omega \wedge( \psi_{\hfrak^\ast}  %+ \frac{1}{t} (\chi_\pfrak\wedge e) 
    + \star_r [S,\chi_\hfrak] )\\
    &- \frac{1}{2} \star_r (d_\omega S+\frac{1}{2}e^2) [\chi_{\hfrak},\chi_{\hfrak}] 
    %- \frac{1}{t}d_\omega e \cdot \chi_{\hfrak}\cdot\chi_{\pfrak} 
     %- \frac{1}{2}(F_\omega)_{r} (\chi_{\pfrak}\wedge \chi_{\pfrak})\\
    %%
    %%
    -\psi_{\hfrak^\ast}  \wedge d_\omega\chi_{\hfrak}
   - 
   \psi_{\hfrak} \wedge (  [S,\chi_{\hfrak}]
   %+ e\wedge\chi_\pfrak
   )
   %- \psi_\pfrak\wedge ( 
   %d_\omega \chi_\pfrak 
   %- \chi_{\hfrak} \cdot e )\\
   + \psi_\pfrak\cdot 
   \chi_{\hfrak} \cdot e \\
   &-\psi_{\hfrak}\wedge \star^{-1}_r {(\psi_{\hfrak^\ast}) } 
   - \frac{t}{2} \psi_\pfrak\wedge \psi_\pfrak\\
   -&\oint_{\p S}
   \psi_{\hfrak} \chi_{\hfrak^\ast} + \psi_{\hfrak^\ast} \chi_{\hfrak}
    + \frac{1}{2} q_{\hfrak^\ast} [\chi_{\hfrak},\chi_{\hfrak}]  + q_\hfrak [\chi_{\hfrak},\chi_{\hfrak^\ast}] 
\end{aligned}
\end{equation}
We notice that in contrast to the BF-BB case, the $\chi$-frames have not been removed from the phase space, a priori. The only part that has been removed are the field-dependent $\chi_\hfrak$-rotations proportional to the generator. 
However, these are \textit{always} in the kernel, and do not really have much meaning by themselves. Consider just the pure Yang-Mills part for the BF theory case, 
\begin{equation}
   \delta B \chi_\hfrak - \frac{1}{2} B [\chi_{\hfrak},\chi_{\hfrak}],
\end{equation}
then by the same calculation and logic as above, if we pick a 2-form $w$, and $B = \gamma w$, then $I_V\chi_\hfrak\propto (a+b\star)\gamma$ will be in the kernel. However, ordinarily we would not consider this as a meaningful kernel to remove. Instead, we will see $\chi_\hfrak$ as a genuine corner degree of freedom that exists in gravity, but not in the topological case.
If we \textit{do} fix it artificially, we end up with something more BF-BB like,
\begin{equation}
    \begin{aligned}
        \Omega_S &\approx \int_S
     \delta S\wedge  \psi_{\hfrak^\ast}
    + \delta e \wedge \psi_\pfrak
    +\delta\omega \wedge \psi_{\hfrak^\ast}  
   \\
   &-\psi_{\hfrak}\wedge \star^{-1}_r {(\psi_{\hfrak^\ast}) } 
   - \frac{t}{2} \psi_\pfrak\wedge \psi_\pfrak
\end{aligned}
\end{equation}
From which we can infer the Poisson relation $\{\omega\wedge\mu,S\wedge\nu\}=\mu\wedge \star^{-1}_r \nu, \{e,e\}=t\eta$ that we already had for the kinematical phase space. Summarising, we can see that the corner symplectic form does have the expected structure involving $e,S,\omega$, the only nontrivial difference to BF-BB being that a priori the $\chi_\hfrak$-frame remains.%, depending on the shape of the solution space of the kernel equations. 
\\This frame, apparently, encodes a crucial difference between the transformation behaviours of bulk and corner fields. The corner charge generating Lorentz transformations coming from the bulk is $\frac{1}{2}\star_\beta e^2$; but this cannot generate the usual Lorentz transformations on $e$ in the symplectic structure $\delta e \delta e$. The symplectic structure here is then consistent with these facts by adding in an additional degree of freedom in $\chi_\hfrak$, which keeps track of the Lorentz transformations enacted by $\frac{1}{2}\star_\beta e^2$, and also loosens the symplectic structure enough for the two wanted properties to be possible at the same time.\\
Let us, finally, remove the $\psi$-frames from the phase space. This involves setting (the variation of) the generator of $\psi$-shifts to zero, making their values arbitrary. This is the same procedure that leads, in the BF-BB case, to the Chern-Simons/Atiyah-Bott symplectic form on the corner. The generator is, from the BF-BB side,
\begin{equation}
    \acal^{op}:= \phi \acal \phi^{-1} + \phi d \phi^{-1} - \Phi(\phi q \phi^{-1}).
\end{equation}
This gives the relations
\begin{equation}
    \psi_\pfrak = \frac{1}{t}( \delta e + \chi_\hfrak\cdot e )
\end{equation}
\begin{equation}
     \psi_{\hfrak} = \star_r(\delta\omega - d_\omega \chi_\hfrak)
\end{equation}
\begin{equation}
    \psi_{\hfrak^\ast}=\star_r(\delta S - [S,\chi_\hfrak]).
\end{equation}
Imposing them on the symplectic form yields a symplectic form that only involves $e,S,\omega, \chi_\hfrak$:
\begin{equation}
\begin{aligned}
        \Omega_S \approx 
    &\int_S \delta S\wedge\star_r\delta\omega + \frac{1}{2t}\delta e \delta e + \delta(\frac{1}{2}\star_\beta e^2 \chi_\hfrak) %- d( S_{r} \frac{1}{2}[\chi_\hfrak,\chi_\hfrak] )
    \\
    -&\oint_{\p S}
    \;(\delta\omega-d\chi_\hfrak) \star_r \chi_{\hfrak^\ast} 
    + \delta S \star_r \chi_\hfrak\\
    &+  (q_{\hfrak^\ast}- \star_r S) \frac{1}{2}[\chi_{\hfrak},\chi_{\hfrak}]  + (q_\hfrak-\star_r \omega) [\chi_{\hfrak},\chi_{\hfrak^\ast}] 
\end{aligned}
\end{equation}
So in fact, as expected, the resulting symplectic form is simply the Atiyah-Bott one for $\omega,e,S$ with the pairing given by $\Phi^{-1}$, but with the addition of the $\chi_\hfrak$-frame, which is still an independent degree of freedom. It is conjugate to the bulk Lorentz charge $\frac{1}{2}\star_\beta e^2$, as one might expect. The puncture, in turn, once again carries a mixture of $\acal$ and $q$ degrees of freedom. These are technically further restricted by the condition $\acal^{op}=0$, but we will not need this here.\\
It seems reasonable, though somewhat ad-hoc, to also include the other bulk gravitational corner charge, the Brown-York momentum, $p^I= -(\star_\beta\omega)^{IJ}\wedge e_J $ in an analogous fashion. This would reintroduce the $\chi_\pfrak$ frame and yield a symplectic form
\begin{equation}
    \Omega_S' := 
    \int_S \delta S\wedge\star_r\delta\omega + \frac{1}{2t}\delta e \delta e + \delta(\frac{1}{2}(\star_\beta e)^2 \chi_\hfrak) -\delta (p_I \chi_{\pfrak}^I).
\end{equation}
Due to the non-closedness of the BF-BB corner symplectic form, it is not clear to us how one would \textit{derive} such a modification, however natural it may appear after the fact. The interpretation, at least, is relatively clear: the corner phase space keeps track of the degrees of freedom which, from the bulk slice $\s$ point of view are gauge edge modes, encoded in the reference frames $\chi_\hfrak,\chi_\pfrak$. These are then realised as quadratic charges in the corner Poisson algebra of $\omega,e,S$, which in principle have their own, separate generators for gauge transformations, not given by the corner charges coming from the bulk. 

\section{Corner algebra in the presence of second-class constraints}
\label{App:SecondClassSchematic}
Here, we will try to understand the formalities of extracting a corner algebra when second-class constraints are present. First, we expose the problem in the notation of integrals of local forms. We then switch to classic index-based notation to understand the rough structure of corner algebras when the Dirac bracket is used.\\

Consider the following logic: We are given a family of functions $F_\alpha$, with $\alpha$ local parameters, of the form
\begin{equation}
    F_\alpha = \int_\s f_\alpha= \int_\s c_\alpha + d q_\alpha.
\end{equation}
For $\alpha$ with vanishing support on corners, we regard $F_\alpha$ as a constraint, i.e. $c_\alpha\approx 0$ and $q_\alpha$ as corner aspects. Their algebra takes the generic form
\begin{equation}
    \{F_\alpha, F_{\bar\alpha}\} = F_{[\alpha,\bar\alpha]} + Z_{\alpha,\bar\alpha}
\end{equation}
where $Z_{\alpha,\bar\alpha}$ is a family of on-shell nonvanishing phase space functions, encoding the second class part of the constraints. Ideally, this takes the local form
\begin{equation}
     Z_{\alpha,\bar\alpha}=\int_\s z_{\alpha,\bar\alpha}.
\end{equation}
Now, for generic $\alpha$ with nonvanishing support on corners, instead we might find an algebra
\begin{equation}
    \{F_\alpha, F_{\bar\alpha}\} = F_{[\alpha,\bar\alpha]} + Z_{\alpha,\bar\alpha} + \oint_{\p\s} w_{\alpha,\bar\alpha}
\end{equation}
with additional codimension 2 terms $w_{\alpha,\bar\alpha}$. A good example of this is in Chern-Simons, where $z_{\alpha,\bar\alpha}=0$ and $w_{\alpha,\bar\alpha}= \bar\alpha d\alpha$.\\
The problem in question arises because $z_{\alpha,\bar\alpha}\neq 0$ on-shell, preventing us from localising the algebra onto the on-shell codimension 2 charges. I.e. second-class constraints make the corner algebra `leak into the bulk'. From the perspective of the corner, this looks rather strange, as if the Poisson bracket between two charges suddenly involves an additional spatial dimension. However, this is perhaps a categorical mistake: If we perform the logic of the corner algebra here, then evaluating this algebra on-shell for constraints only gives
\begin{equation}
    \{F_\alpha=0, F_{\bar\alpha}=0\} \approx Z_{\alpha,\bar\alpha} \neq 0
\end{equation}
which is the classic mistake of using the wrong Poisson bracket that Dirac showed how to avoid. The logic is simple: we need to reduce to the Poisson subspace where the second-class constraints hold, and use the Poisson bracket there, so the Dirac bracket $\{|- ,- |\} $. Then, the Dirac bracket equivalent of the above discussion is again effectively first-class only,
\begin{equation}
    \{|F_\alpha, F_{\bar\alpha}|\} = F_{[\alpha,\bar\alpha]'}  + \oint_{\p\s} w_{\alpha,\bar\alpha}'
\end{equation}
and the corner algebra logic goes through. 

\paragraph{Generic corner algebra structure}\,\\
Now, let us study how this looks like in general. For this, we choose to switch to a matrix-based notation so that we can use the usual formula for the Dirac bracket. Let $\Phi_{\bar{i}}, X_{\bar{a}}$ denote two sets of phase space functions. Among the sets, we identify subsets $\Phi_i=\phi_i$ and $X_a=\chi_a$, which are respectively first class and second class constraints that we wish to impose. We assume that the algebra of these families is
\begin{equation}
    \{\Phi_{\bar i} , \Phi_{\bar j} \} =f_{\bar i\bar j}^{\bar k} \Phi_{\bar k} + \tilde f_{\bar i\bar j}^{\bar c}X_{\bar c} + u_{\bar i \bar j}
\end{equation}
\begin{equation}
    \{\Phi_{\bar i} , X_{\bar a} \} =g_{\bar i\bar a}^{\bar k} \Phi_{\bar k}+\tilde g_{\bar i\bar a}^{\bar c} X_{\bar c}  + v_{\bar i \bar a}
\end{equation}
\begin{equation}
    \{X_{\bar a} , X_{\bar b} \} = h_{\bar a\bar b}^{\bar k} \Phi_{\bar k}+ \tilde h_{\bar a\bar b}^{\bar c} X_{\bar c}  + Z_{\bar a \bar b} + w_{\bar a \bar b}
\end{equation}
in which we require that when we restrict the indices $\bar i, \bar a$ etc to the index subset of the constraints, then the functions $u, v,w$ vanish identically. This algebra structure mirrors the split between bulk constraints and corner charges: When local parameters $\alpha\hat{=} \bar i$ have trivial support on the boundary, the generator $F_\alpha \hat = \Phi_i$ becomes a constraint. The functions $u,v,w$ emulate corner-supported terms that of course will not appear in any of the constraint Poisson brackets. This is because otherwise constraints would not form a closed subalgebra. \\
In particular, the algebra of the constraints is 
\begin{equation}
\begin{aligned}
    \{\phi_{ i} , \phi_{ j} \} &=f_{ i j}^{ k} \phi_{ k} + \tilde f_{ i j}^{ c}\chi_{ c} \\
    \{\phi_{ i} , \chi_{ a} \} &=g_{ i a}^{ k} \phi_{ k}+\tilde g_{ i a}^{ c} \chi_{ c}  \\
    \{\chi_{ a} , \chi_{ b} \} &= h_{ a b}^{ k} \phi_{ k}+ \tilde h_{ a b}^{ c} \chi_{ c}  + Z_{ a  b}.
\end{aligned}
\end{equation}
In the main text, in the codimension 1 algebra of generators, we find that the first class set is $\Phi_{\bar i} = (H_\alpha,J_\beta, M_\mu) $, and the second-class set is $ X_{\bar a} = (I_\phi, L_\rho, K_\nu ) $. For purely bulk-supported parameters, the functions are constraints, but separating the index set cleanly is not immediate. The situation there simplifies in that $\tilde f^{\bar c}_{\bar i\bar j}=0=u_{\bar i \bar j}$. $Z$ is an antisymmetric matrix between the second-class constraints and contains only algebraic ultralocal expressions, i.e. no derivatives, but depends linearly on the fields $c,\tau,e$. Additionally, apart from an influence by the central terms $Z$, the $\Phi,X$ are all observables, which leads to simplifications in the on-shell algebra.\\
Let now $M^{ab}$ be the inverse of the matrix $Z_{ab}$.
The Dirac bracket is then
\begin{equation}
    \{| F , G |\} = \{ F, G\} - \{F, \chi_a\} M^{ab} \{\chi_b, G\}
\end{equation}
and we can insert the functions $\Phi,X$ into this to see the resulting algebra of corner charges. First, the first-class sector,
\begin{equation}
\begin{aligned}
        \{|\Phi_{\bar i} , \Phi_{\bar j} |\} &=f_{\bar i\bar j}^{\bar k} \Phi_{\bar k} + \tilde f_{\bar i\bar j}^{\bar c}X_{\bar c} + u_{\bar i \bar j}
    + M^{ab} 
    (g_{\bar i  a}^{\bar k} \Phi_{\bar k}+\tilde g_{\bar i a}^{\bar c} X_{\bar c}  )
    (g_{\bar j  b}^{ \bar l} \Phi_{\bar l}+\tilde g_{\bar j b}^{\bar d} X_{\bar d}  )
\end{aligned}
\end{equation}
where we see that we get extensions which are in principle quadratic in the original functions $\Phi,X$. The second-class sector is similar,
\begin{equation}
\begin{aligned}
        \{| X_{\bar a} , X_{\bar b} |\} &= h_{\bar a\bar b}^{\bar k} \Phi_{\bar k}+ \tilde h_{\bar a\bar b}^{\bar c} X_{\bar c}+ Z_{\bar a \bar b} + w_{\bar a \bar b}\\
    &+ M^{cd}
     (h_{\bar a  c}^{\bar k} \Phi_{\bar k}+ \tilde h_{\bar a c}^{\bar e} X_{\bar e}  + Z_{\bar a  c} )
     (h_{\bar b  d}^{\bar l} \Phi_{\bar l}+ \tilde h_{\bar b d}^{\bar f} X_{\bar f}  + Z_{\bar b d} )\\
\end{aligned}
\end{equation}
where in particular appears the combination $ Z_{\bar a \bar b} -  Z_{\bar a  c}M^{cd} Z_{d \bar b }$, which vanishes when either of $\bar a,\bar b$ are in the constraint index set, and is thus corner-supported. If furthermore $Z$ is algebraic and ultralocal, this expression vanishes. There are other terms involving products of $Z$, like $X_{\bar e} Z_{\bar b d}$. These are potentially bulk terms, but we need to go on-shell if we care about the corner algebra. In this case, the function paired with $Z$ localises on the corner if the charges we care about are observables. Finally, the mixed sector is
\begin{equation}
    \begin{aligned}
       \{| \Phi_{\bar i} , X_{\bar a} |\} &= g_{\bar i\bar a}^{\bar k} \Phi_{\bar k}+\tilde g_{\bar i\bar a}^{\bar c} X_{\bar c}  + v_{\bar i \bar a} \\
       &+ M^{cd} 
       (g_{\bar i c}^{\bar k} \Phi_{\bar k}+\tilde g_{\bar i c}^{\bar e} X_{\bar e}  )
       (
       h_{\bar a d}^{\bar l} \Phi_{\bar l}+ \tilde h_{\bar a d}^{\bar f} X_{\bar f}  + Z_{\bar a d}
       )
    \end{aligned}
\end{equation}
Overall, we see that the actual on-shell structure of the algebra of generators is involved and, strikingly, takes us out of the linear span of $\Phi,X$. Of course, this need not always be the case, and one can certainly engineer examples in which the resulting algebra of functions is perfectly closed.

\paragraph{Example: Yang-Mills}\,\\
A simple example to apply this to is Yang-Mills theory, equipped with a Coulomb gauge fixing. The variables in question are $(A,E)$, two Lie algebra-valued 1-forms canoically conjugated to each other in the symplectic form
\begin{equation}
    \Omega = \int_\s \ast \delta E\wedge \delta A
\end{equation}
with the spatial Hodge star $\ast$. We can define two families of functions\footnote{In usual spherical coordinates their on-shell values correspond to $E^r$ and $A^r$ respectively. Note that within any 4D solution, $E^r = \p^t A^r - \p^r A^t$, so the two form a `canonical pair' on the timelike boundary: variable and its time derivative. }
\begin{equation}
    C_\alpha = \int_\s \ast E\wedge d_A\alpha\qquad L_\beta = \int_\s \ast A \wedge d_A \beta
\end{equation}
which impose, in the usual logic as constraints, the Yang-Mills/Gauss constraint $d_A \ast E=0$ and a covariant Coulomb gauge\footnote{On a phase space extended by the Lagrange multiplier $A_t$, it is possible to interpret this generator as being associated with the transformation $(E,A_t)\mapsto(E+ \nabla f, A_t - \p_t f)$, which is a redundancy when $E=\p_t A - \nabla A_t$, so when the fields are extensible into solutions of the equations of motion. In this way, the constraint should really read $d_A \ast_3 A = \pi_{A_t}$, but this does not change the discussion.} $d_A\ast A =0$. 
The Yang-Mills generator satisfies the simple algebra
\begin{equation}
    \{C_\alpha, C_{\bar\alpha}\}=C_{[\alpha,\bar\alpha]}
\end{equation}
which suggests a certain type of corner algebra again, if we do not gauge fix. These two constraints are by design second class, meaning
\begin{equation}
    \{C_\alpha, L_\beta\} = \int_\s \ast d_A\alpha\wedge d_A \beta = \int_\s - \alpha \, d_A\ast d_A\beta + \oint_{\p\s} \alpha \ast d_A \beta
\end{equation}
does not vanish generically. It does, for the subset of parameters satisfying the covariant Laplace equation $d_A\ast d_A \beta=0$; these are residual gauge freedoms, but are \textit{absent} for pure bulk-supported parameters $\alpha,\beta$ as these Laplace-type equations do not usually have nontrivial solutions when their boundary value vanishes. So, among the constraints with bulk parameter $\dot \alpha$, this bulk term is never zero, and the constraints are fully second class. As for the remaining functions with nonzero boundary parameter, we can single out a representative subfamily with parameters $\alpha_h$ which are harmonic in the bulk. All functions with the same boundary parameter will differ by a constraint and thus represent the same function on the on-shell phase space. 
\begin{equation}
    \{C_{\alpha_h+\dot\alpha}, L_{\beta_h+\dot\beta}\} = -\int_\s  \dot\alpha \, d_A\ast d_A\dot \beta + \oint_{\p\s} \alpha_h \ast d_A \beta_h
\end{equation}
We thus have a clean separation between constraint and charges, and can construct the Dirac bracket. (Alternatively, we can forget about the bulk-supported $\dot\beta$ and use only $C_{\dot\alpha}$ as first-class constraints.)
For this, select a basis of the $\dot\alpha, \dot\beta\rightarrow \dot f_a$, i.e, the eigenbasis of the covariant Laplacian, with eigenvalues $-\lambda_a$, normalised appropriately. Then, the second class matrix is the standard symplectic matrix, with the eigenvalues $\lambda_a\geq \lambda_{min}>0$ placed appropriately,
\begin{equation}
    Z_{ab} = \{C_{f_a}, L_{f_b}\} =\delta_{ab} \lambda_a.
\end{equation}
Invertibility is a given, and so we can find the Dirac bracket directly. We then calculate the resulting algebra of charges. We only care about the $\alpha_h$ parameters, for which we get
\begin{equation}
   \{|C_{\alpha_h}, L_{\beta_h}|\}
   =
    \oint_{\p\s} \alpha_h \ast d_A \beta_h
\end{equation}
with \textit{no} correction terms, because we have separated the constraints from the charges in such a way that 
\begin{equation}
    \{C_{\alpha_h}, L_{\dot\beta}\}=0=\{C_{\dot\alpha}, L_{\beta_h}\}
\end{equation}
i.e. they are Poisson commuting. Similarly, $\{C_\alpha, C_{\bar\alpha}\}=C_{[\alpha,\bar\alpha]}$ is unchanged, just like the algebra of the $L$, which commute. So, in this example, we can manage to actually avoid all modifications due to the Dirac bracket for the charges of interest. This leads to a corner algebra 
\begin{equation}
    \begin{aligned}
        \{\ast E\alpha ,\ast E\bar\alpha \}_S &= \ast E[\alpha,\bar\alpha]\\
        \{\ast E\alpha ,\ast A\beta \}_S &= \alpha \ast d_A \beta = \alpha D^A_r \beta \, d^2x \\
        \{\ast A\alpha ,\ast A\beta \}_S &= 0
    \end{aligned}
\end{equation}
Or in other words schematically $\{E^r,E^r\} = E^r$, $\{E^r,A^r\} = A^r + \p^r$, 
which does, in fact, give a good corner algebra which includes as a subalgebra the one of $\ast E$ that could be foretold even without the gauge fixing $L$. In fact, this is perfectly analogous to the algebra \eqref{eq:3DGravPoissonAlg} found in 3D gravity: If taken over all valued of the radial parameter $r$, it is a Kac-Moody algebra of functions $\rbb \mapsto \gfrak\ltimes \gfrak^\ast $, in other words a Loop algebra (really, a `line' algebra) of the trivial Drinfel'd double over $\gfrak$, the Lie algebra of the Yang-Mills theory. The angular direction of the 3D gravity example is here replaced by the radial direction $r$, and so in principle there is a radial line algebra per point on $\p\s$. This example shows that it is possible to make sense of second-class constrained systems in terms of the corner algebra logic, if enough technical control is available. 

\paragraph{Reprise: corner algebra under assumptions}\,\\
Say now we assume a further split of the families $\Phi,X$ similar to the one we made above, i.e. we can cleanly separate into charges $(Q_\alpha,P_\mu) $ and constraints $(\phi_i,\chi_a)$, such that the charges Poisson commute with the constraints. More precisely, we will ask for the sensible requirement that the charges are observables: They should weakly Poisson commute with the constraints. It is always possible to achieve this with the $\chi_a$ via suitable modification of the functions under consideration without affecting the on-shell value. Therefore, we require that the algebra
\begin{equation}
    \{Q_\alpha , \phi_{ j} \} =f_{\alpha  j}^{\bar k} \Phi_{\bar k} + \tilde f_{\alpha  j}^{\bar c}X_{\bar c} 
\end{equation}
\begin{equation}
    \{Q_\alpha , \chi_a \} =g_{\alpha  a}^{\bar k} \Phi_{\bar k}+\tilde g_{\alpha  a}^{\bar c} X_{\bar c}  
\end{equation}
\begin{equation}
    \{\phi_{i} , P_\mu \} =g_{ i\mu}^{\bar k} \Phi_{\bar k}+\tilde g_{i \mu}^{\bar c} X_{\bar c}  
\end{equation}
\begin{equation}
    \{P_\mu , \chi_{ b} \} = h_{\mu  b}^{\bar k} \Phi_{\bar k}+ \tilde h_{\mu  b}^{\bar c} X_{\bar c}  + Z_{\alpha  b} 
\end{equation}
reduces to
\begin{equation}
    \{Q_\alpha , \phi_{ j} \} =f_{\alpha  j}^{ k} \phi_{ k} + \tilde f_{\alpha  j}^{c}\chi_{ c} 
\end{equation}
\begin{equation}
    \{\phi_{i} , P_\mu \} =g_{ i\mu}^{k} \phi_{k}+\tilde g_{i \mu}^{c} \chi_{c}  
\end{equation}
\begin{equation}
    \{Q_\alpha , \chi_a \} =g_{\alpha  a}^{k} \phi_{k}+\tilde g_{\alpha  a}^{c} \chi_{c}  
\end{equation}
\begin{equation}
    \{P_\mu , \chi_{ b} \} = h_{\mu  b}^{k} \phi_{k}+ \tilde h_{\mu  b}^{c} \chi_{c} 
\end{equation}
for the second class term $Z_{\alpha  b} $ to disappear, we may need to modify the charge $P_\mu$ by terms involving the second class constraints. The 3rd and 4th row can always be achieved by such modifications; the first and second are nontrivial requirements on the way we select $Q,P$.\\
The assumption we make boils down to the statement that the mixed structure functions between charges and constraints do not produce charges, i.e. $f_{\bar i  j}^{\bar k} =f_{\bar i  j}^{ k} $, together with $Z_{\bar a  b}=0$. This is the statement that constraints, among all the functions, do not just form a subalgebra, but also a Lie algebra ideal. In the Dirac brackets from before, these conditions have the consequence that on-shell, the brackets reduce to the kinematical Poisson brackets:
\begin{equation}
\begin{aligned}
        \{|\Phi_{\bar i} , \Phi_{\bar j} |\} &\approx f_{\bar i\bar j}^{\bar k} \Phi_{\bar k} + \tilde f_{\bar i\bar j}^{\bar c}X_{\bar c} + u_{\bar i \bar j}\\
    \{| X_{\bar a} , X_{\bar b} |\} &\approx h_{\bar a\bar b}^{\bar k} \Phi_{\bar k}+ \tilde h_{\bar a\bar b}^{\bar c} X_{\bar c}+ Z_{\bar a \bar b} + w_{\bar a \bar b}\\
    %%%
    %%%
    \{| \Phi_{\bar i} , X_{\bar a} |\} &\approx g_{\bar i\bar a}^{\bar k} \Phi_{\bar k}+\tilde g_{\bar i\bar a}^{\bar c} X_{\bar c}  + v_{\bar i \bar a} 
\end{aligned}
\end{equation}
This is because all the correction terms are of the form $\{F,\chi_a\}$, and thus when these brackets weakly vanish, the Dirac bracket agrees on-shell with the kinematical Poisson bracket.

\bibliography{MyLibrary}

\end{document}